\documentstyle[12pt,psfig]{report}

 
\begin{document}
\addtolength{\baselineskip}{0.5\baselineskip}
\begin{titlepage}
\begin{center}
{\Large {\textbf {STUDY OF DISSIPATIVE DYNAMICS IN FISSION OF HOT
NUCLEI USING\\ LANGEVIN EQUATION}}}
 \vskip 2.2in
  \textbf{ THESIS SUBMITTED TO\\
 JADAVPUR UNIVERSITY\\
 FOR THE DEGREE OF\\
 DOCTOR OF PHILOSOPHY IN SCIENCE\\
 (PHYSICS)}

\vskip 2.2in \large {by\\
\vskip 0.1in
 \textbf{
GARGI CHAUDHURI\\
\vskip 0.2in
\textit{Variable Energy Cyclotron Centre\\
Department of Atomic Energy\\
1/AF Bidhannagar, Kolkata-700 064}}\\
\vskip 0.1in \textbf{ July 2004 }}
\end{center}
\end{titlepage}

\setcounter{page}{0} \pagenumbering{roman}
\newpage
{~}\\\\\\\\\\\\\\\\\\\\\\
\begin{center} {\Large{ \emph{To the memory of my uncle\\ \textbf
{Bimal kumar Chaudhuri}}}}
\end{center} 




\newpage
\centerline{\large{\bf CERTIFICATE FROM THE SUPERVISOR}}
 \vskip 0.5in

\noindent This is to certify that the thesis entitled
{\large{\bf`` Study of dissipative dynamics in fission of hot
nuclei using Langevin equation"}} submitted by {\large{\bf Smt.
Gargi Chaudhuri}} who got her name registered on 16.02.2001 for
the award of {\large{\bf Ph.D.(Science) degree}} of {\large{\bf
Jadavpur University}}, is absolutely based upon her own work under
the supervision of {\large{\bf Dr. Santanu Pal, Variable Energy
Cyclotron Centre}}, and that neither this thesis nor any part of
it has been submitted for any degree/diploma or any other academic
award anywhere before. \vskip 2.5in

\hspace{4cm} {\bf (Signature of the Supervisor \& date with
official seal)}

\newpage
\centerline{\Large{\bf Acknowledgements}}
 \vskip 0.4in

{\it {It gives me great pleasure to acknowledge my indebtness to
Dr. Santanu Pal, my thesis supervisor, for his invaluable guidance
and encouragement ever since I took up theoretical physics as my
career. He has shown me the path and lighted it for me through his
useful suggestions, thorough discussions and fruitful criticism,
thus helping me to contribute a tiny bit to this vast sea of
nuclear physics. The completion of this thesis owes very much to
his support and his keen interest and involvement in the work.\par
I take this opportunity to express my sincere gratitude to Prof.
Bikash Sinha, Director, Variable Energy Cyclotron Centre(VECC) and
Dr. Jadu Nath De, former Head, Physics Group, VECC, for being
instrumental in my joining the theoretical physics division of
this centre. I owe a lot to Dr. Jadu Nath De for being extremely
caring and motivating throughout the period of this work. I am
grateful to our Director as well to Dr. Dinesh Kumar Srivastava,
Head, Physics Group, VECC, for providing the congenial atmosphere
and full fledged facility which helped me immensely during my
work. I remember with gratitude the valuable discussions with Dr.
Asish Kumar Dhara which inspired me to probe deeper into the
subject.\par I remember with deep respect my teachers of Jadavpur
University as well as Indian Institute of Science, Bangalore, who
have inspired me to enjoy Physics and share the mystery of nature
with all other researchers like me. That I am still continuing
with Physics owes a great deal to all of them.\par
 I am
thankful to the Computer Division of VECC for providing advanced
computing facilities which helped a lot in successful completion
of my work. I also thank the members of Physics Group office as
well as the Director's office for their cooperation at various
stages. Last but not the least, I should thank all the members of
VECC library for providing the necessary help throughout.\par I
remember with  great pleasure all my friends in VECC, both past
and present, whose company and friendship have refreshed and
energized me during my research tenure. I would like to make
special mention of  Dr. Ranjana Goswami, Mr. Partha Ghosh, Dr.
Tapas Sil, and Dr. Debasish Bhowmick all of whom shared with me my
moments of joy and sorrow. I convey my sincere thanks to all my
colleagues in VECC who have made my thesis tenure memorable.\par
At this juncture, I am very much reminded of Somshubhro who has
been my most intimate friend  ever since I chose Physics as my
career in B.Sc. My interest to continue in Physics owes a lot to
the discussions we had during our university days to clear the
doubts regarding different aspects of this great subject. I
remember with pride his support and concern for me both as my
friend and as my husband and his constant advices to be always
hardworking and sincere in my effort. I remember with deep
affection the support and company of my brother Balarko and the
technical suggestions given by him while writing  this thesis.\par
I  fondly remember the cheerful face of my little daughter Jhelum
who have been my constant source of energy and delight. The
completion of this thesis have somewhat deprived her of the due
attention and time for which I feel guilty. I am really at a loss
of words to express my obligation to my parents who have supported
me all through, and have taken up the greater chunk of my
responsibilities allowing me enough opportunity to proceed
smoothly with my work.  This thesis owes most to them. At this
moment I  very much remember with gratitude that both my father
and my husband always insisted on giving my research work the
first priority and encouraged me to produce my best. \par Today on
the verge of climbing a vital step of my career, I gratefully
remember the blessings and affection of all my well wishers
(specially my aunt) since my childhood which have contributed in
shaping up my life and my career. I also gratefully acknowledge
the encouraging words of my parent-in-laws in my journey towards
the goal.
\par On completion of this thesis, when my hope gets
realized, I could visualize the extreme delight of my departed
uncle who cherished the dream more than me and would have been the
happiest person on this little achievement. I very much remember
his unlimited love, his deep concern, his constant encouragement
throughout my academic career and the moral boosts I received from
him on every little success that I achieved. I consider myself
very fortunate to have him as my greatest well wisher.  I feel
extremely sad that he is no more by my side to share my moments of
joy. I cannot put this thesis in his hands today; I can only
dedicate it to his sacred memory.}}

\newpage
\centerline{\large{\bf PREFACE}}
 \vskip 0.5in

The advent of high energy heavy ion beams in various energy
  ranges has led to the discovery
  of a number of significant  nuclear phenomena. The fission of highly excited
 compound nuclei formed in heavy ion induced fusion reactions has
 emerged as a topic of considerable interest in the recent years.
 Multiplicity measurements of light particles and photons strongly
 suggest that fission is a much slower process for hot nuclei than
 that determined from the statistical model of Bohr and Wheeler
 based on phase space arguments.
  This led to the introduction of dynamical
effects, specially the concept of nuclear friction in the
description of fission of hot nuclei. Dissipative dynamical models
based on the Langevin equation were developed and were applied
successfully for fission dynamics of highly excited heavy nuclei.
However, Wall Friction(WF), the standard version of nuclear
friction when incorporated in the Langevin dynamical model was not
able to reproduce simultaneously experimental data for both
prescission neutron multiplicity ($n_{pre}$) and fission
probability ($p_f$). Consequently, an empirical reduction in the
strength of the wall friction was found necessary to reproduce the
experimental numbers by many workers. Interestingly, a
modification of the wall friction was proposed recently where the
reduction was achieved microscopically. This modified version is
known as the chaos weighted wall friction(CWWF) which takes into
account non-integrability of single particle motion. The work in
my thesis aims at using this strongly shape dependent version of
friction (CWWF) in the Langevin dynamical model coupled with
particle and gamma evaporation in order to verify to what extent
it can account for the experimental data of fission of hot nuclei.
The present endeavour is an effort to obtain a clear physical
picture of nuclear dissipation which in turn will help in solving
many open problems related to collective motion, and in
particular, nuclear fission. An important application of current
interest could be the theoretical prediction of survival
probability of superheavy elements against fission which depends
sensitively on nuclear dissipation on the fission path.
\newpage
 \underline{\large{\bf{\it Outline of the thesis:}}}\\

 The work to be presented in this thesis is divided into seven
chapters and seven appendices. Chapter 1 gives an overview of the
subject, where the relevant literature is reviewed briefly. In
Chapter 2 our model for Langevin dynamics of nuclear fission will
be described in details. The origin of  the different inputs used
in our calculation, namely, potential, inertia and level density
parameter will be discussed. The chaos weighted wall friction
which is used for nuclear dissipation in the dynamics and will be
tested for the first time will be described  elaborately in this
chapter. Chapter 3 contains the procedure of solving the Langevin
equation for strongly shape dependent friction in order to
calculate fission width which are subsequently utilized in further
calculations. The fission widths are calculated using both the
wall friction and the chaos weighted wall friction. In Chapter 4,
the different steps of the combined dynamical and statistical
model which couples particle and $\gamma$ evaporation with
Langevin dynamics, is described. The excitation functions of the
prescission neutron multiplicity and the fission probability
calculated from the model using both the versions of the friction
are compared with the experimental data for a number of nuclei in
this chapter. Chapter 5 is devoted to the calculation of the
evaporation residue cross section excitation function as a probe
for nuclear friction which is compared with experimental data.
Chapter 6 discusses in details the effects of transients in
nuclear fission on prescission neutron multiplicity. Chapter 7
contains the thesis summary, conclusions and future directions of
our work. Some details of the formulation and the computation are
given in the different appendices. The cumulative references for
all the chapters are given at the end. \\

 Based on the work
presented in this thesis, following papers have been published in
refereed international journals. The listings at {\it
http://arXiv.org} are given at the end of each reference.
\newpage
\underline{\large{\bf  List of Publications: }}\\

\begin{enumerate}
\item Fission widths of hot nuclei from Langevin dynamics,\\
 \textit{Gargi Chaudhuri} and Santanu Pal,   Phys.Rev. {\bf C63}
 (2001) 064603;  nucl-th/0101037

\item Prescission neutron multiplicity and fission probability
from
Langevin dynamics of nuclear fission,\\
 \textit{Gargi Chaudhuri}
and Santanu Pal,   Phys.Rev. {\bf C65}
 (2002) 054612;  nucl-th/0105010

\item Effect of transients in nuclear fission on multiplicity of
prescission neutrons,\\
 \textit{Gargi Chaudhuri} and Santanu Pal,   Eur.Phys.J. {\bf A14}
  (2002) 287-294;  nucl-th/0204052

\item Evaporation residue cross-sections as a probe for nuclear
dissipation in the fission channel of a hot rotating nucleus,\\
\textit{Gargi Chaudhuri} and Santanu Pal,   Eur.Phys.J. {\bf A18}
(2003) 9-15;  nucl-th/0306003
\end{enumerate}
\vspace{2.2cm}

\hspace{11cm} {\large{\bf Gargi Chaudhuri}}\\
\indent \indent {\large{\bf Variable Energy Cyclotron Centre}},\\
\indent \indent {\bf 1/AF, Bidhannagar, Kolkata, India.}

\newpage
\tableofcontents
\newpage
\setcounter{page}{0}
\pagenumbering{arabic}
\pagestyle{myheadings}
\markright{}

\chapter{Introduction}

\par The study of large scale nuclear dynamics (e.g. deep inelastic
collisions and fission) initiated by energetic heavy ion beams
above the Coulomb barrier is an active area of research in nuclear
physics and presents a number of theoretical challenges. In
particular, one of the exciting aspects
 in such studies is that in these energetic nuclear reactions,
 concepts of non-equilibrium statistical
physics, such as dissipation or thermalization, are extended to
small and dense Fermion systems.  The energy scale under
consideration here ranges from that above the Coulomb barrier and
extends up to the Fermi energy domain($\sim$ 30 MeV/nucleon). The
availability of heavy-ion beams in various energy ranges and the
emergence of exclusive measurements in different experiments
(which provide more insight into the nuclear dynamics than the
inclusive experiments) motivated the development of theoretical
approaches such as transport theories and the time dependent
Hartee-Fock(TDHF) theory. In particular, the transport
theories\cite{Weid1,Noren2} were developed using the general
framework of nonequilibrium statistical physics, the kinetic
theory and the stochastic methods. In such descriptions
dissipation, i.e., the irreversible flow of energy between the
collective and intrinsic degrees of freedom of the system and the
associated fluctuations play very important role. An estimate of
the nuclear dissipation or friction was first made from the
analysis of deep inelastic collision and heavy ion induced fusion
experimental data with classical trajectory
models\cite{De,Gross1}. However the results turned out to be
widely varying even by orders of magnitude. The advent of
exclusive measurements and sophisticated experimental techniques
as well as the development of improved theoretical models
contributed in narrowing down the range of magnitude of nuclear
friction to a great extent. A comparison of the experimental
results with the phenomenological Langevin or Fokker-Planck models
has allowed to extract the key parameters entering these
description, namely the nuclear friction. On the other hand, the
microscopic derivation of the nuclear friction coefficient has
attracted a large amount of theoretical effort. Over the years,
different microscopic as well as phenomenological attempts have
been made to derive the nuclear friction coefficient but any
unambiguous prescription for nuclear friction is yet to be
achieved. In this thesis, we shall be concerned with a detailed
study of the fission dynamics of highly excited nuclei formed in
heavy-ion collisions using a theoretical model of nuclear
dissipation(a modified version of the wall friction). Our main aim
in this work is to test this theoretical model as a candidate for
nuclear friction without any tunable parameter. Such a nuclear
friction has immense applicability in predicting survival
probability of superheavy nuclei against nuclear fission and
production cross section of fission fragments in ISOL-type
radioactive ion beam facilities.
\section{Overview}
\subsection{General}
 In order to present an overview of the advances of the many body
 aspects of nuclear dynamics, it is illuminating to begin with
one of the most fundamental contributions to this field credited
to Niels Bohr, whose
 work had a profound impact on the post-1930s development of
nuclear physics. The pioneering contribution to this field is the
``compound nucleus theory'' of Bohr which has a fascinating appeal
in many aspects of nuclear dynamics. The basic idea of the
compound  nucleus model is a strong and intimate coupling of all
the nucleons with each other. This subsequently led to the
development of the  nuclear liquid drop model by Bohr and
Kalckar\cite{Kal}. It was this model which enabled Meitner and
Frisch\cite{Meit} to explain why a nucleus may undergo fission,
and this led Bohr and Wheeler\cite{Wheel} to develop their
celebrated formula for a first quantitative description of the
decay rate of nuclear fission. This work also laid the foundation
for the concept of nuclear collective motion. The standard
analysis of induced nuclear fission is based on the Bohr-Wheeler
formula for the fission width $\Gamma_{BW}$ which depends on the
ratio of the phase spaces available at the saddle point to that at
the ground state and is given by the following expression.
 \begin{equation}
\Gamma_{BW}=\frac{\hbar}{2\pi\rho(E^{*})}\int_0^{E^{*}-E_f}d\varepsilon
\rho^{*}(E^*-E_f-\varepsilon),\label{Bohr0}
\end{equation}
\noindent where $E^{*}$ is the excitation energy, $E_f$ is the
height of the fission barrier, $\varepsilon$ is the kinetic
energy, and $\rho^{*}$ is the density of levels of the compound
nucleus at the saddle point which arises from excitations of the
intrinsic degrees of freedom only and $\rho$ denote the level
densities of the fissioning nucleus at the ground state. A
simplified expression is obtained with the Fermi gas model for the
level densities which is given by $\rho(E) \sim e^{2\sqrt{aE}}$,
the constant temperature approximation $E = aT^{2}$ and the
condition $E^*\gg E_f$
\begin{equation}
\Gamma_{BW}=\frac{\hbar T}{2\pi}e^{-E_f/T}.\label{Bohr}
\end{equation}
\noindent where $a$ is the usual level density parameter.
 It yields the fission width as a
function of the fission barrier height ${E_f}$ and the nuclear
temperature T. This description of nuclear fission does not invoke
any dynamical features and hence is independent of the nuclear
friction. It is however interesting to note that it is mentioned
in the addendum of Bohr's paper\cite{Kal} that ``non-viscous fluid
can hardly be maintained in view of the close coupling between the
motions of the individual nuclear particles.''
Kramers\cite{Kramers} took up this point and derived a formula for
the fission decay rate in which a correction factor(K) appeared to
the Bohr-Wheeler expression, which was governed by nuclear
friction. Kramers formula for fission width
($\Gamma_K=K\Gamma_{BW}$) is related to that of Bohr-Wheeler
($\Gamma_{BW}$) by the factor $K= [\sqrt{{\tilde{\beta}}^{2} + 1}
-\tilde{\beta}]$, where $\tilde{\beta}$ is proportional to the
nuclear friction coefficient $\eta$. Kramers pictured the
collective motion as a transport process in collective phase
space. He dealt with the general problem of Brownian motion in a
heat bath in the presence of a potential barrier. The importance
of Kramers idea was realized much later in nuclear physics with
the advent of heavy ion accelerators with which nuclear systems
could be excited to much higher energies and it was thus realized
that the energy stored in the collective motion can be dissipated.
\par
 Different microscopic theories were developed to describe nuclear
collective motion.
  This began in the early fifties
with the unified model of Bohr and Mottelson \cite{Mottel}. Small
amplitude collective motion like the normal vibration modes of a
nucleus were explained by the random phase
approximation(RPA)\cite{Ring}. However, the RPA is typically a
small amplitude approximation, and cannot describe collective
processes during which nuclear wave function undergoes important
alterations, such as fission, fusion, heavy ion reactions etc.
Microscopic theories  for large amplitude motion of many body
fermion systems are usually based on a mean field description
known as time dependent Hartee-Fock method(TDHF) developed in the
1970s, and its variants like ATDHF \cite{Ring}. The TDHF equation
is a nonlinear equation for the one-body density matrix $\rho$ and
a first-order differential equation in time, which in its simplest
form reads like
\begin{equation}
i\hbar\dot{\rho}=[h,\rho]
\end{equation}
\noindent with $h=t+\Gamma$. $t$ is the kinetic energy, and
$\Gamma$ is the self consistent mean field which depends on the
density of the nucleus.
 It was realized that as
the excitation energies become higher  and is around the regime of
Fermi energy/nucleon, the nucleon-nucleon correlations  become
dominant over the effects of the averaged forces and it would be
thus necessary to look beyond mean field description. Application
of TDHF requires a large mean free path and hence is a good
approximation for the low energy heavy ion collisions. However, at
high collision energies, the mean free path is strongly reduced
due to the large excitation energies involved in the process. Thus
the inclusion of residual two-body collisions in a self consistent
mean-field theory (generalized or extended TDHF) is a natural step
of a more realistic description of heavy-ion collisions at high
excitation energies. However, because of numerical difficulties,
realistic applications of these approaches seem to be difficult
even with the fastest available computers. This has been tried in
various versions but it had soon become clear that all one is able
to do numerically is to solve such equations in a semi-classical
limit, e.g., in the version of the Boltzman-Uehling-Uhlenbeck(BUU)
or Landau-Vlasov equation\cite{Dasgupta}. The BUU equation is as
follows
\begin{eqnarray}
 \frac{\partial f}{\partial
t}+\vec{v}\cdot{\vec{\nabla}}_{r}f-{\vec{\nabla}}_{r}U\cdot
{\vec{\nabla}}_{p}f=-\frac{1}{{(2\pi)}^6}\int
d^3p_2d^3p_{2^\prime}
d\Omega\frac{d\sigma}{d\Omega}v_{12}\nonumber\\
&\hspace{-5.5cm}\times\{[f_1f_2(1-f_{1^\prime})
(1-f_{2^\prime})-f_{1^\prime}f_{2^\prime}(1-f_1)(1-f_2)]\nonumber\\
&\hspace{-8.7cm}\times{(2\pi)}^3{\delta}^3
(\vec{p_1}+\vec{p_2}-\vec{p_{1^\prime}}-\vec{p_{2^\prime}})\}.
\end{eqnarray}
\noindent where the right hand side is the collision integral
including the Pauli blocking and when it is set equal to zero, one
obtains the Vlasov equation. \par The macroscopic variables behind
TDHF are the ones which relate to matrix elements of the one-body
density operator. A much simpler version is given if one
parameterizes the time evolution of the mean field by time
dependent shapes and provided it is possible to complement such a
description with the dynamics of a conjugate momentum, one may
view this motion as a transport process in collective phase space.
This gave rise to the revival of the theoretical studies based on
the original works of Kramers who viewed the collective dynamics
on similar lines. There have been (a) applications of linear
response theory to formulate a theoretical description appropriate
for heavy-ion collisions by Hofmann and
Siemens\cite{Siemens1,Siemens2}, (b) the applications of the
methods of spectral distributions or the random matrix model by
the groups of N\"{o}renberg \cite{Noren1} and
Weidenm\"{u}ller\cite{Weid1,Noren1,Nemes} and (c) the suggestion
of Norenberg to model relative motion of two heavy ions by way of
a ``dissipative, diabatic dynamics(DDD)'' \cite{Noren2}. The
linear response theory aims at describing large-scale collective
motion on the basis of a locally harmonic approximation. This
approximation is exploited to define propagators and to derive
equations for their dynamics in small areas of phase space and
over time lapses which are small on a macroscopic scale. In the
work of Weidenm\"{u}ller {\it et al.}\cite{Nemes}, the statistical
properties of matrix elements which couple the collective degrees
of nuclear motion with the intrinsic degrees of freedom, are
evaluated in an adiabatic approximation. A random-matrix model is
used for the residual interaction. The basic idea of
DDD\cite{Nor1,Nor2} is that the energy dissipation in slow
collective nuclear motion is viewed as a combined effect of a
diabatic production of particle-hole excitations, leading to a
conservative storage of collective energy, and a subsequent
equilibration due to residual two-body collisions. The effective
equation of motion for the collective degree of freedom contains a
time retardation in the dissipative term and allows for a
simultaneous description of two different attitudes of nuclear
matter. The elastic response of heavy nuclei for `fast' collective
motion switches over to pure friction for very `slow' collective
motion. A first application of the diabatic dynamical approach is
made for the quadrupole motion within a diabatic deformed harmonic
oscillator basis. \par The concept of friction and the associated
statistical fluctuations play an important role in many areas of
physics, chemistry and biology, when one is dealing with transport
processes. The equations which are usually applied are master
equations, Fokker-Planck equations and Langevin equations. The
books of Risken \cite{Risken}, Van Kampen\cite{van}, and the
article of Hanggi {\it et al.}\cite{Talkner} provide complete
mathematical background of the subject. With the discovery of
deep-inelastic processes in heavy-ion collisions, the concept of
friction was introduced in the description of complex nuclear
reactions, where it was impossible to follow all involved degrees
of freedom explicitly. First, models with classical trajectories
which are determined by conservative and frictional forces have
been developed\cite{Gross1}. Then, Norenberg \cite{Noren3}
introduced a Fokker-Planck equation for the description of charge
transfer as a diffusive process in deep-inelastic heavy-ion
collisions. Subsequently, multi-dimensional Fokker-Planck
equations were applied by many authors in order to describe
deep-inelastic differential cross sections with respect to the
scattering angle, energy loss, and to mass and charge transfer
variables.\par It has always been a challenge to extend Kramers'
result of the decay of a metastable system to the quantal regime.
It was only in the 80's that one began to understand how to
incorporate quantum effects and among the vast literature
available on ``dissipative tunneling'' one may refer to
\cite{Talkner,Hanggi,Leggett}. Common to all these approaches is
the application of the technique of path integrals for imaginary
time propagation. For nuclear fission and nuclear
multifragmentation, a formulation with real time propagation is
much more appropriate. This has been achieved by a suitable
application of linear response theory. The transport equation
obtained is similar in structure to that of Kramers', with only
the diffusive terms being modified to take account of the quantum
effects\cite{Hofmann3}. The modification of Kramers' equation
using quantal diffusion coefficients is also studied in Ref.
\cite{Ingold}. In Ref. \cite{Thoma}, the decay of a metastable
system is described by extending Kramers' method to the quantal
regime. It is seen that the quantum corrections to the decay rate
would lead to an increase of the later \cite{Hofmann4}. This
effect is significant for temperatures of the order of 1 MeV or
less. \vspace{2cm}
\subsection{Nuclear Fission}
 Nuclear fission is one of the earliest and most
thoroughly studied of all nuclear phenomena. Fission is the most
prominent and classic example of `slow' large-scale collective
motion in nuclear physics.
  The standard
statistical model of Bohr and Wheeler \cite{Wheel} was sufficient
for a long time to describe the observed effects of nuclear
fission till the availability of high energy heavy-ion beams. A
spate of experimental data from heavy-ion induced reaction
studies, carried out in the last two decades have resulted in the
interesting observation of unexpectedly large pre-scission yields
of charged particles \cite{Vaz}, neutrons \cite{Hinde1}, and giant
dipole resonance(GDR) decay $\gamma$ rays \cite{Chakra} from the
compound system before fission. The standard statistical model
 was found to underestimate the prescission yields of particles and
$\gamma$-rays, the discrepancy being large at excitation energies
greater than 50 MeV.  The underestimation of prescission particles
at high excitation energies by the statistical model led one to
think that sufficient time is not available for the particles to
evaporate prior to fission. In other words, the fission width
calculated on the basis of phase space arguments is overestimated
in statistical model at high excitation energies. At lower
excitation energies the standard statistical model calculations
hold good because in this energy regime, the particle multiplicity
has negligible dependence on the fission width and hence the
simplified arguments used in statistical model was sufficient to
reproduce the particle multiplicities. However, with the increase
in the excitation energies, fission width increases and becomes
comparable to the particle emission widths and the dependence of
the particle multiplicities on fission width becomes significant.
This realization motivated more rigorous calculation of the
fission width invoking dynamical effects at higher excitation
energies and led one to look beyond the standard statistical
model. The experimental data revealed that fission of hot nuclei
is a slower process than that predicted by the statistical model.
The need for a slowing down mechanism naturally suggests one to
consider the effects of nuclear friction on fission lifetime and
this inspired the use of a transport description of fission since
it includes the dynamical features not contained in the
statistical model. This gave rise to the revival of the
theoretical studies based on the original works of Kramers who
considered induced nuclear fission as a transport process of the
fission degree of freedom over the fission barrier as a
consequence of thermal fluctuations.
 Dissipative dynamical models for
fission of hot nuclei based on the transport theory were
subsequently developed.
\section{Dissipative dynamical model of nuclear fission}
\subsection{ Introduction}
The goal of any transport theory is to reduce the description of
the time evolution of a complex system to that of a small subset
of its degrees of freedom. The dynamics of the residual set is not
 explicitly considered though their effect is taken into account
 in some average sense.
 Often the
first class of variables is referred to as the collective or
``macroscopic'' ones, whereas the rest  of the degrees of freedom
is referred to as the ``intrinsic'' system. The notion macroscopic
indicates that in many cases these variables are chosen to
represent quantities whose dynamics can be visualized as a
transport of matter, total charge etc. A macroscopic description
of fission dynamics is based on the idea that the gross features
of the fissioning nucleus can be described in terms of a small
number of variables called the collective variables or the
collective degrees of freedom. At nuclear excitations which give
rise to temperatures up to a few MeV , the dominant collective
modes relevant for nuclear fission are expected to be those
involving changes in the nuclear shape,  and the coordinates of
the nuclear surface itself provide a natural set of collective
variables.
 In the  transport theory which is also referred to as the
dissipative dynamical model, the dynamics associated
 with the fission degree
of freedom(collective motion) with a large inertial mass is
considered to be similar to that of a  massive Brownian particle
floating in a viscous heat bath under the action of a potential
field. The rest of the nuclear system comprising of a large number
of intrinsic degrees of freedom (assumed to be in thermal
equilibrium) is identified with the heat bath. It is also assumed
that the impact of Brownian particle dynamics on the heat bath is
insignificant. It is of great importance, however, to understand
how the heat bath influences the dynamical macroscopic object. In
fact, the introduction of the heat bath makes the dynamics of the
Brownian particle irreversible and will exhibit fluctuations in
observable quantities.  Fluctuations arise in the theoretical
description, because attention is focussed entirely on a few
degrees of freedom (the collective variables), and the loss of
information caused by disregarding the many other degrees of
freedom manifests as sizeable fluctuations in physical
observables. In most cases the inertial mass associated with the
collective degree of freedom is large enough so that its dynamics
is governed entirely by the laws of classical physics. This
separation of the whole system into a Brownian particle and a heat
bath relies on the basic assumption that the equilibration time of
the intrinsic degrees of freedom($\tau_{equ}$) is much shorter
than the typical time scale of collective motions($\tau_{coll}$),
i.e, the time over which the collective variables change
significantly. The separation of these two time scales allows the
decomposition of the Hamiltonian into a collective part(describing
the shape degrees of freedom) and a intrinsic part (describing the
intrinsic degrees of freedom). Moreover, if one assumes that the
intrinsic motion loses memory very quickly, one can easily derive
transport equations for the collective degrees of freedom. If
$\tau_{Poincar\acute{e}}$ is the time it takes the entire system
to return to a point very close to its original position in phase
space (Poincar$\acute{e}$ recurrence time) then it should be much
greater than the time scale for collective motion so that the
collective dynamics is irreversible. Thus the time scales
governing the behavior of an equilibrating system must obey the
following inequalities for a transport description to be viable.
\begin{equation}
\tau_{equ} \ll \tau_{coll} \ll \tau_{Poincar\acute{e}}.\label{1a}
\end{equation}
The domain of applicability of transport theories has been
extensively discussed in the case of deep inelastic heavy-ion
reactions in Ref. \cite{Weid1}. Later it was found that transport
theories can also be applied for describing competitive decay of
composite nuclear systems \cite{Grange1}. The crucial parameters
in a diffusion model for fission are the nuclear friction $\eta$,
which gives the strength of the coupling between the fission and
the intrinsic degrees freedom, and the diffusion constant $D$,
related to each other by the Einstein relation. A diffusion model
is applicable to fission when the internal equilibration time
$t_{equ}$ of the heat bath is small compared to
 the characteristic time of the diffusion process itself(related to
$\eta^{-1}$), and to $\tau_f$(=$\hbar/\Gamma_{f}$) and
$\tau_n$(=$\hbar/\Gamma_{n}$), where $\Gamma_{f}$($\Gamma_{n}$)are
the fission (neutron) widths at the excitation energies under
consideration. On the basis of microscopic considerations, simple
estimates of these time scales(leading to $t_{equ}\simeq 3 \times
10^{-22}$ sec) suggest a diffusion model is applicable for
$\frac{\eta}{m} \leq 3\times 10^{21}$
 sec$^{-1}$ \cite{Weid2} ($m$ being the mass of the Brownian particle),
  and for excitation energies of 100 MeV
 or more. We shall assume that the transport(diffusion) equation will be
 applicable to nuclear fission at high excitation energies in
 the dissipative dynamical model. \par
 The description of the intrinsic modes of
excitation in terms of a heat bath has two consequences. First,
energy flows irreversibly from the collective motion into the
intrinsic excitation and manifests as a friction force in the
collective dynamics. Second, the fact that the dynamics of the
intrinsic degrees of freedom, collectively represented by the
temperature T, are uncorrelated gives rise to random features in
the coupling between the heat bath and the collective motion. As a
consequence energy is exchanged randomly in both directions in a
fine time scale, though the net flow  is into the heat bath over a
larger time scale.
 Thus, the
time development of the collective variable has a random
character. This is analogous to that of a Brownian particle which
collides with gas molecules having a Maxwellian velocity
distribution. The Brownian particle undergoes a random walk and is
slowed down but on a staggering path. The motion of a Brownian
particle in an external force field which essentially is the model
of fission dynamics considered here, can be described by two
alternative but equivalent mathematical formulations, which will
be briefly described in the following subsections.
\subsection{Fokker-Planck equation}
The Fokker-Planck equation and the Langevin equation are the two
equivalent descriptions of a Brownian particle in a heat bath. The
Fokker-Planck equation can be derived starting from the Langevin
equation\cite{Abe1}. One begins with the Liouville equation which
describes the conservation of probability, i.e., with the
continuity equation for probability,
\begin{equation}
\frac{\partial}{\partial t}f(p,t)=-\frac{\partial}{\partial
p}(\dot{p}(t)\cdot f(p,t)), \label{Lio}
\end{equation}
\noindent where $f$ is the distribution function in momentum
space. The Langevin equation describing the motion of a Brownian
particle of mass $m$ (to be described in detail in the next
subsection) in the presence of a potential $V$ and friction
coefficient $\eta$ reads as follows
\begin{equation}
{dp \over dt}= -\frac{\eta}{m}p +R(t)-\nabla V \label{Lan}
\end{equation}
Substituting for $\dot{p}$ from Eq. \ref{Lan} and integrating Eq.
\ref{Lio} between $t$ and $t+\Delta t$, ($\Delta t$ is much larger
than the time scale of the random force $R(t)$) leads to
\begin{equation}
f(p,t+\Delta t)=\left[1+\int_t^{t+\Delta
t}dt_1\Omega(p,t_1)+\int_t^{t+\Delta
t}dt_1\int_t^{t_1}dt_2\Omega(p,t_1)\Omega(p,t_2)+\dots\right]f(p,t)
\end{equation}
\noindent where
\begin{equation}
\Omega(p,t)= \frac{\partial}{\partial p}(\frac{\eta}{m}p-R(t) +
\nabla V)
\end{equation}
\noindent Taking  an average over all possible realizations of the
random force $R(t)$ and using the properties of $R(t)$ (to be
discussed in the next subsection), in the limit $\Delta t
\rightarrow 0$, it can be shown that
\begin{equation}
{\partial \over \partial t}f(r,p;t)+\frac{(p\cdot
\nabla_{r})}{m}f(r,p;t)-(\nabla_{r}V\cdot
\nabla_{p})f(r,p;t)=\nabla_{p}(\frac{\eta}{m}pP(r,p;t))+
\frac{\nabla_{p}^{2}}{2}(Df(r,p;t)).\label{1b}
\end{equation}
\noindent where $D$ is the mean square strength of the random
force. This equation is known as the Fokker-Planck equation or the
Kramers equation. The fact that it is possible to derive the
Fokker-Planck equation from the Langevin equation (using the
continuity equation or the master equation) clarifies the relation
 between the two equations and establishes their equivalence.
\par
 The
Fokker-Planck equation is a probabilistic dynamical description
and it deals with the time-evolution of the distribution function
of the Brownian particle. The probability distribution f(r,p,t)
for finding the particle at a point(r,p) in classical phase space
is obtained by solving the above Fokker-Planck equation.
Kramers(1940) applied it to the decay rate of nuclear fission. He
obtained the equilibrium solution of the above equation and
derived the quasi stationary fission width from it which is given
by the following expression.
\begin{equation}
\Gamma_K=\frac{\hbar\omega_1}{2\pi}\{{[1+{(\frac{\eta}{2m\omega_0})}^2]}^{1/2}-
\frac{\eta}{2m\omega_0}\}\cdot\exp{(-E_f/T)}.
\end{equation}
Here $\omega_0$ and $\omega_1$ are the oscillator frequencies of
the parabola osculating the nuclear potential in the first minimum
and at the saddle respectively. In the limit of small $\eta$,
$\Gamma_K$ reduces to the transition state expression given by
\begin{equation}
\Gamma_{BW}= \frac{\hbar\omega_1}{2\pi}\exp{(-E_f/T)}.\label{Kram}
\end{equation}
 \noindent The Kramers width $\Gamma_{K}$ is related
to the Bohr-Wheeler width $\Gamma_{BW}$ (Eq. \ref{Bohr}) through
the Kramers factor $K$(also called reduction factor) which is
given by $\{{[1+{(\frac{\eta}{2m\omega_0})}^2]}^{1/2}-
\frac{\eta}{2m\omega_0}\}$. In Eq. \ref{Kram}, $\Gamma_{BW}$ is
the Bohr-Wheeler width corrected for the presence of collective
vibrations\cite{VM} in the potential pocket not taken into account
in the density of levels $\rho(E^*)$ in Eq. \ref{Bohr0} (refer
section 1.1.1). The Kramers factor depends on the nuclear friction
coefficient $\eta$ and is interpreted as a restriction in phase
space around saddle point due to friction. It is thus remarkable
that the importance of friction in nuclear dynamics was
anticipated by Niels Bohr in 1939, and H. A. Kramers correctly
predicted a reduction of fission width which was experimentally
confirmed after about 50 years. Between 1940 and the beginning of
the 80s Kramers approach did not attract much attention in the
context of nuclear fission.  This happened because the simple
Bohr-Wheeler formula worked well, at least within the
uncertainties of the fission barrier height and of the level
density parameter. Forty years later, in the eighties,
Weidenmuller and his group \cite{Grange0} followed the line of
approach of Kramers and adopted the diffusion model to investigate
how the quasistationary flow over the fission barrier is attained.
Their study was motivated by the experimental findings
\cite{Gavron2} which seemed inconsistent with the Bohr-Wheeler
prediction in showing an excess of evaporated neutrons. They
succeeded in getting the time dependent solution of the two
dimensional Fokker-Planck equation after making a number of
simplifying assumptions and obtained the time dependent fission
width $\Gamma_f(t)$ by calculating the probability current through
the saddle point. Their work first showed that for finite values
of the friction coefficient $\eta$, there is a time $\tau$ which
elapses between the start of the induced fission process and the
attainment of the stationarity condition. This time $\tau$ depends
on $\eta$ and during this time fission is suppressed. The larger
the value of $\tau$, more will be the time for evaporation and
more strongly will particle and gamma evaporation compete with the
fission process. Their study first established the importance of
`transients', i.e., those processes which occur before the
quasistaionary flow over the barrier is attained. They showed that
the fission probability $P_f$ is modified compared to the
Bohr-Wheeler formula in two ways: (i) $P_f$ suffers an overall
reduction in the stationary fission rate due to friction
(reduction factor K of Kramers) (ii) the inclusion of transients
reduces $P_f$ further, particularly at higher excitation energies.
Both these effects will significantly increase the neutron
emission as demanded by the experimental data. It was also shown
that the entire fission process becomes a transient when there is
no fission barrier \cite{Grange2}. The detailed study of
transients in nuclear fission were considered in a series of
publications.  We shall discuss our own contribution to this topic
in
 chapter 6. Dynamical studies of induced fission
with the Fokker-Planck or Kramers equations have also been studied
by other groups\cite{Zhen,Delag} investigating the reduction of
the Bohr-Wheeler width by the Kramers factor as well as the
existence of the transient time. These findings stimulated refined
measurements of the multiplicities of neutron, light charged
particles and photons\cite{Hinde1,Hinde2}. Theoretical
developments were made for a proper description of the competitive
decays of particle evaporation and
fission\cite{Delag,Dietrich,Strum}, an effect which becomes
especially important when one considers the fission of hot nuclei.
Multi-dimensional Fokker-Planck equations were subsequently
applied to the description of nuclear fission\cite{Adeev8}. \par
Analytic solutions of the Fokker-Planck equations were  initially
restricted to the use of the quasi-linear method, in which the
driving terms are expanded to the lowest order and only the first
and second moments of the Fokker-Planck equation together with a
Gaussian ansatz are used to calculate the distribution function at
large times, from which the cross sections can be obtained.
However it turns out that in many cases the Gaussian ansatz is not
a good approximation. The multi-dimensional Fokker-Planck
equations for deep-inelastic collisions and induced fission  can
be solved numerically with grid methods which is an exact
procedure but turns out to be extremely difficult even with
present day computers. Modelling the same problem in terms of the
equivalent Langevin equations, and solving these equations by
Monte-Carlo sampling, is a more practicable way for obtaining more
accurate solutions than with the Gaussian ansatz. Suggestions were
made to apply Langevin equation in nuclear physics in Refs.
\cite{Rand,Rehm}. The first calculations using Langevin equation
were performed later, for deep-inelastic procesess by Barbosa {\it
et al.} \cite{Barbosa}, for fission by Abe {\it et al.}
\cite{Abe1} and for fusion by Fr\"{o}brich\cite{Frob1}. Since then
a large volume of work have been reported , which have applied
Langevin equation with the aim to describe data for deep-inelastic
heavy-ion collisions, fusion, and heavy-ion induced fission.
\subsection{Langevin equation}
 Langevin approach
which is an alternative description of the Brownian motion was
first proposed by Y. Abe \cite{Abe1} as a phenomenological
framework to describe nuclear fission dynamics.  The Fokker-Planck
equation deals with the time evolution of the distribution
function(in classical phase space) of the Brownian particle while
the Langevin equation deals directly with the time evolution of
the Brownian particle and hence is much more intuitive. The two
approaches describe different aspect of the dynamics but they are
equivalent with respect to their physical content. The motion of a
Brownian particle under the action of a external force field as
given by the one-dimensional Langevin equation (\ref{Lan}) can be
written as follows,
\begin{equation}
{dp \over dt}= F(t) + H(t)\label{1c}
\end{equation}
where $F(t)$ is the external force and $H(t)$ is given by
\begin {equation}
H(t)= -\frac{\eta}{m}p+ R(t)\label{1d}
\end{equation}
 The coupling of the collective motion with the heat bath
 is described by $H(t)$. It
has two parts; a slowly varying part which describes the average
effect of heat bath on the particle and is called the friction
force ($\frac{\eta}{m} p$), and the rapidly fluctuating part
$R(t)$ which has no precise functional dependence on t. Since it
depends on the instantaneous effects of collisions of the Brownian
particle with the molecules of the heat bath, $R(t)$ is a
random(stochastic) force with its mean value zero and with a
specific probability distribution. It is further
assumed\cite{Abe1} that $R(t)$ has an infinitely short time
correlation, i.e. it describes a Markovian process. Therefore
$R(t)$ is completely characterized by the following moments,
\begin{eqnarray}
\langle R(t) \rangle             &=& 0 ,\nonumber\\
\langle R(t)R(t^{\prime})\rangle &=&
2D\delta(t-t^{\prime}).\label{1e}
\end{eqnarray}

\noindent where $D$ is the diffusion coefficient and is related to
the friction coefficient $\eta$ (to be described later).
 It should be noted that Langevin equation is
different from ordinary differential equations as it contains a
stochastic term R(t). In order to calculate physical quantities
such as mean values of observables from such a stochastic
equation, one has to deal with a sufficiently large ensemble of
trajectories for a true realization of the stochastic force. The
physical description of Brownian motion is therefore contained in
a large number of stochastic trajectories rather than in a single
trajectory, as would be the case for the solution of a
deterministic equation of motion. The Kramers equation or the
Fokker-Planck equation is a partial differential equation which
can be solved analytically under simplifying assumptions whereas
the Langevin equation is a stochastic differential equation and
therefore not amenable to analytic treatment. This is possibly the
reason why the Langevin approach was not used in nuclear
applications for a long time, while the Fokker-Planck equation was
preferred for applications in heavy-ion collisions, especially for
the deep-inelastic processes. Further, the Fokker-Planck equation
or the Langevin equation are to be solved numerically for
practical applications to nuclear collective motions where more
than one degree of freedom are involved and the transport
coefficients(friction, inertia) are coordinate dependent.
Numerically, the Langevin equation is more straightforward to
handle for a number of reasons. Firstly, it is easier to
accommodate more degrees of freedom in this ordinary differential
equation. On the other hand, the Fokker-Planck equation is a
partial differential equation and adding  more degrees of freedom
generates a multidimensional partial differential equation, the
solution of which is very time consuming even with modern
supercomputers. The multiple reduplication of the trajectory
calculation(Langevin approach) is the price one has to pay to
avoid solving a partial differential equation in many degrees of
freedom. Secondly, the solution of the Langevin equations by
Monte-Carlo sampling of trajectories is numerically more stable
than the approximate methods available for a direct solution of
Fokker-Planck equation\cite{Abe2}. Moreover, the Langevin equation
can be extended to include non-Markovian processes as well
\cite{Abe2}. It may also be mentioned that there is a quantal
version of the Langevin equation based on which a full-fledged
transport theory has been formulated in \cite{Eckern} within a
quasi-classical approach. By virtue of its intuitiveness,
generality, and other practical advantages, Langevin approach is
preferred to that of Fokker-Planck and is mostly followed in the
recent years. \par Both the friction coefficient $\eta$ and the
random force $R(t)$ arise due to coupling of the collective
dynamics with the intrinsic motion of the system. Since they have
the same microscopic origin, they are expected to be correlated.
In fact in his famous analysis of Brownian motion, Einstein showed
in 1905 that the friction coefficient $\eta$ and the diffusion
constant $D$ (related to $R(t)$ by Eq. \ref{1e}) are related to
each other. This is intuitively understandable since both of these
constants describe different aspects of the same physical process
- the exchange of momentum and energy between the collective
variable and the heat bath. The argument is universal and applies
as well to nuclear systems. It can be shown\cite{Abe2} that there
is a relation between them called the `fluctuation-dissipation
theorem' which reads as follows
\begin{equation}
D=\eta T \label{1f}
\end{equation}
\noindent where $T$ is the temperature of the heat bath.
  This relation
is also supported from a phenomenological analysis.  As time $t$
approaches infinity,  the Brownian particle is expected to be in
equilibrium with the heat bath and the average kinetic energy (for
one dimensional motion) becomes equal to $T/2$ ($T$ is in units of
energy). Using the Langevin equation (\ref{1c}) for a free
Brownian particle ($F(t)$ = 0) and using the properties of the
random force given by Eq. (\ref{1e}), the average kinetic energy
of the Brownian particle is calculated as follows
\begin{equation}
\frac{\langle p^2\rangle}{2m}=2D/{4\eta}+\frac{\langle
{p(0)}^2\rangle}{m^2}e^{-\frac{\eta}{m} t}.
\end{equation}
\noindent As $t \rightarrow \infty$, one gets $2D/{4\eta} = T/2$,
which yields $D = \eta T$. Substituting in Eq. (\ref{1e}), one
finally gets
\begin{equation}
\langle R(t)R(t^{\prime})\rangle = 2\eta T\delta(t-t^{\prime}).
\end{equation}
The above relation connects the mean square strength of the
stochastic force with the friction coefficient.
 This fluctuation-dissipation theorem points
out the cause-effect relationship between the stochastic and
dissipative component of the dynamics. It also implies that any
dissipation is always associated with fluctuations and vice versa.
\par In the dissipative dynamical model of nuclear fission discussed
previously, it is assumed that the fission of hot nuclei involves
two distinct time scales; one being associated with the slow
motion of the fission degrees of freedom and the other with the
rapid motion of the intrinsic degrees of freedom. The time
evolution of the macroscopic(collective) coordinate may be viewed
as the slow motion in comparison with the agitation of the
individual particles(microscopic motion) of the bath. A Markovian
Langevin approach is valid as long as a clear separation between
these two time scales is possible. However, when the collective
motion is faster and hence the two time scales become comparable,
one has to generalize the Langevin equation to allow for a finite
memory and the process becomes non-Markovian\cite{Abe2}. For fast
collective motion, the generalized Langevin equation reads as
\begin{equation}
{dp \over dt} = F(t) -\int^{t}
dt^{\prime}\eta(t-t^{\prime})p(t^{\prime}) +R(t)
\end{equation}
The friction kernel here is non local in time.
 This implies that the friction $\eta$ have a memory time, i.e,
 the friction depends on the past stages of the collective motion.
It is therefore also called a retarded friction. The time
correlation of the stochastic force is generalized accordingly and
is given by the following equation.
\begin{equation}
\langle R(t)R(t^{\prime})\rangle = 2\eta(t-t^{\prime})T.
\end{equation}
Thus the random force does not have a white noise(vanishing
correlation time) but a colored (finite correlation time) one. The
correlation property also states that there is a memory time
$\epsilon$ (also called the correlation time) within which the
stochastic variable $R(t^\prime)$ at time $t^\prime$ influences
the variable $R(t)$ at time $t$. For slow collective motion the
memory effects can be neglected, the correlation time vanishes and
we have a time-local friction force. The dynamics is then said to
be ``$\delta$- correlated'' or Markovian. Nuclear collective
motion is studied within the framework of ``linear response
approach" to examine whether it is Markovian or not\cite{Yamaji3}.
\par Another distinguishing feature of nuclear collective dynamics
from that of a Brownian particle is the fact that whereas in
Brownian motion, the large bath of oscillators influences the
motion of the Brownian particle, the bath itself is not affected
by its coupling to the collective motion (in particular, its
temperature remains constant). However, this is not strictly valid
for a nuclear system. In deep-inelastic collisions or during the
fission process, we assume that the bath represents the intrinsic
degrees of freedom of the nuclei. Here again, the thermal
capacity(intrinsic nuclear excitation $\sim$ 100 MeV) of the heat
bath though much larger than the collective kinetic energy of the
fission degree of freedom($\sim$ 10 MeV), the variation in the
temperature of the bath due to energy flow from the collective
mode(friction) cannot be neglected. In order to conserve total
energy, the net kinetic energy loss of the Brownian
particle(fission degrees of freedom) manifests as energy gain(rise
in temperature $T$) in the heat bath. Thus the fluctuation
strength coefficient $D (= \eta T)$ which determines the strength
of the Langevin(random) force is not constant, but is continually
re-adjusted as the bath heats up. The assumption underlying this
scheme is that the internal system equilibrates quickly, i.e. its
equilibration time is smaller than the correlation time
$\varepsilon$, and also smaller than the time scale of macroscopic
collective motion. The above assumption thus implies that the
Langevin dynamics can be applied with confidence for slow
collective motion of a highly excited nuclear system. This is best
fulfilled in fission of highly excited large compound nuclei.
Hence we shall assume in our work Langevin equations with a
phenomenological Markovian friction term and it is understood that
the temperature and therefore also the fluctuation strength of the
Langevin force, change with time, but at a rate which is slow on
the scale of the equilibration and the correlation times.

\section{Nuclear Dissipation}
\subsection{Introduction}
In the early literature a brief remark is made in the famous paper
of Kramers\cite{Kramers} to the effect that friction might play a
role in the nuclear fission rate. Strutinsky\cite{VM} mentions
friction in connection with fission when discussing solutions of
Kramers equation. But for a long time the statistical model for
fission and particle evaporation developed by Bohr and
Wheeler\cite{Wheel} and Weisskopf\cite{Weiss} and developed
subsequently into computer codes by Puhlhofer\cite{Puhl},
Blann\cite{Blann} and others were sufficient to describe fission
data. Pre-scission particle multiplicities were not measured at
that time. The status changed dramatically in the 1980s when
measurements reveal enhanced neutron multiplicities as compared to
statistical model code\cite{Gavron2,Hils}. This work was
accompanied by theoretical investigations based on the
Fokker-Planck equation by Grange and Weidenm\"{u}ller
\cite{Grange1,Weid2,Grange0} predicting reduced fission
probabilities due to friction effects which should also influence
emission of neutrons\cite{Grange1,Grange2,Hass2,Weid3}. The
increased neutron multiplicities were further studied by different
groups \cite{Zank,Gavron1,Grange3,Hinde3} and values for friction
coefficients were obtained to fit experimental
data\cite{Zank,Grange3}. Experimental evidence of fission as a
slow and highly dissipative process came from the pre-scission
multiplicities of neutron\cite{Hinde2}, charged
particles\cite{Lestone1}, and $\gamma$ rays\cite{Chakra}. These
experiments suggest collective motion to be overdamped, possibly
providing an answer to the question raised by Kramers as early as
1940 in his seminal paper\cite{Kramers}, namely, `` Is nuclear
friction abnormally small or abnormally large''. It was found that
the pre-scission neutron multiplicities increase more rapidly with
bombarding energy than the statistical model predictions, no
matter how one varies the parameter of the model, i.e., the
fission barrier, the level density parameter and the spin
distribution, within physically reasonable limits\cite{Newton1}.
It was strongly established that it was not adequate to treat
fission of hot nuclei along the lines of statistical model without
dissipation. Thoennessen and Bertsch \cite{Thoe1} studied
different systems and found the systematics of the threshold
excitation energy when statistical model starts losing its
validity. This data presents to the theorist the problem of
understanding the dissipation and how it depends on excitation
energy. The excess yield of particles and $\gamma$-rays from heavy
compound systems were analyzed by incorporating the nuclear
friction parameter and transient effects allowing for the build up
of the fission flux. General reviews of the experiments and also
surveys on theoretical models for their interpretation can be
found in the articles of Newton\cite{Newton2}, Hilscher and
Rossner\cite{Ross}, Hinde\cite{Hinde5} and with emphasis on
pre-scission giant dipole $\gamma$-emission, in the article of
Paul and Thoennessen\cite{Paul}.\par It was thus well established
that a dissipative force operates in the dynamics of a fissioning
nucleus. In the dissipative dynamical model, where induced nuclear
fission is viewed as a diffusion process of the fission degree of
freedom over the fission barrier, nuclear friction is interpreted
as the average effect of the interaction of the slow collective
motion with already thermalized intrinsic degrees of
freedom(mostly comprising of uncorrelated particle-hole
excitations). The dynamical behavior of large-amplitude collective
motion, such as those occurring in fission and heavy ion
reactions, depend crucially upon the rate at which energy of
collective motion is dissipated into internal single particle
excitation energies, as well as upon the mechanism by which the
dissipation proceeds.
Dissipation affects the dynamical motion primarily by\\
 (1) increasing the time required to go from one shape to another
 which results in enhancement of prescission particle emission,\\
 (2) heating the system at the expense of collective kinetic energy which manifests
in fission fragment kinetic energy distribution,\\
 (3) introducing fluctuations in a natural way which results in fluctuations
 around the mean path in multi-dimensional deformation space which in turn
introduces fluctuations in different experimental observables.\\
 Despite these effects on the nuclear dynamics, unambiguous extraction of the
 strength of the nuclear friction was not possible from experimental data in the earlier
years($\sim 80's$) essentially because the experimental data were
not very sensitive to the details of the nuclear friction. However
it is only recently($\sim$ last 10 years) that considerable
progress has been made, mainly from new experimental measurements
such as prescission neutron multiplicities and evaporation-residue
cross-section and the choice of the range of nuclear friction to
fit data has narrowed down substantially. \par Theoretical work on
the detailed nature of the nuclear friction, either
phenomenologically or from specific microscopic models, has made
considerable progress in the recent years.
  In \cite{Hilsch}, a compilation of data on the
magnitude of dissipation has been presented. In \cite{Paul} and
\cite{Back}, information on T-dependence of dissipation has been
extracted from comparison with experimental findings.  The
microscopic structure of the friction coefficient has been studied
together with fluctuations in the collective variable within
microscopic transport theories based on random matrix
approach\cite{Noren5,Noren6}, the one-body dissipation
model\cite{Gross2,Blocki1}, and the linear
response\cite{Siemens1,Siemens2}. From the seventies, several
attempts have been made to derive dissipation coefficient for
nuclear friction theoretically but a complete theoretical
understanding of the dissipative force in fission dynamics is yet
to be developed. The results obtained in various one-body or
two-body viscosity models differ very much in the strength and
coordinate dependence and also with respect to its dependence on
the temperature.
  They sometimes
differ by an order of magnitude, a feature which not only reflects
the complexity of the problem, but also urges for finding the
solution.
\subsection{Probes of nuclear friction in heavy-ion induced fission}
Friction in the fission process is expected to manifest itself in
a number of observables as we have discussed in the previous
sub-section.  Friction  affects fission probability (fission cross
section/compound nucleus formation cross section) which in turn
will directly affect the pre-scission particle (particularly
neutron) and $\gamma$ multiplicity. Therefore, the measurement of
prefission particle and GDR $\gamma$-ray multiplicities provide
suitable clocks to probe fission time scale and nuclear
dissipation. In particular, neutrons are expected to work as a
clock to measure fission time scale, because of their short life.
In order to analyze the pre-scission neutron data with the
statistical model, a long `delay time'$(\approx 5\times 10^{-20})$
\cite{Hinde1,Hinde2} was initially introduced during which fission
was suppressed. This delay time has been interpreted as a
transient time during which the fission degree of freedom attains
quasistationary distribution in phase space. This time interval
depends on the strength of the friction force. Therefore the
nuclear friction coefficient can be deduced by analyzing the
prescission multiplicities using Fokker-Planck or Langevin
equation. Secondly, friction is expected to influence the
distributions of the total kinetic energy, mass and charge of the
fission fragments. These distributions of fission fragments is
related to the dynamics of fission and analyzing these data one
can further probe nuclear dissipative forces. \par From the
theoretical side, Strumburger {\it et al.} \cite{Strum} have
combined a Fokker-Planck description with rate equations and
analyzed data for pre-scission light particle multiplicities. Nix
and his collaborators\cite{Davies1,Davies2} introduced friction in
classical equations of motion in order to describe kinetic
energies of fission fragments.  Weidenm\"{u}ller and coworkers
\cite{Pauli1,Pauli2} investigated also the effect of friction on
the width of the kinetic energy distribution. Adeev and
collaborators used multi-dimensional Fokker-Planck equations
\cite{Adeev1,Adeev2} in order to describe the variances of
mass\cite{Adeev3,Adeev4}, energy\cite{Adeev5,Adeev6} and
charge\cite{Adeev7} distributions of the fission fragments. Work
concerning the Fokker-Planck description of fission fragment
distributions is reviewed in Ref. \cite{Adeev8}. \par Langevin
approach was first proposed by Abe {\it et al.} \cite{Abe1} as an
intuitive phenomenological framework to describe nuclear
dissipative phenomena such as heavy-ion reactions and fission.
Fission dynamics of hot nuclei were investigated by Abe and
others\cite{Abe2} using the two-dimensional Langevin equation
including particle evaporation. Both the calculated number of
pre-scission neutrons and  the average total kinetic energy of
fission fragments were found to be consistent with experimental
values using one-body dissipation. Detailed studies of Langevin
dynamics with a combined dynamical and statistical model (CDSM)
were made and the influence of friction on prescission neutron,
charged-particle and $\gamma$-multiplicities, on the energy
spectra of these particles, on fission time distributions, and on
evaporation and fission cross sections were investigated by
Fr\"obrich and his collaborators \cite{Frob}. Their
phenomenological analysis yielded a strong deformation dependent
nuclear friction. They also concluded from their study that
evaporation residue cross section is a very sensitive probe for
nuclear friction\cite{Frob2}. Hence more precise measurements of
evaporation residue cross sections would help to discriminate
between the different versions of friction used in the analysis of
fission data. Similar conclusions were also drawn by other workers
in the recent years\cite{Brinkmann}. In Ref. \cite{Lazarev}, the
so called `long-lifetime fission component' or LLFC was proposed
as a new probe of dynamical effects in heavy-ion induced fission
and it was concluded that measurements of LLFC for heavy systems
can provide decisive information about the strength of nuclear
friction for compact configurations in fission. Giant dipole
resonance(GDR) $\gamma$ was used as a probe to study the viscosity
of saddle-to scission motion in hot $^{240}$Cf and a measure of
the saddle to scission time was extracted from the prescission
$\gamma$ yield\cite{Dioszegi}.
\subsection{ Origin and nature of nuclear dissipation}
 In theoretical models for nuclear friction, two kinds of
dissipation mechanisms are generally considered: one is the
wall-and-window one-body dissipation and the other is the
hydrodynamical two-body dissipation. In the wall friction, the
intrinsic motion of the nucleons is assumed to be described by the
extreme single-particle model of the nucleus whereas its
collective dynamics is described by its shape evolution. The
nucleons within the nuclear volume are assumed not to collide with
themselves but they undergo collision with the moving nuclear
surface(`wall') and thereby damps the surface
motion\cite{Blocki1}. The irreversible feature of friction comes
out after suitable averaging is carried out. A similar picture is
used in the linear response theory approach to nuclear
friction\cite{Siemens1}. There the `wall' is replaced by the shell
model potential, the nucleons move in quantum states and are
allowed to `scatter' from one another. Details of this theory can
be found in \cite{Hofmann1}, together with numerical computations
of the transport coefficients and their temperature dependence on
the basis of ``locally harmonic approximation''. The basic
assumption
 here consists of the hypothesis that close to $Q_0$, ($Q$ is the
 coordinate corresponding to the shape
degree of freedom, $Q_0$ can be any fixed value of the coordinate
which the system may reach) and for a small time interval $\delta
t$, the actual $Q(t)$ can be approximately described by the
`harmonic' motion associated with a properly defined osculating
oscillator. This approximation implies the expansion of the
Hamiltonian $\hat{H}(Q)$ keeping terms up to second order i.e., up
to $(Q-Q_0)^2$. The condition imposed on the time scale is
$\tau\ll\delta t\ll\tau_{coll}$,($\tau_{coll}$ and $\tau$ are the
collective and nucleonic time scales) which guarantees that within
$\delta t$ collective motion does not drive the system too far
away from $Q_0$. The assumption is that the collective motion is
sufficiently slow such that the large scale motion can be
linearized locally. The effect of the coupling term (between the
collective and intrinsic motion) which is given by
$(Q-Q_0)\left(\frac{\partial H}{\partial Q}\right)$, is treated by
the linear response theory. The response function
$\tilde{\chi}(t)$ measures the response of the system of nucleons
to the coupling and the transport coefficients follow after
evaluating the moments in time of the response function by Fourier
transforms. The limitation of this procedure is that the transport
coefficients should not vary too much with the collective variable
$Q$. The variation of the transport coefficients  (friction
$\eta$, inertia $m$ etc) with temperature and shape for average
fission dynamics is studied using this model of linear response
theory\cite{Yamaji2}. It has been shown in \cite{Hofmann2} that
the friction coefficient obtained within linear response theory(in
the zero frequency limit) becomes close to the one of wall
friction after applying smoothing procedures in the sense of
Strutinsky method. This feature goes along very nicely with the
claim that wall friction represents the macroscopic limit for a
system of independent particles. The transition from ``independent
particle motion to collisional dominance" in view of the linear
response approach is looked at in Ref. \cite{Magner2}.
\par There are theories for which friction shows a
`hydrodynamical' behavior, in the sense of being proportional to a
relaxation time $\tau_{intr}$ of nucleonic motion and thus to
$T^{-2}$, $T$ being the nuclear temperature. This concept is used
in the theory of ``dissipative diabatic dynamics'' proposed in
\cite{Noren3} which is based on the assumption that nuclear
collective motion happens predominantly diabatically and is used
for the entrance phase of a heavy ion collision. In \cite{Ivan1},
 the von Neumann equation had been applied to the deformed shell
 model, complimented by a collision term in relaxation time
 approximation. For the previous two models, the association to
 hydrodynamics is only given somewhat loosely through the
 proportionality factor $T^{-2}$ in the friction coefficient, or
 components of it. Hydrodynamical viscosity in the proper sense of
 ``collisional dominance'' is found whenever the nucleonic dynamics
 is described by transport equations like the Landau-Vlasov equation with
 the collision term.  There is a recent work \cite{Magner}, which
 combines the use of such an equation with a special treatment of
 the surface by way of collective variables. In \cite{Bush}, a
 model has been presented in which collective dynamics itself is
 governed by two-body collisions, rather than by the picture of a
 time dependent mean-field. Microscopic calculations of the
diffusion coefficient\cite{Cha} (assuming purely diffusive motion
up to the saddle point) and the friction constant\cite{Boiley}
(from microscopically derived Langevin equation as applied to
thermally induced nuclear fission) resulted in too strong
dissipation for nuclear collective motions.
\par An attempt to account for
 both one-body and two-body mechanisms of friction was made in Ref.
 \cite{Ivan1,Ivan3}  within the so-called relaxation time
 approximation(RTA), in which the time dependent mean field theory
 is extended by an account of the collision integral in linear
 order in the deviation of the density matrix from some
 equilibrium distribution. The two components of friction obtained
 within the relaxation time approximation show the temperature
 dependence which is characteristic for one- and two-body
 dissipation. The non-diagonal component is very small for
 temperatures below 2 MeV; it increases with temperature and
 reaches a kind of plateau at a temperature of the order of 2-4
 MeV depending on the specific choice of the single-particle
 potential. The absolute value in the plateau region is very close
 to the wall friction for a sharp edge(infinitely deep square well)
 potential and a few times smaller in the case of a very diffuse
 (harmonic oscillator) potential and thus is found to depend on
 the diffuseness of the potential.  The diagonal component is
 proportional to the relaxation time and in this way is similar to
 the two-body viscosity. However, the proportionality factor is
 too large, which causes some doubt as to whether the RTA can be
 applied to describe friction in the case of large scale
 collective motion. \par It has also
been noticed that at very small temperatures, pairing correlations
require dissipation to vanish. It needs to be stressed that a
small damping strength at small temperatures may have quite
drastic implications. If the dissipation strength falls below some
limit, the nature of the dissipation process would change
completely. Then the dissipation is too weak to warrant relaxation
to quasiequilibrium. This not only violates Kramers formula but
also the Bohr-Wheeler formula becomes inapplicable\cite{Rummel}.
So far no method exists how to incorporate collective quantum
effects.
\par The models described above encompass the whole range of
assumptions one may
 make for nuclear dynamics, from pure independent particle model
 to the ones which are entirely governed by collisions. We should
now briefly review the standard one and two-body dissipation in
terms of their comparison with experimental data.
\subsection{One body dissipation vs. two body viscosity}
 The models of hydrodynamical
viscosity \cite{Davies1} are based on the assumption that nuclear
dissipation arises from individual two body collisions of
nucleons. It was further observed that two-body viscosity hinders
the formation of a neck in nuclear fission. This leads to more
elongated scission configuration and consequently to a smaller
kinetic energy of the fission fragments. Davies {\it et al.}
\cite{Davies1} deduced the value $0.015\pm 0.005$ TP = $9 \pm 3
\times 10^{-24}$ MeVs/fm$^{3}$ for the viscosity coefficient $\mu$
by analyzing the mean total kinetic energies of fission fragments
with the Newtonian equation for the mean trajectory. It was
observed that the mean kinetic energy of the fission fragments is
not very sensitive to the details of the dissipative forces and
both one and two body dissipations in classical dynamical
calculations have been found to describe systematics of
experimental mean kinetic energies. It was however concluded from
extensive experimental data that the hydrodynamical two body
viscosity cannot give consistent explanation of both neutron
multiplicity and fission fragment kinetic energy distribution. A
strong ($\mu= 0.20 TP$ or larger) two-body viscosity is required
to reproduce the observed neutron multiplicity. However, the total
kinetic energy calculated with this value of $\mu$ is far smaller
than given by the Viola systematics. A consistent explanation of
neutron multiplicities and fragment kinetic energies indeed
support the one-body friction and not the two-body
viscosity\cite{Wada}.  Studies of macroscopic nuclear dynamics
such as those encountered in low-energy collisions between two
heavy nuclei or nuclear fission have also established that
one-body dissipation is the most important mechanism for
collective kinetic energy damping. Gross \cite{Gross2} first
pioneered the concept of a one body mechanism which considered the
transfer of energy from the motion of nuclear surface to the
nucleon motion as a result of frequent collisions of the nucleons
with the nuclear surface. One-body mechanism is expected to be the
main process at low nuclear excitation energies(temperatures  up
to a few MeV) because nucleon-nucleon collisions are suppressed by
the Pauli principle by limiting the phase space into which the
nucleons can scatter. When the excitation of the nucleus is not
too high, the mean free path of the nucleons is greater than the
nuclear dimensions and hence two-body processes are less favored
(short mean free path assumption implicit in ordinary two body
viscosity is not valid in this energy range) compared to one-body
processes in this long mean path dominated mean field
regime(independent particle model of nucleus considered). The
analogous classical system is therefore a Knudsen gas confined
within a container, rather than a short mean-free-path fluid
dominated by two-body interactions. Some estimates made about the
time between two subsequent collisions of a particle with the wall
\cite{Weid1} gives $\tau_{wall} \cong 1.6 \times 10^{-22} s$
whereas the time between two subsequent collisions of a single
particle with another particle was estimated to be $16 \times
10^{-22}s \cong 10.\tau_{wall}$. This seems to favor collisions
with the wall as the main process of energy dissipation. These
theoretical arguments supported by the experimental observations
led to the conclusion that one-body dissipation is the dominant
mechanism for energy dissipation in nuclear fission when the
excitation energy is not too high (much below the Fermi energy
domain). However, two-body collisions are expected to gain more
importance at higher temperatures. The importance of one body
dissipation motivated the derivation of one-body friction by
microscopic theories. The proper quantal description of one-body
dynamics is the time-dependent Hartee-Fock(TDHF) where
single-particle wave functions describing the nucleons evolve
through a Schrodinger-like equation containing the nuclear mean
field. Despite the exact nature of the TDHF solution to one-body
dynamics,  the need for calculational simplicity demands a
macroscopic description involving a small number of explicit
degrees of freedom.
\subsection{Wall and Window Friction}
 Blocki {\it et al.}
\cite{Blocki1} derived a simple expression(in a classical
picture), namely, the ``wall formula''(WF) for one body
dissipation. According to the formula, the rate of  collective
energy dissipation is given as
\begin{equation}
\dot{E}_{WF}(t)=\rho_m \bar{v} \int \dot{n}^2 d\sigma,
\end{equation}
where $\dot{n}$ is the normal component of the surface velocity at
the surface element $d\sigma$, while the nuclear mass density and
the average nucleon speed inside the nucleus are denoted by
$\rho_m$ and $\bar{v}$ respectively.
 The time dependent mean field
nuclear potential is identified with the `wall' and the net energy
dissipation from the wall(collective degree) to the nucleons
through their interaction is given by the wall friction. Typical
estimates were made of the characteristic time scale of the
one-body dissipation theory resulting from balancing typical
inertial and dissipative terms in the equations of motion. It
turned out to be in the range of $(0.7-1.3) \times 10^{-22}$ sec
for mass numbers between 50 and 250\cite{Blocki1}. These damping
times are intrinsically short compared to many characteristic
collective time scales, which suggest that one-body energy
dissipation may often dominate collective nuclear dynamics.
\par The wall friction was also obtained from a formal theory of
one body nuclear dissipation which is based on
  classical linear response technique \cite{Randrup} applied to a
 Thomas-Fermi description of the nucleus  and expressions
  for the collective kinetic energy  and the rate of
  energy dissipation for slow collective motion were
  identified. These
   quantities are characterized by mass and dissipation kernels,
considering the nucleus as a large system of independent nucleons
contained within a leptodermous time-independent single-particle
potential.
    The rate of dissipation is expressed as a double surface
  integral involving the normal surface velocity at different points,
coupled via a dissipation kernel. This kernel is simply related to
the imaginary part of the single-particle Green function for the
nuclear potential. In the large nucleus limit, these kernels were
shown
    to be independent of the surface-diffuseness of the single particle
    potential and to be simply dependent on the nuclear temperature.
     For a given nuclear shape, the kernels were expressed in terms of the
     classical trajectories for nucleons within the nucleus, and
 are therefore sensitive functionals of the nuclear shape. In the limit of
velocity fields varying slowly over the nuclear surface, the
classical one-body friction or wall friction is obtained. It was
     demonstrated that these results could also be derived by taking the
     stationary-phase (large nucleus) limit of an entirely quantal
     formulation. The one-body mass and dissipation coefficients differed
      significantly from those of incompressible, irrotational hydrodynamics.
       Yamaji \textit{et al.}
 \cite{Yamaji1}, also obtained a friction coefficient comparable
with wall friction using the linear response theory. More
recently, in \cite{Skalski}, the concept of wall friction has been
reexamined by performing computer simulations to follow the
particles of a gas  in a container when the shape of the container
undergoes  harmonic vibrations driven by an external force. Both a
classical gas as well as a quantum system having the typical
nuclear dimensions have been considered. They concluded from their
study that there is a minimal collective speed above which the
wall friction is applicable. \par  The ``window friction" was
formulated which accounts for the role of nucleon exchange through
a neck in a dinuclear system\cite{Blocki1}. When the two halves
 of a nucleus are in relative motion due to leftward and
 rightward drift, any particle passing through the window will
 damp the motion because of the momentum transferred between the systems.
This gives rise to an effective dissipation coefficient which is
termed as window friction. The wall friction in conjunction with
the window friction was found to be quite successful in
reproducing a large volume of experimental data of damped heavy
ion collisions\cite{Blocki2}, fusion{De}, and
fission\cite{Blocki1, Sierk1}. However, the damping widths of the
giant resonances calculated from the wall friction turned out be
rather unsatisfactory when compared with experimental
data\cite{Nix,Myers}.\par
 An extensive application of
the Langevin equation to study one-body friction was made by
Frobrich and Gontchar \cite{Frob}. A combined dynamical and
statistical model for fission was employed in their calculations
and it was first shown by them that wall friction fails to
reproduce simultaneously excitation functions for pre-scission
neutron multiplicity and fission probability. A detailed
comparison of the calculated fission probability and pre-scission
neutron multiplicity excitation functions led to a
phenomenological shape dependent nuclear friction. The
phenomenological friction turned out to be considerably smaller
than the standard wall friction value for nuclear friction for
compact shapes of the fissioning nucleus whereas a strong increase
of the friction was found to be necessary at large deformations.
 Earlier, Nix and Sierk\cite{Sierk2,Sierk3} also suggested in their
analysis of mean fragment kinetic energy data that the dissipation
is about 4 times weaker than that predicted by the
wall-plus-window formula of one-body dissipation. Thus the
different experimental observations insisted on a reduction of
strength of the wall friction and hence its modification.
\subsection{Modification of wall friction}
 The dynamics of independent particles
in time-dependent cavities has been extensively studied by Blocki
and his coworkers\cite{Skalski,BSS3,BBS1,BJS,BBS2,BSS2,MSB,BBSS}.
Considering classical particles in vibrating cavities of various
shapes, a strong correlation between chaos in classical phase
space and the efficiency of energy transfer from collective to
intrinsic motion was numerically observed\cite{BSS3}. It has been
argued in \cite{Skalski,BSS3} that the wall friction in its
original form should be applied only for systems for which the
particle motion shows fully chaotic behavior. Hence the wall
friction needs to be modified to make it applicable for those
systems which are partially chaotic. In this regard it is thus
necessary to discuss briefly the relevance of chaos to nuclear
dissipation. \par  One of the major themes of contemporary science
is the study of order to chaos transition in dynamical systems.
Nuclear dynamics is known to exhibit chaotic features and the most
prominent one is that given by Wigner's law for the distribution
of levels of the compound nucleus, as seen in neutron
resonances\cite{Bohigas}. Statistically significant agreement
between measured level spacing fluctuations and Wigner's random
matrix model was established and it was concluded that the order
to chaos transition from an integrable(regular) to a chaotic
system is reflected in the level spacing distribution by a smooth
transition from the Poisson distribution to the Gaussian
orthogonal ensemble(GOE) distributions. It was also conjectured
that the fluctuation properties of generic quantum systems, which
in the classical limit are fully chaotic, coincide with those of
GOE. The close agreement between the GOE prediction and
fluctuation properties of nuclear levels suggests that the nucleus
is a chaotic system, at least at excitation energies above several
MeV. The transition from ordered to chaotic nucleonic motions in
the nuclear mean-field potential is reflected in the disappearance
of shell effects in nuclear masses and deformations, and in the
transition from an elastic, through an elastoplastic, to a
dissipative behavior of the nucleus in response to shape
changes\cite{Swiat}. In general, a nuclear system is neither fully
integrable nor fully chaotic and the elatsoplastic behavior
expected in this intermediate regime was utilized in modification
of the wall friction which is valid in  the fully chaotic regime.
\par The wall friction was originally derived for idealized
systems employing a number of simplifying assumptions such as
approximating the nuclear surface by a
 rigid wall and
considering only adiabatic collective motions. The validity of
these assumptions were scrutinized in the framework of
random-phase approximation (RPA) damping and it was shown that in
the limiting situation where the above assumptions are valid, RPA
damping coincides with the wall friction \cite{Yann,Griffin}. It
was subsequently realized that it is possible to improve upon the
dissipation rate given by the wall friction by examining its
various assumptions more critically. One of the important
assumptions of the wall friction concerns the randomization of the
particle motion. It is  usually assumed that successive collisions
of a nucleon with the one-body potential gives rise to a velocity
distribution which is completely random \cite{Blocki1}. In other
words, a complete mixing in the classical phase space of the
particle motion is required. This condition is satisfied for
one-body potentials whose shapes are rather irregular. It was
realized earlier \cite{Blocki1,Koonin} that any deviation from
this randomization assumption would give rise to a reduced
strength of the wall friction. This happens because the energy
transferred to a particle from a time-dependent wall could be
partly reversible if the motion of the particle is not completely
random. \par A modification of the wall friction has been proposed
recently \cite{Pal1} in which the full randomization assumption is
relaxed in order to make it applicable to systems in which
particle motion is not fully randomized on successive reflections.
This modified version of the wall friction is known as the
``chaos-weighted wall friction''(CWWF), where reduction in
strength of the wall friction is achieved through chaos
considerations. In the present work, the modified version of the
wall friction known as the chaos weighted wall friction will be
used in Langevin dynamical calculations for nuclear fission in
order to verify to what extent it can account for the experimental
data.
\section{Motivation of the work}
It is now apparent from the discussion in the previous sections
that a proper understanding of nuclear dissipation is an important
topic of contemporary nuclear physics. While the other inputs  to
the fission dynamics like potential, inertia can be obtained from
standard nuclear models, the strength of the dissipative force is
still not an unambiguously defined quantity and is often fixed
empirically  to fit experimental data. A clear physical picture of
friction is yet to be developed and the present work is an effort
in this direction. Both friction and random force depends on the
dissipative properties of nuclei which is hence a very important
input of Langevin dynamical calculations. The emphasis of the
thesis will be on the choice of a dissipative force, based on
physical arguments, which can be used in a dynamical description
of nuclear fission. A thorough understanding of the mechanism of
dissipation for nuclear systems as well as its shape and
temperature dependence will help in explaining the  experimental
data for pre-scission particle and $\gamma$ multiplicities in
nuclear fission, evaporation residue cross-section, fission
fragment kinetic energy and mass distribution. An improved
knowledge of nuclear friction will also help in the search for
superheavy elements. The synthesis of superheavy elements(SHE) by
the cold or warm fusion of heavy target projectile combinations is
a challenging problem in the recent years for both experimental as
well as theoretical physicists. The residue cross sections of SHEs
depends sensitively on both fusion probability as well as the
survival probability of the compound nucleus. The fusion
probability in turn depends on the fusion hindrance which depends
on the dissipation of collective energy of the amalgamated system
which has to overcome a conditional saddle in order to reach the
spherical shape, i.e., the compound nucleus. This problem of
overcoming a barrier under energy dissipation requires a thorough
knowledge of the nuclear friction for an appropriate dynamical
description. The survival probability\cite{Abe4} of the compound
nucleus against fission depends on the fission probability which
in turn depends critically on the time scale of fission. Fission
width or fission time scale depends very much on the rate at which
energy of collective motion gets dissipated. Thus a proper
understanding of nuclear friction is very crucial for theoretical
predictions of stability of SHEs against fission. The cross
section of formation of radioactive nuclei as fission fragments
also depends sensitively on nuclear dissipation on the fission
path. Therefore the main motivation of the work contained in the
present thesis is to critically examine the usefulness of CWWF as
a theoretical model of nuclear friction which can reproduce
experimental data without any tuning of the input parameters.
\section{Scope of the work}
 The main concern of this thesis will be application of
Langevin dynamics to nuclear fission. The diffuse surface liquid
drop model with Yukawa-plus exponential folding\cite{Brack} would
be used to calculate the nuclear potential for the evolving
nuclear shapes. Chaos weighted wall friction (CWWF), would be used
is used for nuclear friction in the Langevin equation. The other
inputs to the dynamical calculation,
 namely the inertia, particle and $\gamma$ widths, etc  are taken from
 standard models of nuclear physics. A special feature of the
 present work is that there is no free parameter in the entire
 calculation. In the statistical branch ,
the input fission widths are the Kramers limit whose systematics
 would be obtained by solving the Langevin equation in a separate
 procedure. The combined dynamical plus statistical calculation
would be performed for different nuclei ranging from high fission
barrier($\sim$ 10 MeV) to almost vanishing fission barrier and
different observables like fission probability ($P_f$),
prescission neutron multiplicity ($\nu_{pre}$) and evaporation
residue cross section ($\sigma_{ER}$) obtained from the
calculation are compared with experimental data. The primary aim
of this work is to verify to what extent the chaos weighted wall
friction can account for the experimental data. \par The different
inputs used in our dynamical model, in particular the chaos
weighted wall friction, are described in details in the next
chapter. Chapter 3 sketches the steps involved in solving the
Langevin equation to calculate fission width. In chapter 4, the
 combined dynamical and statistical model for
fission of hot nuclei is discussed. The results of the
calculations, i.e, $\nu_{pre}$, $P_f$ are compared with
experimental data for a number of nuclei in the same chapter.
$\sigma_{ER}$ is calculated from the model and compared with
experimental data in Chapter 5. The importance of transients in
nuclear fission is elaborated in Chapter 6. The last chapter
summarizes the entire work, and presents the conclusions and the
future prospects of the work.

\chapter{Langevin Dynamics of fission: Formulation of the model}

\section{Introduction}

The dynamical time evolution of the fission process from an
initially formed compound nucleus (with a more or less compact
shape) to the saddle and scission configurations and the
simultaneous emission of light particles during this deformation
process constitute a complex problem as we have discussed in the
previous chapter. In the absence of a complete microscopic ab
initio theory of such a dynamical process, a classical description
of the evolution of the collective coordinates is often found
useful and consequently in such descriptions, collective
parameters appear (collective mass, friction and diffusion
coefficients) which depend on the collective coordinates. A
classical description of the fission dynamics of a heavy and
highly excited nuclei is usually made on the ground that the De
Broglie wavelength associated with the fission degree of freedom
is much smaller than the nuclear dimensions, e.g, the ratio $\sim$
$0.1$ for a typical kinetic energy of $10$ MeV in the fission
degree of freedom and for a typical compound nuclear mass of $A
\sim 200$. The level density of such a compound nucleus at typical
excitation energies of a couple of tens of MeV is also extremely
large thus allowing for a classical description of its motion. It
should therefore follow that the quality of a theoretical
description will depend largely on a pertinent choice of the
collective coordinates and the degree of realism of the underlying
theory used to determine the collective parameters. While
modelling the dynamics of nuclear fission, the Fokker-Planck
equation was initially used more frequently though the application
of the Langevin equation was found to be more convenient in the
later works\cite{Abe2,Frob}. Apart from being more intuitive and
general, the Langevin equation is easier to handle numerically and
this motivated us to follow the Langevin dynamical approach for
nuclear fission in our work. In order to implement the Langevin
description for nuclear fission, it is necessary to specify the
nuclear potential energy, collective kinetic energy, and the rate
of energy dissipation in terms of the nuclear shape and its rate
of change with time. Great advances have been made in the past to
calculate the nuclear potential energy of deformation. Methods
ranging from purely macroscopic through microscopic-macroscopic to
exclusively microscopic are now accurate to within 1-2 MeV. In
contrast to this development, not much significant progress has
been made in our understanding of nuclear dissipation. However,
one of the most important inputs to such Langevin dynamical
calculations is the dissipative property of the nucleus since it
accounts for both the dissipative and the random forces acting on
the fission degrees of freedom. While the other inputs to the
Langevin equation such as the potential and inertia can be fixed
from standard nuclear models, the strength of the dissipative
force is still not an unambiguously defined quantity and is often
fixed empirically in order to fit  the experimental data. The
emphasis of this chapter will be on the choice of a dissipative
force, based on physical arguments, which can be used in a
dynamical description of nuclear fission. However, we shall first
describe the shape parametrization of the nucleus as well as the
other inputs to our dissipative dynamical model such as potential,
inertia and level density parameter in this chapter.

\section{Nuclear shape}
Fission is a multi-dimensional process, in the sense that a number
of deformation degrees of freedom can be involved. Therefore, any
reasonable dynamical model would require a number of parameters to
describe  the evolution of the nuclear shape. In a complete
dynamic description of the process, these parameters would appear
as the generalized coordinates. It is thus natural to resort to
parameterizing the nuclear shape in terms of a few collective
variables and making assumptions about the flow of matter in the
nuclear interior. The utility of a given shape parametrization
depends on how closely it approximates the shapes through which
the real system evolves and on how conveniently the three key
quantities, namely, the potential energy, the rate of energy
dissipation, and the collective kinetic energy, can be evaluated
for a given shape. For a dynamical calculation of nuclear fission,
it is normally assumed that the shape of the compound nucleus
remains axially symmetric. Different shape parameterizations have
been used in the literature which are mostly restricted to
elongation, neck and the mass asymmetry coordinates .  Cassinian
ovaloids \cite{Rabotnov,Cherd,Tillack}, Legendre-polynomial
parametrization \cite{Wada,Carjan}, ``funny hills''
parametrization\cite{Brack} are some of the commonly used shape
parametrizations  developed in order to specify the collective
coordinates for a dynamical description of nuclear fission. It
should however be noted that the computation time increases fast
with the increasing number of collective coordinates in the
Langevin equations. \par In the present work we will use the well
known ``funny hills'' parameters $\{c,h,\alpha \}$
 as suggested by Brack {\it et al.} \cite{Brack} which has been
 found to describe fission dynamics successfully in the past.
   The parameter $\alpha$ describes the asymmetry of
the shape in the $z$ direction. Since we will mainly be concerned
with fission of hot nuclei where symmetric division of nuclei is
the dominant mode of decay, we will consider only symmetric
fission in our calculations. Moreover, the asymmetry parameter is
mainly essential to calculate the mass and the kinetic energy
distribution of the fission fragments. Since we shall be concerned
with analysis of prescission neutron multiplicity and fission
probability data, we will use $\alpha = 0$ in our work. The
collective coordinate  $c$ corresponds to the elongation degree of
freedom of the nucleus and is related to the dimensionless fission
coordinate $q$, which is half the distance between the center of
masses of the future fission fragments divided by the radius of
the compound nucleus $R$, by the following relation,
\begin{equation}
q(c,h)=(3c/8)(1+{2 \over 15}(2h+(c-1)/2)c^{3})\label{2a}
\end{equation}
\noindent where $h$ corresponds to the neck degree of freedom.\par
   The surface of a nucleus  in cylindrical coordinates using the
 parameters $c$ and $h$ is given by,
 \newpage
\begin{eqnarray}
\rho^2(z) &=& \left(1 - \frac{z^2}{c_o^2}\right)(a_oc_o^2 +
b_oz^2), \hspace{2.5cm}  b_o \geq 0,a_o \geq 0 \nonumber\\
          &=&  \left(1 - \frac{z^2}{c_o^2}\right)\left(a_oc_o^2
\exp{(\frac{b_oc_oz^{2}}{R^{3}})}\right),  \hspace{1cm}  b_o <
0,a_o \geq 0\label{2b}
\end{eqnarray}
\noindent where $z$ is the coordinate along the symmetry axis and
$\rho$ is the radial coordinate of the nuclear surface. The
quantities $a_o$, $b_o$ and $c_o$ are defined by means of the
shape parameters $c$ and $h$ as
\begin{eqnarray}
c_o &=& cR, \nonumber\\
b_o &=& \frac{c-1}{2} + 2h.
\end{eqnarray}
\noindent where $R = 1.16 A^{1 \over 3}$, $A$ being the mass
number of the compound nucleus. $a_o$ and $b_o$ for $a_o \geq 0$
are related by\cite{Hasse}
\begin{eqnarray}
a_o &=& \frac{1}{c^3} - \frac{b_o}{5},\hspace{3cm} b_o \geq 0, \nonumber\\
     &=& -\frac{4}{3}\frac{b_o}{e^{p}+\left(1+\frac{1}{2p}\right)(\sqrt{-\pi
p}) erf(\sqrt{-p})}\hspace{0.5cm} b_o < 0
\end{eqnarray}
\noindent where $p=b_oc^3$ and $erf(x)$ is the error function.
 The two definitions join smoothly for  small absolute values of $b_o$.
\vspace{1.0cm}
\begin{figure}[h!]
\centerline{\psfig{figure=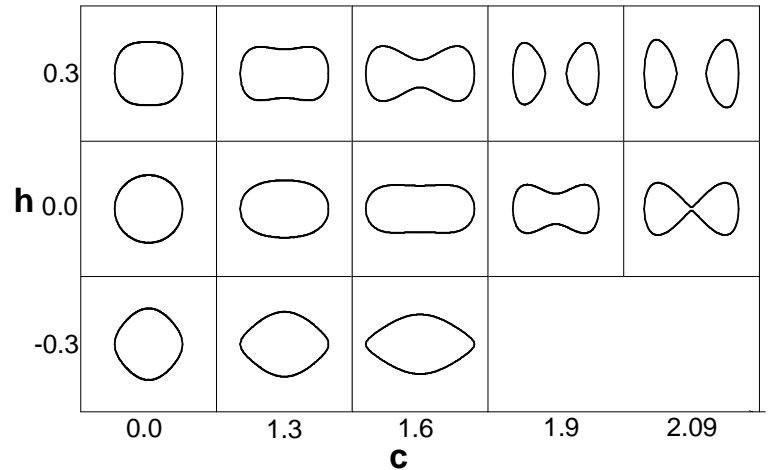,width=10cm}} \caption
{\label{b1}Shapes of the nucleus for different values of $c$ and
$h$ ($\alpha = 0$.)}
\end{figure}
  The volume is kept constant in the above
parametrization for all variations of the nuclear shape. The total
length of the longer axis of the density distribution, in units of
$R$, simply equals $2c$. The parameter $h$ describes the variation
of the thickness of the neck without changing the length $2c$ of
the nucleus, and is chosen in such a way that the $h=0$ line fits
approximately the bottom of the liquid drop valley. Positive
values of $h$ implies that the neck formation starts for a lower
value of $c$ as compared to the the case of $h=0$ and hence
scission of the nucleus into two fragments also takes place for a
lower value of $c$.
 Inclusion of
the neck degree of freedom is thus expected to accelerate the
fission process. When $b_o=0$, one has a set of oblate $(a_o
> 1)$, and prolate $(a_o < 1)$ ellipsoids. When $\alpha=0$, one obtains a family of
symmetric shapes ranging from the spherical shape$(a_o =1,b_o=0)$
to two fragment shapes $(a_o < 0, b_o > 0)$. For $b_o \geq 0, a_o
\leq 0$, $a_o$ and $b_o$ are connected by the following
expression\cite{Hasse},
\begin{equation}
\frac{1}{c^3}=a_o+\frac{b_o}{5}+(b_o+\frac{a_o}{5}){(-a_o/b_o)}^{3/2}.
\end{equation}
 This parametrization describes separated shapes when $h
 \geq((5/2c^3)-1/4(c-1))$.

\section{Langevin equation for fission}
The Langevin equation is the equation of motion of a Brownian
particle in a viscous medium placed in an external potential field
and is essentially given by the Euler-Lagrange equation where a
dissipative force and a random force are included in the force
balance equation. In the case of nuclear fission, the fission
degree of freedom is considered as the Brownian particle while the
friction and random forces arise out of the interaction of the
fission degrees of freedom with the rest of the nuclear degrees of
freedom, as we have discussed in the previous chapter. The two
dimensional Langevin equation in $(c,h)$ coordinates has the
following form\cite{Wada}
\begin{eqnarray}
\frac{dp_i}{dt} &=& -\frac{p_jp_k}{2} \frac{\partial}{\partial
q_i}{(m^{-1})}_{jk} - \frac{\partial F(q_i)}{\partial q_i} -
\eta_{ij}{(m^{-1})}_{jk}p_k + g_{ij}\Gamma_j(t), \nonumber\\
\frac{dq_i}{dt} &=& {(m^{-1})}_{ij}p_j. \label{2d}
\end{eqnarray}
\noindent with summation from 1 to 2 (c \& h) over repeated
indices; $q_i$ corresponds to $c$ $\&$ $h$ and $p_i$ corresponds
to $p_c$ $\&$ $p_h$. $F(q_i)$ is the free energy of the system and
$m_{ij}(q_i)$ and $\eta_{ij}(q_i)$ are the shape-dependent
collective inertia and dissipation tensors, respectively. The
random force $R_i(t)$ represents the random part of the
interaction between the fission degrees of freedom and the
intrinsic degrees of freedom (considered as a thermal bath in the
present picture) and is given by the following equation
\cite{Abe1,Abe2},
\begin{equation}
R_i(t)=g_{ik}\Gamma_k(t). \label{2d1}
\end{equation}
The  time-correlation property of $R_i(t)$ is given the following
relation,
\begin{equation}
\langle R_i(t)R_j(t^{\prime})\rangle=2D_{ij}\delta(t-t^{\prime}).
\label{2d2}
\end{equation}
\noindent where the strength of the random force is assumed to
satisfy Einstein relation  (fluctuation-dissipation theorem) which
reads as follows
\begin{equation}
 D_{ij}=\eta_{ij}T. \label{2d3}
 \end{equation}
 \noindent where $T$ is the temperature of the compound nucleus.
 It is assumed that \cite{Abe1,Abe2}
 \begin{equation}
 \langle
 \Gamma_k(t)\Gamma_l(t^{\prime})\rangle=2\delta_{kl}\delta(t-t^{\prime}). \label{2d4}
 \end{equation}
 Comparing the above equations (Eqs. \ref{2d1} to \ref{2d4}), it follows that
 \begin{equation}
 g_{ik}g_{jk}=\eta_{ij}T.
 \end{equation}
 The conservative force is usually specified from an appropriate
 nuclear model while the friction force is treated as a
 phenomenological quantity.
   These different
inputs to the Langevin equation as chosen in our dissipative
dynamical model will be described in detail in the following
sections.

\subsection{Potential}
The potential energy  $V(c,h)$ enters into our calculation through
its dependence on the deformation coordinates $c$ and $h$. It
could in principle be obtained from a microscopic mean-field
calculation at a finite temperature.  This type of Hartee-Fock
calculation using a reasonable effective nucleon-nucleon
interaction of the Skyrme type or Gogny type at every point in the
multidimensional deformation space, demands tremendous computer
time even with the most powerful computers and hence performing
such a computation for the present purpose is not attempted.
 To perform the same kind of
calculation even on the level of a self consistent semiclassical
approximation like the Extended Thomas-Fermi (ETF) method
\cite{Guet} at finite temperature \cite{Bartel}, which would
describe the average nuclear structure without shell oscillations,
would also be far too time consuming with advanced computers. We
have  therefore used a still simpler semiclassical approach where
the deformation dependent potential energy is obtained from the
finite range liquid drop model \cite{Sierk0} with the
parametrization of Myers and Swiatecki \cite{Swiat1}. In the
rotating liquid drop model\cite{Cohen}, the nucleus is assumed to
be formed of an incompressible fluid with a constant charge
density and a sharp surface, which rotates as a rigid body. There
are three important contributions to the deformation-dependent
potential energy in the liquid drop model: surface tension energy
(arising from saturating short range nuclear forces) which tends
to minimize the surface area of the nucleus,  repulsive Coulomb
energy(arising from mutual repulsion of protons) which tends to
distort or disrupt the nucleus, and rotational energy which also
favours disruption because large moments of inertia are
energetically favoured. The basic assumption in the liquid drop
model is that the surface thickness and the range of the force
should be much smaller than any geometrical parameter of the
configuration under consideration. This assumption breaks down in
the highly deformed shapes of a fissioning nucleus with small neck
where the neck dimension becomes comparable to small surface
thickness. In these cases the finite range of the nuclear force
and the diffuse surface lead to reduction in the energy which must
be taken into account\cite{Krappe}. The following changes are
therefore incorporated in the finite-range liquid drop model
relative to the liquid drop model, namely (1) the surface energy
of the liquid drop model is replaced by the
Yukawa-plus-exponential nuclear energy, which models effects of
the finite range of the nuclear force, nuclear saturation, and the
finite surface thickness of real nuclei\cite{Krappe}; (2) the
Coulomb energy is calculated for a charge distribution with a
realistic surface diffuseness\cite{Davies3}; and (3) the
rotational moments of inertia are calculated for rigidly rotating
nuclei with realistic surface density profiles\cite{Davies3}. The
different contributions of the deformation dependent potential
energy in the finite range rotating liquid drop model are
described
 briefly  as follows.
\newpage
(i) \underline{Yakawa-plus-exponential nuclear energy} :\\ The
surface energy of the liquid drop model suffers from several
deficiencies in attempting to describe real nuclei. The most
important of these is the neglect of proximity effects; that is
there is an unrealistically high surface energy for strongly
deformed shapes  and an absence of attraction between separated
nuclei in the liquid-drop model\cite{Sierk0}. One important step
for obtaining an improved macroscopic nuclear energy is the
Yukawa-plus-exponential double folding potential\cite{Krappe}.
With this technique, using one additional parameter (the range of
the potential) compared to the liquid-drop model, one can describe
suitably heavy-ion scattering potentials, fusion barriers for
light and medium-mass nuclei, the lower fission barriers observed
in nuclei with $A\leq 200$, and also satisfy the condition for
nuclear saturation\cite{Krappe,Moller}. The better reproduction of
the fission barriers with this Yukawa-plus-exponential potential
motivated us to use it in our calculation. The Yukawa-plus
exponential nuclear energy may be written as
\begin{equation}
E_{n} = -{c_{s} \over 8\pi^{2}r_{0}^{2}a^{3}} \int d^{3}r\int
d^{3}r^{\prime} \left[{\sigma \over a} -2\right]{e^{-\sigma/a}
\over \sigma} \label{2j}.
\end{equation}
where $\sigma= \vec{r}-\vec{r^{\prime}}$, $c_{s} = a_{s}(1-
\kappa_{s} I^{2})$ and $I \equiv (N-Z)/A$ is the neutron-proton
asymmetry. The integrals are over the volume of a sharp surfaced
nucleus. The range $a$ is the one additional parameter  of this
modification of the liquid drop model.  The value of $r_{0}$ is
determined from average charge radii of nuclei found in
electron-scattering experiments, $a$ is determined from heavy-ion
scattering experiments, while the surface energy and surface
asymmetry constants $a_{s}$ and $\kappa_{s}$ are determined from
fitting the macroscopic fission barriers of nuclei with mass
numbers from 109 to 252 at low angular
momentum\cite{Krappe,Moller}. The values of the constants used
here are as follows\cite{Sierk0}:
\begin{eqnarray}
r_{0}&=& 1.16 fm ,  \nonumber\\
 a    &=& 0.68 fm , \nonumber\\
a_{s}&=& 21.13MeV,\nonumber\\
 \kappa_{s}&=&2.3.\nonumber
\end{eqnarray}

 (ii) \underline{Coulomb energy} :\\
 The sharp surfaced charge distribution of a nucleus is made diffuse by
  folding a Yukawa
 function with range $a_{c}$ over a liquid drop distribution, and the
 Coulomb energy of the liquid drop model is modified by a
 Yukawa-plus-exponential function (almost similar in form as
 in the case of surface energy),
 proportional to $ \int d^{3}r\int
d^{3}r^{\prime} \left[1+{\sigma \over
2a_{c}}\right]{e^{-\sigma/a_{c }} \over \sigma}$. The range
parameter $a_{c}$ is chosen to be 0.704 fm. The diffuse surface
correction lowers the Coulomb energy since charge is spread over a
greater effective volume when the surface is made diffuse.
 The six dimensional integrals in Coulomb and surface energy are
 reduced to three dimensional integrals by Fourier transform
 techniques. The axial symmetry of the shape parametrization is
 also utilized effectively in simplifying these integrals. The
 method used for evaluating the nuclear potential (surface and
  Coulomb) is briefly
 described in Appendix A.

(iii) \underline{Rotational energy} : \\
The rotational energy of a nucleus is given by $
E_{R}={L^{2}\hbar^{2} \over 2I}$ where $I$ is the largest of the
principal-axis  rigid body moments of inertia. For a matter
distribution made diffuse by folding a Yukawa function over a
sharp-surfaced one, the  rigid body moment of inertia is modified
by the term $4M_{0}a_{M}^{2}$, where $a_{M}$ is the range
parameter of the folding function. The same diffuseness parameter
is used for both the charge and the matter distribution and
$a_{M}=a_{C}=0.704$ fm\cite{Sierk0}. \par The fission barriers
calculated from this model of potential energy have been found to
be within 1 MeV (for the angular momentum values which are sampled
in such experiments) of those which optimally reproduce fission
and evaporation-residue cross-sections for a variety of nuclei
with masses ranging from 150 to above 200\cite{Sierk0}. Langevin
dynamical calculations of fission fragment mass distribution in
fission of excited nuclei is reported in Ref. \cite{Kosenko} using
two liquid drop models(LDM's): the LDM with the sharp surface of
the nucleus and the finite range LDM and it is seen that the
fission fragment mass distributions and their variances calculated
with finite-range LDM are in much better agreement with
experimental data.
\begin{figure}[h!]
\centerline{\psfig{figure=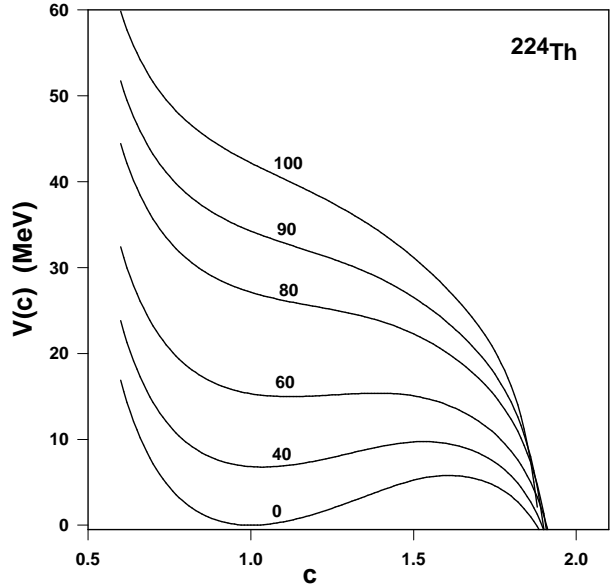,width=8cm}} \caption
{\label{b2}Potential energy $V(c)$ which includes the surface
energy, Coloumb energy and the rotational energy(as explained in
the text) for different angular momentum $l$ (marked in the
figure) in units of $\hbar$.}
\end{figure}
\par The deformation dependent potential energy
$V(c)$ which includes contributions from (i) surface energy (ii)
Coulomb energy and (iii) rotational energy is plotted as function
of the deformation coordinate $c$ for different values of angular
momentum $l$ as marked in Fig. \ref{b2}. It is seen from the
figure that the fission barrier decreases with increasing values
of angular momentum $l$ and gradually vanishes for higher values
of $l$. We have not included any shell and pairing effects in our
potential. Compound nuclei formed with large angular momentum in
heavy-ion collisions of highly excited nuclei ($E^* \sim 100$ MeV)
will generally also have high internal excitation energy. For
sufficiently high internal energies, shell and pairing effects are
very small and therefore can be neglected for all practical
purposes. The de-excitation of a compound nucleus by particle and
gamma emission may  however lead to a daughter nucleus with a
lower excitation energy where the quantum effects like pairing
correlations and shell effects start becoming important.  But at
such low excitation energies, fission cross section is also very
low and also the neutron emission threshold is not reached. The
nucleus predominantly cools by photon emission in this energy
regime. Since  we will mainly be concerned with prescission
neutron multiplicity and fission probability, neglecting shell
effects in the fission dynamics calculation is expected not to
introduce any serious error in our calculation.
\subsection{Level density parameter}
 The level density parameter is an
important input for our calculations. Fr\"obrich {\it et al.} made
an extensive study of different parameterizations  available for
this crucial quantity and finally considered the form given by
Ignatyuk {\it et al.} to be the most appropriate for the fission
process. We shall use the following level density parameter due to
Ignatyuk {\it et al.} \cite{Igna} which incorporates the nuclear
shell structure at low excitation energy and goes smoothly to the
liquid drop behavior at high excitation energy. In Ignatyuk's
approach the level density parameter is itself taken as a smooth
function of mass but with an energy dependent factor which
introduces the shell structure explicitly:
\begin{eqnarray}
a(E_{int})=\bar{a}(1+\frac{f(E_{int})}{E_{int}} {\delta M} ),
\label{2g}
\end{eqnarray}
\noindent with
\begin{eqnarray}
f(E_{int})=1-exp(-E_{int}/E_{D}) \nonumber
\end{eqnarray}
\noindent where  $\bar{a}$  is  the  liquid  drop  level  density
parameter, $E_{D}$ determines the rate at which the shell effects
disappear at high  excitations,  and  $\delta  M$  is  the  shell
correction  given  by the difference between the experimental and
liquid drop  masses,  $(\delta  M=M_{exp}-M_{LDM}  )$.  We  shall
further   use  the  shape-dependent  liquid  drop  level  density
parameter as function of elongation coordinate $c$ given as
\cite{Balian} (leaving out the curvature corrections),
\begin{eqnarray}
\bar{a}(c)=a_{v}A+a_{s}A^{\frac{2}{3}}B_{s}(c) \label{2h}
\end{eqnarray}
\noindent The choice of the values for the parameters $a_{v}$,
$a_{s}$ and the  dimensionless  surface  area $B_{s}$  by
  Fr\"obrich {\it et al.}\cite{Frob1} was motivated by the fact
that when using a stronger deformation dependence of the level
density parameter, e.g. that of Ref. \cite{Toke}, it was not
possible to find a universal, i.e. for all systems, the same
friction parameter $\eta$. Hence they selected among the different
possibilities the weakest coordinate dependence which is
consistent with data. Those values correspond to $a_{v}= 0.073
MeV^{-1}$ and $a_{s}=0.095 MeV^{-1}$\cite{Igna} which we shall use
in our work. The parametrization used for $B_{s}$ is of the
following form\cite{Frob4}:
\begin{eqnarray}
 B_{s}&=& 1+0.4(\frac{64}{9})(q-0.375)^{2} \hspace{1.1cm} (\mbox{if}\hspace{.5cm}  q < 0.452),\nonumber\\
      &=& 0.983 + 0.439(q-0.375) \hspace{1cm}  (\mbox{if} \hspace{.5cm}  q \geq 0.452).\label{2i}
\end{eqnarray}
where $q(c,h)$ is related to c \& h by Eq. \ref{2a}.
\begin{figure}[t!]
\centerline{\psfig{figure=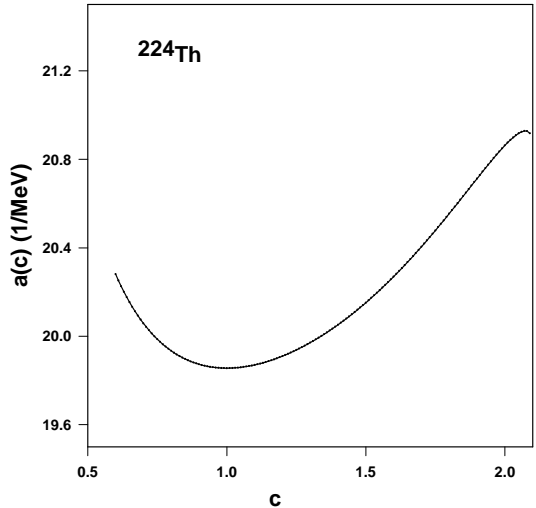,width=7cm}} \caption
{\label{b3}Variation of level density parameter $a$ with
elongation $c$.}
\end{figure}
 The variation
of the level density parameter $a(c)$ with the elongation
coordinate $c$ is shown in Fig. \ref{b3}. The fission rates turn
out to be sensitive to the detailed coordinate dependence of the
level density parameter. The temperature of the system is also
extracted using the level density parameter which in turn dictates
the emission of light particles. Thus emission rates therefore
also depend on the level density parameter.\par In  Ref.
\cite{Karpov}, the level-density parameter and the Helmholtz free
energy are calculated using the generalized finite-range liquid
drop model(LDM). The finite-range LDM  based on the
Yukawa-plus-exponential potential was generalized by
Krappe\cite{Krappe1} to describe the temperature dependence of the
nuclear free energy. This dependence is obtained by fitting the
results of the temperature-dependent Thomas-Fermi
calculation\cite{Guet1} with a finite-range formula. Based on
these calculations, the level-density parameter was approximated
by a leptodermous-type expression. The coefficients of this
expansion are in good agreement with those obtained earlier by
Ignatyuk {\it et al.}\cite{Igna}. The results of Langevin
dynamical calculations of the mean prescission neutron
multiplicity and fission probability are practically the same for
 the level-density parameter calculated with Ignatyuk's
coefficients as well as the one calculated using the generalized
finite-range liquid-drop model\cite{Karpov}. This fact establishes
the validity of our dynamical calculations performed with
Ignatyuk's level-density parameter.
\subsection{Free Energy}
The driving potential for a hot thermodynamic system such as the
excited nuclei has to be the free
energy\cite{Hofmann3,Frob4,Bohr2,Hofmann5} which can easily be
seen from the following arguments. The total energy change is
given by $dE_{tot} = TdS-Kdq$ where $Kdq$ is the work done and
$dS$ is the change in entropy. Using the relation $E_{tot}=F+TS$
in this formula, one obtains $K=-{(\partial F(q,T)/\partial
q)}_{T}$, i.e. the driving force $K$ is the negative gradient of
the free energy  $F$ with respect to the fission coordinate $q$ at
a fixed temperature $T$.
\begin{figure}[ht!]
\vspace{0.5cm}
\centerline{\psfig{figure=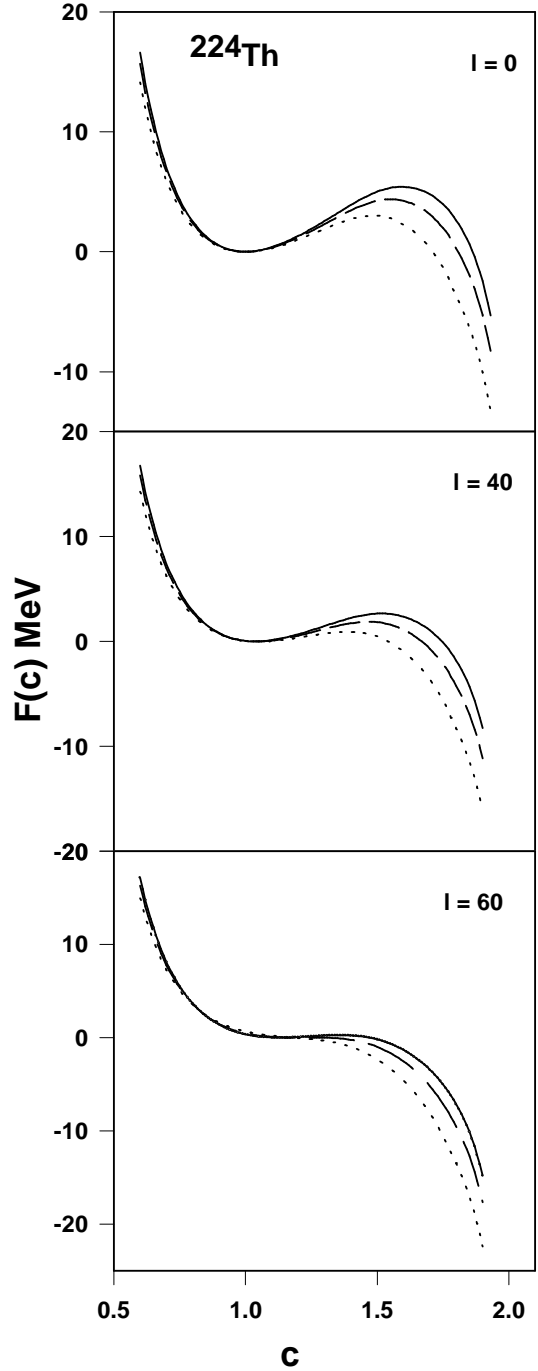,width=0.45\textwidth}} \caption
{\label{b4}Free energy $F(c)$ for three different temperatures 1
MeV(solid line), 2 MeV(dashed line) and 3 MeV(dotted line). The
plot is repeated for three different angular momentum $l$ (marked
in the figure) in units of $\hbar$.}
\end{figure}
 Considering the nucleus as a non
interacting Fermi gas, the following expression will be used for
free energy $F$ (as function of the elongation coordinate $c$),
\begin{equation}
 F(c,T) = V(c) - a(c) T^2, \label{2e}
\end{equation}
\noindent where $T$ is the temperature of the system, $V(c)$ is
the potential energy and $a(c)$ is the coordinate dependent level
density parameter. The driving force is thus given by
\begin{equation}
K=-{(\partial F(c,T)/\partial c)}_{T}=
-dV(c)/dc+(da(c)/dc)T^{2}\label{2f}
\end{equation}
i.e. it consists of the usual conservative force $-dV(c)/dc$ plus
a term which comes from the thermodynamical properties of the
fissioning nucleus, which enter via the level density parameter
$a(c)$, whose deformation dependence is now essential. The
properties of the heat bath enter in the description via the
temperature $T$, which is calculated from the internal energy
$E^{*}$ and the level density parameter $a$ by the Fermi gas
relation $T=\sqrt{E^*/a}$. Free energy $F(c)$ is plotted in Fig.
\ref{b4} for three different temperatures. It is seen from the
figure that the fission barrier decreases with increasing
temperature for a fixed value of angular momentum $l$. The plot is
repeated for three different angular momentum $l$ as seen from the
figure. It is also seen from comparison of Figs. \ref{b2} and
\ref{b4} that fission barrier in the free energy profile is lower
than that in the potential energy profile and this will have
significant impact in the calculation of different observables in
fission dynamics of hot nuclei where free energy will be used to
generate the driving force.
\subsection{Inertia}
We will make the Werner Wheeler approximation\cite{Davies1,Nix0}
for incompressible irrotational flow to calculate the collective
inertia term $m_{ij}$. We shall follow the work of Davies. {\it et
al.}\cite{Davies1} for this purpose. The total kinetic energy of
the system is given as
\begin{equation}
T = {1 \over 2}\rho_{m} \int v^{2} d^{3}r \label{2k}.
\end{equation}
 We
specialize here to axially symmetric shapes, for which the
velocity is given in cylindrical coordinates by
\begin{equation}
\vec{v} = \dot{\rho}\hat{e_{\rho}} + \dot{z}\hat{e_{z}}\label{2l},
\end{equation}
 \noindent where $\hat{e_{\rho}}$ and $\hat{e_{z}}$ are unit vectors in
$\rho$ and $z$ directions, respectively. The Werner-Wheeler method
is equivalent to assuming that $\dot{z}$ is independent of $\rho$
and $\dot{\rho}$ depends linearly on $\rho$, i.e.,$\dot{z} =
\sum_{i} A_{i}(z;q)\dot{q_{i}}$ and $\dot{\rho} ={\rho \over P}
\sum_{i} B_{i}(z;q)\dot{q_{i}}$, where $\dot{q_{i}}$ are the
generalized velocities and correspond to $\dot{c}$ and $\dot{h}$
in our case. $P=P(z;q)$ is the value of $\rho$ on the surface of
the shape at the position $z$. By virtue of the equation of
continuity, the velocity field $\vec{v}$ for an incompressible
fluid satisfies $\nabla\cdot \vec{v} =0$ and using this relation
it can be shown that the expansion coefficients $B_{i}$ and
$A_{i}$ are related by the equation $ B_{i} = -{1 \over
2}P{\partial A_{i} \over \partial z}$. The collective kinetic
energy of the system depends on the generalized velocities as
\begin{equation}
T= {1 \over 2}\sum_{i,j}m_{ij}(q)\dot{q_{i}}\dot{q_{j}}
\label{2m},
\end{equation}
 where $q$
denotes the generalized coordinates that specify the shape of a
system and corresponds to $c$ and $h$ in our case. Substituting
the expressions for $\dot{\rho}$ and $\dot{z}$ in Eq. \ref{2m} and
comparing Eq. \ref{2k} with Eq. \ref{2m} , we obtain for the
elements of the inertia tensor the result
\begin{equation}
m_{ij}= \pi\rho_{m} \int_{z_{min}}^{z_{max}} P^{2}(A_{i}A_{j} +{1
\over 8}P^{2}A_{i}^{\prime}A_{j}^{\prime})dz \label{2n},
\end{equation}
where the primes denote differentiation with respect to z. The
expansion coefficients $A_{i}$ are determined from the condition
that for an incompressible fluid the total(convective) time
derivative of any fluid volume must vanish. The formula for
$A_{i}(z:q)$ is given by the following expression
\begin{equation}
A_{i}(z;q) = {1 \over P^{2}(z;q)}{\partial \over \partial q}
\int_{z_{min}}^{z} P^{2}(z^{\prime};q)dz^{\prime}\label{2o}.
\end{equation}
\begin{figure}[t!]
\centerline{\psfig{figure=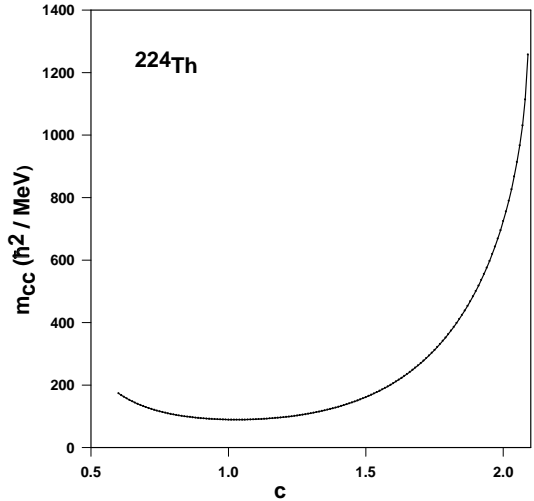,width=7cm}} \caption
{\label{b5}Variation of inertia $m_{cc}$ with elongation $c$.}
\end{figure}
This inertia tensor is  then calculated for different values of
the collective coordinate $c$ and $h$. $m_{ij}$ has three
components, namely, $m_{cc}$, $m_{hh}$ and $m_{ch}$. Fig. \ref{b5}
shows the variation of the component $m_{cc}$ with elongation $c$.
It needs mentioning that our calculations will be essentially in
one dimension($c$ coordinate), and hence we will be concerned
mainly with the component $m_{cc}$, the other two being not needed
for our purpose. We shall denote $m_{cc}$ by $m$ for the sake of
simplicity.
\subsection{Random force R(t)}
The instantaneous random force $R(t)$ plays a very crucial role in
the Langevin description of nuclear fission.  Though the initial
collective  kinetic energy of the fission degrees of freedom may
be lower than the fission barrier, as a result of receiving
incessant random kicks from the fluctuating force $R(t)$, the
fission degrees of freedom can finally pick up enough kinetic
energy to overcome the fission barrier. This random force is
modelled after that of a typical Brownian motion and is assumed to
have a stochastic nature with a Gaussian distribution whose
average is zero \cite{Abe1}. It is further assumed that $R(t)$ has
extremely short correlation time implying that the intrinsic
nuclear dynamics is Markovian. Consequently the strength of the
random force  can be obtained from the fluctuation-dissipation
theorem and the properties of $R(t)$ can be written as,
\begin{eqnarray}
\langle R(t) \rangle &=& 0 ,\nonumber\\
\langle R(t)R(t^{\prime})\rangle &=& 2\eta T \delta(t-t^{\prime})
. \label{2o}
\end{eqnarray}
\noindent where $\eta$ is the strength of the dissipation and $T$
is the nuclear temperature.
\section{One-body dissipation}
\subsection{Wall Friction}
It is already pointed out in the previous chapter that one-body
dissipation is the dominant mode of energy damping in nuclear
fission and is considered to be more successful in describing
fission dynamics than two-body viscosity \cite{Wada}.  At
excitation energies per nucleon much smaller than the Fermi energy
domain, the exclusion principle will severely restrict the phase
space available for two-body collisions and the one-body mechanism
is expected to dominate. Hence we shall use the one-body
dissipation for nuclear friction in the Langevin equation. The
standard prescription for one-body dissipation is the wall
friction which has been introduced in the previous chapter in some
detail. However, for realistic applications, it was found
\cite{Sierk2,Frob3} that the wall friction overestimates the one
body dissipation (as required to fit experimental data for fission
of hot nuclei) by an order of magnitude. In order to reproduce
simultaneously the measured prescission neutron multiplicities and
the variance of the fission fragment mass-energy distribution, the
reduction coefficient($k_s$) of the contribution from a wall
formula has to be decreased at least by half of the one-body
dissipation strength ($k_s$ $\leq$ $0.5$)\cite{Vanin1,Nadtochy1}.
\par In the microscopic derivation of the wall friction it is seen
that the energy transferred to particle motion in a dissipative
process of very short duration (smaller than the interval between
successive collisions between a particle and the wall) is given by
the original wall friction. In other words, energy damping is
given by wall friction irrespective of the shape of the potential
when time available for energy damping is short compared to the
time between successive collisions. When the dissipative process
lasts longer, a part of the energy transfer could be reversible
and the net energy transfer can be shown to depend on the shape of
the potential. One of the important assumptions of the wall
friction concerns the randomization of the particle motion. It is
assumed\cite{Blocki1} that successive collisions of a nucleon with
the one-body potential give rise to a velocity distribution which
is completely random. This is normally satisfied for one body
potentials whose shapes are rather irregular. If the particle
motion is not fully random, energy transferred to a particle from
a time-dependent wall could be partly reversible, resulting in a
reduced strength of the wall friction. The classical wall friction
 was  therefore re-examined in order to
distinguish between the reversible and irreversible energy
transfers. The irreversible energy transfer was identified with
the true one body dissipation. The wall friction was modified to
describe the irreversible energy transfer and the modified wall
friction would be applicable to systems in which particle motion
is not fully randomized. Studies of the nature of dynamical
systems \cite{Martin} demonstrated that even simple systems
possess a rich phase space structure and classical dynamics can be
idealized for the two extreme cases of either fully
regular(integrable) or fully chaotic(nonintegrable) motions. Most
of the systems of practical interest fall in between the above
limiting cases and they are `mixed' systems which display both
regular and chaotic(irregular) features. Arvieu {\it et al.}
\cite{Arvieu} pointed out the importance of topology of phase
space to characterize the motion of a particle moving in a
deformed potential. Blocki {\it et al.} \cite{BSS3} showed a
strong correlation between chaos in classical phase space and
efficiency of energy transfer from collective to intrinsic motion.
It was established by their calculations  that for slow
dissipative processes, wall friction is valid when the intrinsic
dynamics is fully chaotic. The wall friction was therefore
modified to make it applicable to mixed systems i.e., when mixing
in phase space is partial. Following the above line of argument,
we shall describe
 a specific model, namely the Chaos Weighted
Wall Friction, in the following sub-section.
\subsection{Chaos Weighted Wall Friction}
Pal $\&$ Mukhopadhyay\cite{Pal1} introduced a measure of chaos
into the classical linear response theory  for one body
dissipation, developed earlier by Koonin and Randrup \cite{Koonin}
and a scaled version of the wall friction, namely the ``chaos
weighted wall friction" was thus obtained. We shall present a
brief account of this model of one-body friction here. \par
Following the work of Koonin and Randrup, a classical system of
independent particles placed in a container with time-dependent
walls is considered which is described by Hamiltonian $H_0$ and
$H_1(t)$ respectively.
 Under
the linear response approximations which require the validity of a
perturbative treatment and the assumption that the relaxation time
of intrinsic motion is short in comparison with the time scale for
collective motion(adiabatic approximation), the rate of energy
dissipation from the collective motion of the wall can be
expressed as
\begin{equation}
\dot{Q}= -\int dr \int
\frac{dp}{{(2\pi)}^{3}}\dot{H}_{1}(r,p;t)\frac{\partial
f_0(r,p)}{\partial H_0}\left(\int_{0}^{\infty}dt^{\prime}
\dot{H}_{1}(R_{0}(r,p;t^{\prime}),{P}_{0}(r,p
,t^{\prime});t\right)\label{2p}
\end{equation}
This equation corresponds to a physical picture in which a
particle originating from a point $(r,p)$ in phase space
contributes a dissipation rate equal to the product of the initial
impulse received $\dot{H}_{1}(t)$ and the sum of all impulses
received subsequently along its entire (unperturbed) trajectory
$(R_0,P_0)$. In the above expression $f_0$ is the single particle
phase space distribution function governed by the unperturbed
Hamiltonian $H_0$ and the factor ${\partial f_0}/{\partial H_0}$
ensures that for a Fermi-Dirac distribution only particles near
the Fermi surface contribute. It was observed that relaxation of
the adiabatic approximation to realistic collective speeds reduces
the damping by $20\%-30\%$. Therefore the above equation shall
give the leading contribution to one-body damping even when the
collective and intrinsic time scales become comparable.
Considering a leptodermous system in which the nuclear potential
is uniform throughout the volume but rises steeply at the surface,
the above time integral can be written as a sum of the impulses
received by a particle during its successive encounters with the
nuclear surface along its unperturbed trajectory. Separating the
contribution of the first impulse given to a particle near its
point of origin at $t^{\prime} =0$(local part) from those arising
out of the successive reflections from other regions of the
nuclear surface (nonlocal part) Koonin and Randrup\cite{Koonin}
obtained the energy damping rate as
\begin{equation}
\dot{Q}=\dot{Q}_{local}+\dot{Q}_{nonlocal}\label{2q}
\end{equation}
The nonlocal term is determined by the correlation in the velocity
field sampled at successive reflection points of a particle
trajectory in the unperturbed system. This correlation, in turn,
depends on the nature of the velocity field at the cavity surface
as well as on the nature of the particle trajectory. It was found
that the local and the nonlocal parts completely cancel each other
in an integrable system in which particle trajectories are fully
regular. This essentially reflects a regular distribution of the
velocity fields  which results in a strong correlation  when
sampled at reflection points along a regular trajectory. It is
argued that energy transfer to the particle motion described by
the regular part of
 the classical phase space is reversible i.e., energy gained by a
particle from the wall is eventually fed back to the wall when
particle motion is regular. An integrable system is thus
completely non dissipative in this picture and wall friction tends
to the limit of zero energy loss. Noninteracting particles in a
spherical cavity constitute an example of an integrable system in
which particle trajectories are fully regular. \par In the case of
a non integrable system, on the other hand, where particle
dynamics is fully chaotic, the velocity fields sampled by such
trajectories are expected to be highly uncorrelated leading to a
vanishing nonlocal dissipation. This in turn corresponds to a
completely irreversible energy transfer arising from the local
term alone and energy dissipation in an irregular system can thus
be shown to reduce to the wall friction. Particle trajectories in
cavities with octupole and higher multipole deformations follows
the full wall friction limit which confirms that the energy
damping for irregular systems is entirely determined by the local
term. Considering classical particles in vibrating cavities of
various shapes, it was demonstrated in\cite{BSS3,BBS1} that while
the energy transfer is much smaller than the wall friction limit
in a cavity undergoing quadropule vibration, it reaches the wall
friction limit for higher multipole vibrations. Similar
conclusions were also reached \cite{Skalski,BBS2,BSS2,MSB} when
the particle motion was treated quantum mechanically, though the
quantal energy transfers were found to be  somewhat suppressed
compared to the classical ones. It was also noted that if the
interaction time is too short for any possible transfer of
particle energy to the wall after the first collision, the net
energy transfer rate would be given by the local term, or,
equivalently, the wall friction, irrespective of the system being
regular or chaotic.\par Most of the physical systems of interest
 however are neither fully integrable(full regularity) nor fully
nonintegrable(complete chaos). The dynamics of such systems
display both the characteristic features of regularity and chaos
in classical phase space. The measure for the degree of chaos or
nonintegrability for mixed systems is usually defined as the
relative volume of phase space that belongs to chaotic
trajectories. A trajectory is said to be regular when originating
from a given point on the cavity wall and moving in a given
direction, it closes smoothly in phase space. On the other hand,
another trajectory leaving the same point but in a different
direction could be a chaotic one which does not close in the phase
space. It has already been stated that regular trajectories
contribute to zero net dissipation while the dissipation due to
the chaotic trajectories correspond to the full wall friction. The
dissipation rate in mixed systems can therefore be decomposed into
following four terms,
\begin{equation}
\dot{Q}_{mixed}=
\dot{Q}_{local}^{regular}+\dot{Q}_{nonlocal}^{regular}+\dot{Q}_{local}^{chaotic}
+\dot{Q}_{nonlocal}^{chaotic}\label{2r}
\end{equation}
The first two terms on the right side represent the local and
nonlocal contributions to the dissipation due to the trajectories
which are regular and they  are expected to cancel each other as
noted before. The nonlocal term due to the chaotic trajectories
also vanish due to the random nature of the surface velocity
components at successive reflecting points as mentioned
previously. Therefore we are left with the local term due to the
chaotic trajectories and the net dissipation rate amounts to
$\dot{Q}={\dot{Q}}^{chaotic}_{local}$. This term represents the
contribution of the chaotic trajectories alone and can be written
as
\begin{equation}
\dot{Q}=\mu{\dot{Q}}_{wall}
\end{equation}
\noindent where $\mu$ is the fraction of the chaotic trajectories
and ${\dot{Q}}_{wall}=\rho_m \bar{v} \int \dot{n}^2 d\sigma$
($\dot{n}$ is the normal component of the surface velocity at the
surface element $d\sigma$) represents the full strength of the
wall dissipation where all trajectories (regular + chaotic) are
considered. The details of the derivation can be found in Ref.
\cite{Pal1}.\par The wall friction is thus modified by a factor
$\mu$ (chaoticity) which gives the average fraction of
trajectories which are chaotic when sampling is done uniformly
over the surface. In other words, the chaoticity $\mu$ is used to
express the degree of irregularity in the dynamics of the system.
This modified or scaled version of the wall friction is known as
the "chaos-weighted wall friction"(CWWF)\cite{Pal1}. \par The CWWF
coefficient $\eta_{cwwf}$ will therefore be given as
\begin{equation}
\eta_{cwwf}=\mu \eta_{wf} \label{2s}
\end{equation}

\noindent where $\eta_{wf}$ is the friction coefficient as given
by the original wall friction. This modified version is applicable
for any system  lying between a fully regular ($\mu =0$, i.e., no
dissipation) and a fully chaotic one ($\mu =1$, i.e, original wall
friction).  For mixed systems, the dissipation rate depends on the
degree of chaos in single particle motion of the nucleons within
the nuclear volume and it is thus necessary to calculate $\mu$ for
such systems. The chaoticity is a specific property of the
nonintegrability of the nuclear shape.  Thus it is required to be
calculated for all possible shapes of the nucleus up to the
scission configuration. In order to calculate the chaoticiy $\mu$,
it is required to identify a classical trajectory as a regular or
a chaotic one. For conservative Hamiltonian systems, the methods
which are mostly used to
 distinguish between regular and chaotic
trajectories are to investigate\\
 $(a)$ Poincare surfaces of section.\\
 $(b)$ Lyapunov exponents.\\
 In our work we will use the second method to evaluate the chaoticity $\mu$.
\subsection{Chaoticity from Lyapunov exponent}
 One
representative feature of a chaotic trajectory is its sensitivity
to initial conditions and the consequent exponential divergence of
the neighboring trajectories. A typical calculation for chaoticity
proceeds as follows. Two  chaotic trajectories (systems) having
very close initial conditions and governed by the same set of
equations of evolution will eventually fall apart very rapidly as
the time progresses  and will never come back close to each other
(except accidently). Here lies the unpredictability of a chaotic
system, though governed by deterministic equations and hence the
name deterministic chaos in contrast to noise which is statistical
in nature. The initial distance $\delta_{0}$ (a measure of
difference in the initial condition) between two trajectories can
diverge or converge as
\begin{equation}
\delta(t)
 = \delta_{0} \exp({\lambda t}). \label{2t}
\end{equation}

\noindent If the value of $\lambda$ is zero or negative, the
trajectories converge rapidly (integrable system) whereas positive
values imply exponential divergence and chaos. For a system with
$n$ dimensions in phase space there will be $n$ such exponents
corresponding to each dimension. The coefficient $\lambda_{i}$
($i=1$ to $n$) is known as the Lyapunov exponent in the limit when
time tends to infinity and the initial distance $\delta_{0i}$
tends to zero. Hence
\begin{equation}
\lambda_{i}= \lim_{\delta_{0i}\rightarrow 0}\lim_{t\rightarrow
\infty} \ln \left(\frac
{\delta_{i}(t)}{\delta_{0i}(0)}\right)\cdot
\left(\frac{1}{t}\right)\label{2u}.
\end{equation}
 \par
In order to calculate the dissipation according to the chaos
weighted wall friction (CWWF), the chaoticity $\mu$ is required
for all the deformations through which the nuclear shape evolves
with time. The chaoticity for each deformation is obtained by
considering particle trajectories in a cavity with the same
deformation and distinguishing between the regular and chaotic
trajectories. The chaoticity is defined  as the average fraction
of chaotic trajectories by uniformly sampling the trajectories
which originate from the nuclear surface. Hence we calculate it by
considering a large number of (typically 1000 or more)
trajectories whose starting points on the nuclear surface are
chosen at random. The initial coordinates of a classical
trajectory starting from the nuclear surface are chosen by
sampling a suitably defined set of random numbers such that all
initial coordinates follow a uniform distribution over the nuclear
surface. The initial direction of the trajectory is also chosen
randomly and its Lyapunov exponent is obtained by following the
trajectory for a considerable length of time. A particle's
trajectory is specified by giving its initial coordinates $\phi$
and $\theta$ on the nuclear surface and the components $v_x$,
$v_y$, $v_z$ of its velocity $\vec{v}$. For a cavity, a trajectory
is independent of the magnitude of the velocity, and hence four
quantities, namely $\theta$, $\phi$ and the orientation of
$\vec{v}$ (two angles) are sufficient to define the initial
conditions of a trajectory. The magnitude of $v$ may be used as a
convenient unit of velocity. The method\cite{BBSS} of computing
the Lyapunov exponent is based on computing numerically the
average rate of exponential divergence (or convergence) of two
trajectories with nearly identical conditions (differing only by a
small value greater than some predetermined noise threshold), in
the limit when the difference between the two initial conditions
tends to zero and the time over which the averaging is performed
tends to infinity. In general some of the exponents may be
positive and some negative. If positive exponents are present, the
largest of them will eventually dominate the divergence between
trajectories and it will control the exponential instability
leading to chaos. In the case of regular trajectories the exponent
is zero. The procedure consists in evolving numerically two close
trajectories originally separated in phase space by $\delta_0$,
for a given short interval $\tau$, after which the magnitude of
their separation is scaled back to $\delta_0$. The procedure is
then repeated $k$ times. The largest Lyapunov exponent can be
found from the limiting procedure
\begin{equation}
\lambda_{max}= \lim_{\delta_0 \rightarrow 0}\lim_{k \rightarrow
\infty}\frac{1}{k\tau}\sum_{i=1}^{k} \ln
\frac{\delta(\tau)}{\delta_0}.\label{2v}
\end{equation}
where $\delta^2(\tau)= d{\mathbf{p}}^{2} + d{\mathbf{q}}^{2}$ is
the square of the phase space separation between two trajectories
after time $\tau$. The result turns out to be essentially
independent of the direction in phase space of the original
displacement $\delta_0$. In our case the length of the time
interval $\tau$ was chosen to be $R/2v$ ($R$ being the radius of
the spherical system), i.e., $\tau$ =0.5, when time is measured in
units of $R/v$. Each trajectory is identified either as a regular
or as a chaotic one by considering the magnitude of its Lyapunov
exponent and the nature of its variation with time. Operationally,
a trajectory is deemed chaotic if for
  $t/t_{0} = 10^{4}$,($t_{0} = R/v$), $\lambda(t)$ saturates
  to finite values and the value is
greater than $10^{-3}$. The length of duration is found to be
sufficient for this decision since for intervals longer than this
time, Lyapunov exponent is found to be tending rapidly to zero for
regular trajectories. For other trajectories recognized as
chaotic, it stays at much higher and more or less steady value.
With this method, a large number of trajectories(Lyapunov exponent
calculated for each trajectory by following it for the time
$t/t_{0}=10^4$)
 is sampled for each shape of the cavity. Thus after marking
  and counting the chaotic trajectories $N_{ch}$ (those
  trajectories for which $\lambda \neq 0$)
 out of the total number of trajectories sampled, the measure of chaoticity
 $\mu$ for the deformation determined by the ratio of chaotic
 trajectories to the total number (say N) of sampled trajectories
 $\mu$ is given by
 \begin{equation}
 \mu=\frac{N_{ch}}{N} \label{2w}.
 \end{equation}
  The value of $\mu$ changes
from 0 to 1, as the nucleus evolves from a spherical shape to a
highly deformed one.\par The chaos parameter defined so far is a
classically defined quantity which is calculated by sampling
trajectories in the classical phase space. For the corresponding
quantum system, the chaos parameter is obtained from a measure of
the fluctuations of the single particle energy spectrum. For mixed
systems, it has been argued that \cite{Percival} that in the
semiclassical limit a spectrum should consist of regular and
irregular parts that are associated with the classical regular and
irregular regions of phase space.\par
\begin{figure}[b!]
 \centerline{\psfig{figure=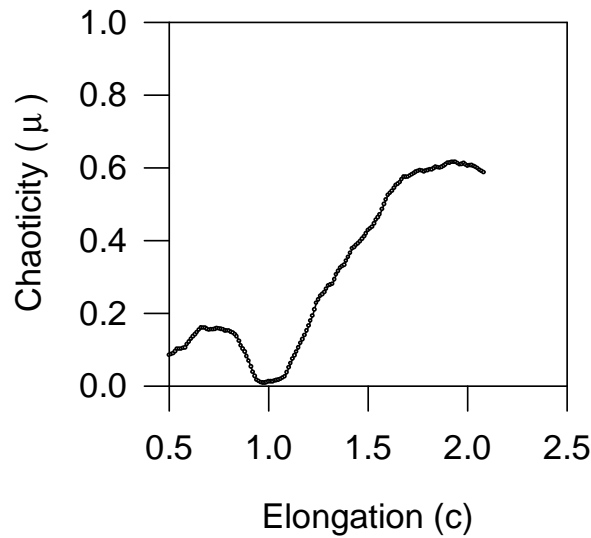,width=0.5\textwidth}} \caption
{\label{b6}Variation of chaoticity with elongation $c$ ($h=0$).}
\end{figure}

  Using the values of chaoticity calculated as above, the CWWF friction was
subsequently found \cite{Pal2,Pal3} to describe satisfactorily the
collective energy damping of cavities containing classical
particles and undergoing time dependent shape evolutions. Thus
suppression of the strength of the wall friction achieved in the
CWWF suggests that lack of full randomization(lack of chaos) in
single particle motion can provide an explanation for reduction in
strength of friction for compact nuclear shapes as required in the
phenomenological friction of Ref. \cite{Frob}. This motivated us
to use the CWWF for nuclear dissipation in fission dynamics which
is the main aim of the present thesis. In the present work, the
chaoticity is calculated over a range of shapes from oblate to the
scission configuration (at c=2.09 where neck radius becomes zero)
at small steps of c, the elongation coordinate. Fig. \ref{b6}
shows the calculated values of the chaoticity which will be
subsequently used to obtain the chaos-weighted wall friction.
Variation of chaoticity with elongation coordinate $c$ is plotted,
while the coordinate $h$ corresponding to the neck degree of
freedom is chosen to be zero ($h=0$). Very small values of
chaoticity for near-spherical shapes($c\sim 1$) implies a strong
suppression of the original wall friction for compact shapes of
the compound nucleus.
 Chaoticity, however increases as
the shape becomes more oblate or changes towards the scission
configuration.
\subsection{ Window Friction  and Center of mass correction to Wall Friction}
The wall friction assumes isotropic velocity distribution of
particles with respect to average drift velocity of the nucleus.
 In the final stages of fission, the
formation of a neck restricts the free passage of particles from
one half of the system to the other and the effect of this
restriction is that particles bombarding surface elements of each
part of the system come mostly from that part and the system is
characterized by a leftward as well as a rightward drift. The
relevant value of $\dot{n}$, the normal component of wall velocity
with respect to the drift velocity, for the left part of the
system is no longer the normal surface velocity with respect to
bulk of the whole gas (which is at rest) but the velocity with
respect to the leftward moving part. This enforces a correction in
the normal velocity of the wall called the center-of-mass motion
correction and the normal velocity of the surface element w.r.t
particles about to strike it will be ($\dot{n} -\dot{D}$), where
$\dot{D}$ is the relative
 velocity of the part under consideration with respect to the
 center-of-mass of the nucleus which is at rest\cite{Sierk1}. \par
The window friction is expected to be effective after a neck is
formed in the nuclear system \cite{Sierk1}. When the two halves
are in relative motion due to leftward and rightward drifts, any
particle passing through the window will damp the motion because
of the momentum transferred between the systems. The dissipation
rate due to window formation is given by the following
expression\cite{Blocki1}
\begin{equation}
\dot{E}_{win}(t)  = {1 \over 4 }\rho_{m}\bar{v} \Delta
\sigma(2{D_{\parallel}}^2+{D_{\perp}}^2), \label{2x}
\end{equation}
\noindent where $\dot{{D}_{\parallel}}$ and $\dot{{D}_{\perp}}$
are the components of $\dot{D}$ along and at right angles to the
normal through the window and $\Delta \sigma$ is the area of the
window. The radius of the neck connecting the two future fragments
should be sufficiently narrow in order to enable a particle that
has crossed the window from one side to the other to remain within
the other fragment for a sufficiently long time. This is necessary
to allow the particle to undergo a sufficient number of collisions
within the other side and make the energy transfer irreversible.
The window friction should be very nominal when neck formation
just begins. Its strength should increase as the neck becomes
narrower, reaching its classical value when the neck radius
becomes much smaller than the typical radii of the fragments. Very
little is known regarding the detailed nature of such a
transition, i.e., at which point to switch on the window friction.
In our calculation, a transition point $c_{win}$ is defined in the
elongation coordinate beyond which window friction will be
switched on. The assumption is that the compound nucleus evolves
into a binary system beyond $c_{win}$ and accordingly correction
terms for the motions of the centers of mass of the two halves
will be added to wall friction and the window friction will be
switched on as well for $c>c_{win}$. It is noted that while the
window friction makes a positive contribution to the wall friction
for $c>c_{win}$, the center of mass motion correction reduces the
friction. These two contributions thus cancel each other to a
certain extent and hence, the resulting wall-and-window friction
is not very sensitive to the choice of the transition point. This
point is explored by the following calculation. The transition
point $c_{win}$ can lie anywhere between $c=1.5$(where neck
formation just begins) and $c=2.08$(scission point). Calculations
for fission probability and prescission neutron multiplicity were
performed with different values for $c_{win}$ beyond 1.5, the
calculated values were in agreement within $5 \%$. Therefore, the
values of $c_{win}$ is not very critical for our purpose. A value
for $c_{win}$ is chosen at the point when neck radius is half the
radius of either of the would be fragments. The value of $c_{win}$
is thus halfway between its lower and the upper limit in terms of
the neck radius.
\subsection{Friction coefficient $\eta$}
 We shall use the following
expressions for the wall-and-window friction coefficients in one
dimension ($\eta=\eta_{cc}$)\cite{Gargi2},
\begin{eqnarray}
\eta_{wf}(c < c_{win})= \eta_{wall}(c < c_{win}), \label{2y}
\end{eqnarray}
\noindent where
\begin{eqnarray}
 \eta_{wall}(c < c_{win})=
{1 \over 2} \pi \rho_m {\bar v} \int_{z_{min}}^{z_{max}} { \left(
\frac{\partial \rho^2}{\partial c} \right)}^2  {\left[\rho^2 +
{\left({1 \over 2}\frac{\partial \rho^2} {\partial
z}\right)}^2\right]}^{-{1 \over 2}} dz, \label{2z}
\end{eqnarray}
\noindent and
\begin{eqnarray}
\eta_{wf}(c \ge c_{win})= \eta_{wall}(c \ge c_{win}) +
\eta_{win}(c \ge c_{win}), \label{2aa}
\end{eqnarray}
\noindent where
\begin{eqnarray}
\eta_{wall}(c \ge c_{win})&=&{1 \over 2} \pi \rho_m {\bar v}
\left\{\int_{z_{min}}^{z_N} {\left( \frac{\partial \rho^2}
{\partial c} + \frac{\partial \rho^2}{\partial z} \frac{\partial
D_1}{\partial c}\right)}^2 {\left[\rho^2 + {\left({1 \over
2}\frac{\partial \rho^2}{\partial z}\right)}^2\right]}^ {-{1 \over
2}}dz\right .\nonumber\\ &&+\left .\int_{z_N}^{z_{max}} {\left(
\frac{\partial \rho^2}{\partial c}+ \frac{\partial \rho^2}
{\partial z} \frac{\partial D_2}{\partial c}\right)}^2
{\left[\rho^2 + {\left({1 \over 2} \frac{\partial \rho^2}{\partial
z}\right)}^2\right]}^ {-{1 \over 2}}dz  \right\}, \label{2bb}
\end{eqnarray}
\noindent and
\begin{equation}
\eta_{win}(c \ge c_{win}) =    {1   \over   2}   \rho_m  {\bar  v}
{\left(\frac{\partial  R}{\partial  c}\right)}^2  \Delta  \sigma.
\label{2cc}
\end{equation}
 In  the  above  equations, $\rho^2$ is given by Eq. \cite{2b},
  $\rho_{m}$ is the mass density of the
nucleus, $\bar{v}$  is  the  average  nucleon  speed  inside  the
nucleus. $\bar{v}$  at zero temperature is defined as
\begin{equation}
\frac{\bar{v}}{c}=\frac{\bar{p}}{mc}=\frac{3}{4}\frac{\hbar}{mc}{(3{\pi}^{2}\rho_m)}^{1/3}.
\label{2dd}
\end{equation}
\noindent with the Fermi momentum $p_F=\hbar
k_F=\hbar{(3{\pi}^{2}\rho_m)}^{1/3}$ and $\rho_m=0.17 fm^{-3}$ is
the  nuclear matter density.\\
 $D_{1}$, $D_{2}$ are the positions of the
centers of mass of the two parts of the fissioning system relative
to the center of mass of the whole system. $z_{min}$ and $z_{max}$
are the two extreme ends of the nuclear shape along the $z$ axis
and $z_{N}$ is  the position of the  neck  plane that divides the
nucleus into two parts. In the window friction coefficient, $R$
$(=\mid D_2-D_1\mid)$ is the distance between centers of mass of
future fragments and $\Delta \sigma $ is the area of the window
between the two parts of the system.\par The  wall  friction
coefficients  given  by (Eqs. \ref{2z} and \ref{2bb}) are obtained
\cite{Blocki1} under the assumption of a fully  chaotic nucleon
motion within the nuclear volume. However, a fully chaotic motion
is achieved only when the nuclear shape is extremely irregular
whereas the nucleon motion is partly  chaotic in  varying  degrees
for  typical nuclear shapes through which a nucleus evolves when
it undergoes fission. We have already argued in the preceding
section that for such cases, the chaos  weighted wall  friction
($\eta_{cwwf}$) should be employed instead of the original   wall
friction.   Accordingly,  we  shall  replace Eqs. \ref{2z}  and
\ref{2bb} by their chaos weighted versions and the chaos-weighted
wall-and-window   friction   (denoted henceforth by $\eta_{cwwf}$)
is subsequently obtained as

\begin{eqnarray}
\eta_{cwwf}(c < c_{win})= \mu (c) \eta_{wall}(c < c_{win}),
\label{2ee}
\end{eqnarray}
\noindent and
\begin{eqnarray}
\eta_{cwwf}(c \ge c_{win})= \mu(c) \eta_{wall}(c \ge c_{win}) +
\eta_{win}(c \ge c_{win}). \label{2ff}
\end{eqnarray}
\vspace{0.8cm}
\begin{figure}[h]
\centerline{\psfig{figure=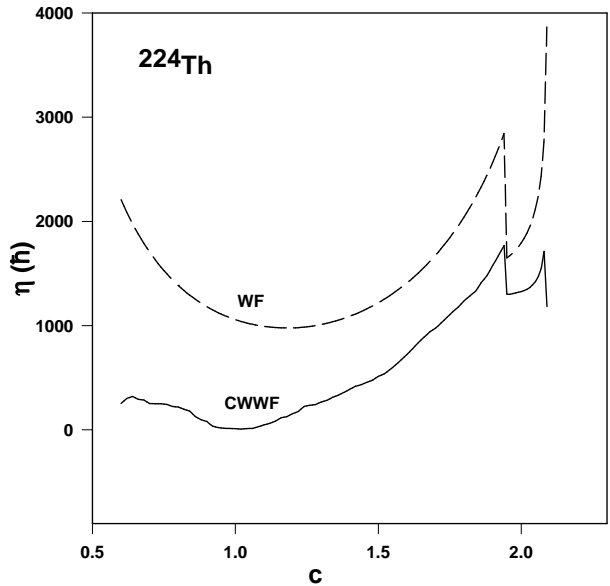,width=8cm}} \caption
{\label{b7}Variation of friction coefficient $\eta$ with
elongation $c$ for chaos weighted wall friction (full line) and
wall friction (dashed line).}
\end{figure}
Fig. \ref{b7} depicts the variation of the friction coefficient
$\eta$ with elongation $c$ for both CWWF and WF.
\begin{figure}[h]
\centerline{\psfig{figure=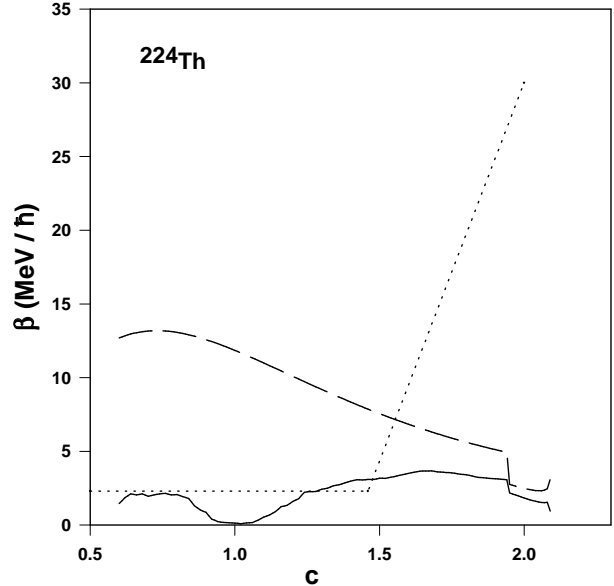,width=8cm}}
 \caption
{\label{b8}Reduced one-body friction coefficient $\beta$ with
chaos weighted wall friction (full line) and wall friction (dashed
line) frictions. The phenomenological reduced coefficient (dotted
line) from Ref.[Frob3] is also shown.}
\end{figure}
Defining  a  quantity  $\beta (c)= \eta (c)/ m(c)$ ($m$
corresponds to the inertia component $m_{cc}$) as the reduced
friction coefficient, its dependence on the elongation coordinate
is shown in Fig. \ref{b8} for both the WF and CWWF  for the
$^{224}$Th nucleus. A strong suppression of the original wall
friction  for compact shapes of the  nucleus can be immediately
noticed in the CWWF. This implies that the friction is very small
for near spherical shapes ($c \sim 1$), the physical picture
behind which is  as follows. A particle moving in a spherical mean
field represents a typical integrable system  and its dynamics is
completely regular. When the boundary of the mean field is set
into motion (as in fission), the  energy gained  by the particle
at one instant as a result of a collision with the moving boundary
is eventually fed back to the boundary motion  in the  course of
later collisions. An integrable system thus becomes completely
nondissipative in this picture  resulting in  a vanishing friction
coefficient. This aspect  has  been investigated extensively on
earlier occasions \cite{Koonin,Pal1} and has been found to be
valid for any generic integrable system. The reduction in the
strength  of the  wall friction  is  shown in Fig. \ref{b8} along
with the phenomenological reduced friction obtained in Ref.
\cite{Frob3} to fit experimental data. The reduction in the
strength of the phenomenological friction is found to be very
similar to that obtained from chaos considerations. Though the
CWWF agrees qualitatively with the phenomenological friction for
$c<1.5$, it is beyond its scope to explain the steep increase of
phenomenological friction for $c>1.5$. We shall discuss  this
point further while presenting the results in chapter 4.\par The
strong shape-dependence of the CWWF can have some interesting
consequences.  In  a dynamical description of fission, a compound
nucleus spends most of its time in undergoing shape  oscillations
in  the  vicinity  of its ground state shape before it eventually
crosses the saddle and proceeds towards the scission point. Since
the  spin  of  a compound nucleus formed at a small excitation
(small temperature) is also small, its ground state shape is
nearly spherical  and  in this region the CWWF  is also small.
Conversely,  higher spin values are mostly populated in  a highly
excited  compound nucleus (high temperature) making its ground
state shape highly deformed and thus it experiences  a stronger
CWWF. Therefore, if one uses a shape-independent friction in a
dynamical model of fission, its strength  has to increase with
increasing temperature in order to give  an equivalent description
to that provided   by the temperature-independent but
shape-dependent CWWF. In fact,  it  was  observed in Ref.
\cite{dioszegi}   that   a shape-dependent  friction fits the
experimental data equally well to that achieved  by  a  strong
temperature-dependent  friction. Since  there  is a physical
justification for shape-dependence in nuclear friction from chaos
considerations, it  is  quite  likely that   the   above  strong
temperature-dependence,  at  least  a substantial part of it, is
of dynamical origin  as  explained  in the  above  and  thus  is
an  artifact  arising  out  of using a shape-independent friction.
\par It  must  be  pointed  out, however, that  one   would expect
a temperature-dependence   of nuclear friction   from   general
considerations such as larger phase space becoming accessible for
particle-hole   excitations at higher   temperatures.   In  a
microscopic model of  nuclear friction  using  nuclear  response
function,  Hofmann  {\it  et al.} \cite{Rummel} have obtained a
nuclear friction which depends upon  temperature  as  $0.6T^{2}$
(leading term).   This  may  be compared  with  the  empirical
temperature-dependent  term  of $3T^{2}$  which  was  found   in
Ref. \cite{dioszegi}.  It therefore appears  that  only  a small
fraction of the empirical temperature-dependence can be accounted
for by the inherent temperature-dependence  of  nuclear  friction
while  the rest of it has a dynamical origin as we have discussed
in the above. In fact it shown in Ref. \cite{Shaw1} that in hot
rotating $^{240}Cf$ where saddle to scission emission dominates
the prescission particle and $\gamma$-ray spectra, the extracted
nuclear  dissipation coefficient is found to be independent of
temperature and large dissipation during the saddle to scission
path provides good fit to the $\gamma$-ray spectra. Similar
conclusion is reached in Ref. \cite{Shaw2} for $^{200}Pb$ where
the friction parameter is found to be smaller inside the saddle
and increases sharply outside the saddle in order to match
experimental data.  In the present work, we shall not consider any
empirical temperature-dependence of the CWWF or WF in order to
study solely the effects of shape-dependence. In what follows, we
shall  use both the WF and CWWF  in a dynamical model of fission
and shall investigate the effect of the reduction in the CWWF
strength on prescission neutron multiplicity, fission probability
as well as evaporation residue cross section.

\chapter{ Fission widths of hot nuclei using Langevin dynamics}

In this chapter, a detailed systematic study of the fission rates
is made using both chaos-weighted wall friction(CWWF) and wall
friction(WF) in the Langevin equation for different spins and
temperatures of the compound nucleus. Similar studies of fission
rates using different versions of friction is not found in the
literature except for the work  of Abe {\it et al.}
\cite{Abe2,Wada1} where time dependent fission rates were
calculated using both the two-body viscosity and the wall friction
in the Langevin equation. The aim of the present study is two
fold. First, the effect of introducing the chaos factor in nuclear
friction parameter on fission rates will be examined at different
excitation energies and spins of the compound nucleus. The effect
of the choice of different scission criteria on fission rates will
also be investigated. The second one concerns a parametric
representation of the numerically obtained  fission width, the
need for which arises as follows. Fission width is an essential
input along with particle and $\gamma$ widths for a statistical
theory in the stationary branch of compound nucleus decay. Kramers
\cite{Kramers} obtained an analytical expression for the
stationary fission width assuming a large separation between the
saddle and scission points and a constant friction. Gontchar {\it
et al.} \cite{Frob4, Frob3} later derived a more general
expression taking the scission point explicitly into account but
still assuming a constant shape independent friction coefficient.
The CWWF however is not constant and is strongly shape dependent
and hence the corresponding stationary fission width cannot be
analytically obtained. Thus it becomes necessary to find a
suitable parametrization of the numerically obtained stationary
fission widths using CWWF in order to use them in the statistical
regime of the compound nucleus decay. In the next section, the
procedure for calculating the fission rates by numerically solving
the Langevin equation  is given.  The results of our calculation
will be presented in section 3.2 while a summary of the chapter
will be given in the last section.

\section{Solving the Langevin equation to calculate fission rate}
\subsection{Inputs to the equation}
 We  have  discussed in details  the Langevin equation
along with the various inputs of our model in the last chapter.
The same definitions and notations will be followed henceforth.
The shape parameters $c,h$  as suggested by Brack {\it et al.}
\cite{Brack} will be taken  as the collective coordinates for the
fission degree of freedom. We shall further assume in the present
work that fission would proceed along the valley of the potential
landscape in $(c,h)$ coordinates though we shall consider  the
Langevin equation in elongation $(c)$ coordinate alone in order to
simplify the computation. Consequently, the one-dimensional
potential $V(c)$ in the Langevin equation will be defined as
$V(c)=V(c,h)$ {\it at valley}. The potential $V(c,h)$ is
calculated over a grid of $(c,h)$ values and the valley of the
minimum potential is located. The potential values along this
valley are used in solving the Langevin equation. Other quantities
such as inertia $m(c)$ and friction $\eta(c)$ will also be
similarly defined. We shall, therefore, proceed by considering $c$
and its conjugate momentum $p$  as the dynamical variables for
fission for our present study and the coupled Langevin equations
in one dimension will be given as

\begin{eqnarray}
\frac{dp}{dt}   &=& -\frac{p^2}{2}   \frac{\partial}{\partial
c}\left({1  \over  m} \right) -
   \frac{\partial F}{\partial c} - \eta \dot c + R(t), \nonumber\\
\frac{dc}{dt} &=& \frac{p}{m} . \label{3a}
\end{eqnarray}

The different inputs to the Langevin dynamics, namely the shape
dependent collective inertia $m$, the friction coefficient $\eta$,
the free energy of the system $F$, and the random force $R(t)$ are
 described in detail in the previous chapter.  The usual wall
friction (WF) as well as its modified version CWWF will both be
used for the friction coefficient $\eta$ in order to study the
effect of introducing the chaos-factor in CWWF. The random force
is given by $R(t)=g\Gamma(t)$, where the diffusion coefficient $D
(=g^2)$ is related to the friction coefficient $\eta$ through the
Einstein relation $D(c) = \eta (c) T$. In this framework the
temperature $T$ is simply a measure of the non-collective part of
the nuclear excitation energy $E_{int}$ and related to the later
by the usual Fermi gas relation $E_{int} = a(c)T^2$, where $a(c)$
is the level density parameter of the considered nucleus at a
nuclear deformation characterized by $c$. The excitation energy
itself is determined by the conservation of the total energy as
will be discussed afterwards.

\subsection{Method of solving the equation}
 The Langevin
equation \cite{Lang} has been applied to many fields of physics.
It was solved on  several occasions for parabolic potential wells
\cite{Uhlen,Chandra}. Recently there has been a
publication\cite{Abe3} which showed a general analytical scheme to
solve multi-dimensional Langevin equations near a saddle point.
Since analytical solutions of the Langevin equation can be derived
for quadratic potentials only, it is mostly handled by numerical
simulations. The ``direct simulation'' method is the most commonly
used method for solving the Langevin equation. In this method,
once the stochastic equation of motion is formulated, it is
straightforward to numerically simulate the process in question,
using a random number generator to supply the noise. By repeating
the simulation with different sequences of random numbers, one
obtains independent realizations of the process in question,
reflecting the statistical distribution of events. This method
however becomes impractical when studying rare outcomes. For
instance, while computing the cross-section for the fusion of two
heavy nuclei, where the vast majority of realizations will end
with the nuclei flying apart, the number of simulations required
to obtain even a handful of fusion events may well be
prohibitively large. Recently a method based on the idea of
importance sampling\cite{Mazonka} has been developed  for
computing the probabilities of rare events for processes described
by Langevin equations. However, for our purpose of studying
fission dynamics by the Langevin equation, the direct simulation
method to solve the equation is quite applicable since fission
probabilities of hot nuclei are not too small. We shall follow
\cite{Abe1} for the purpose.\par The Langevin equation is a
stochastic differential equation, which has a rapidly changing
force in addition to the ordinary one. The random force $R(t)$ has
no well-defined derivatives with respect to time $t$ and hence the
usual methods of solving differential equations such as
Runge-Kutta  algorithm cannot be utilized for solving the Langevin
equation. Therefore it has to be integrated by direct methods. To
integrate  by the iteration method the Langevin equation is
rewritten as follows:
\begin{eqnarray}
\frac{dp}{dt} &=&  H(p,c)+ g\Gamma(t), \nonumber\\
\frac{dc}{dt} &=& \frac{p}{m}  \label{3b}
\end{eqnarray}
 where
 \begin{eqnarray}
 H(p,c) &=&-\frac{p^2}{2} \frac{\partial}{\partial
c}\left({1 \over m} \right) -
   \frac{\partial F}{\partial c} - \eta \dot c, \nonumber\\
      g &=& \sqrt{\eta T}   \label{3c}
\end{eqnarray}

Integrating Eq.~(\ref{3b}) from $t$ to $t+\tau$, we have
\begin{eqnarray}
p(t+\tau)-p(t) &=& \int_{t}^{t+\tau}
dt^{\prime}H(p(t^{\prime}),c(t^{\prime})) + g\int_{t}^{t+\tau}
dt^{\prime} \Gamma(t^{\prime}) \nonumber\\
               &\simeq& \tau H(p(t),c(t)) + g\tilde{\Gamma_{1}}(t),\nonumber\\
c(t+\tau)-c(t) &=& {1 \over m} \int_{t}^{t+\tau} dt^{\prime}
p(t^{\prime}) \nonumber\\
               &\simeq& {\tau p(t)\over m } \label{3d}
\end{eqnarray}
By repeating the same procedure $n$ times starting at $t=0$, we
can obtain $p(T)$ and $c(T)$ at time $T=n\tau$. At each step, we
need $\tilde{\Gamma_{1}}(t) =\int_{t}^{t+\tau} \Gamma(t^{\prime})
dt^{\prime} $, which is a sum of Gaussian random numbers, and
thereby is itself a Gaussian random number. Its average and
variance can be calculated with the statistical properties of
$\Gamma(t)$ and they are as follows.
\begin{eqnarray}
\langle\tilde{\Gamma_{1}}(t)\rangle &=&\int_{t}^{t+\tau}
dt^{\prime}\langle \Gamma(t^{\prime})\rangle = 0\nonumber\\
\langle\tilde{\Gamma_{1}}(t)^{2}\rangle &=&\int_{t}^{t+\tau}
dt_{1}\int_{t}^{t+\tau}dt_{2}\langle
\Gamma(t_{1})\Gamma(t_{2})\rangle = 2\tau.
\end{eqnarray}
 Thus we can
describe $\tilde{\Gamma_{1}}(t)$ by a new Gaussian random number
$\omega_{1}(t)$, i.e.
\begin{eqnarray}
\tilde{\Gamma_{1}}(t) &=& \sqrt{\tau}\omega_{1}(t) \label{3e},
\end{eqnarray}
 where
$\omega_{1}(t)$ has the following properties (average and
variance),
 \begin{eqnarray}
\langle \omega_{1} \rangle &=& 0\nonumber\\
\langle{\omega_{1}}^{2}\rangle &=& 2.
\end{eqnarray}
\noindent The method for generation of random numbers following a
particular type of
 probability distribution is described in Appendix B. The same method
 is followed
here as well as in all other cases where Monte-Carlo simulation is
used in the present thesis. Eq.~(\ref{3d}) is the first order
approximation in $\tau$. In a case such as fission one has to
describe the system over a rather long period, which means one has
to repeat the small steps many times. The time step $\tau$ is
restricted by the friction strength as well as the force or the
derivative of the potential. A very small time step of $0.005
\hbar /MeV$ for numerical integration is used in the present work.
The numerical stability of the results is checked by repeating a
few calculations with still smaller time steps. The solution of
Langevin equation of a free Brownian particle obtained by this
method and its comparison with the analytical solution is
presented in Appendix C. The close agreement of the numerical and
the analytical solution confirms validity of the algorithm used in
solving the Langevin equation. The units and dimensions used for
different dynamical variables in the Langevin equation are
described in details in Appendix D.
\subsection{Initial conditions and scission criteria}
\underline{\large{\it{Initial conditions}}}\\
 The initial
distribution of the coordinates and momenta are assumed to be
close to equilibrium and hence the initial values of $(c,p)$ are
chosen from sampling random numbers following the
Maxwell-Boltzmann distribution. Starting with a given total
excitation energy ($E^{*}$) and angular momentum ($l$) of the
compound nucleus, the energy conservation in the following form,
\begin{equation}
E^{*}=E_{int}+V(c)+p^{2}/2m \label{3f}
\end{equation}
 gives
 the intrinsic excitation energy $E_{int}$ and the corresponding nuclear
 temperature \\ $T=(E_{int}/a)^{1/2}$ at each step of the fission
 process(each integration step).
 The centrifugal potential is included in $V(c)$ in the above equation.
 Once the initial conditions are fixed one can integrate the
 system of equations of motion, i.e. Eq.~(\ref{3a}) using their
 finite difference version i.e., Eq.~(\ref{3d}). At each time step
 $[t,t+\tau]$ one draws a random number from a gaussian
 distribution which defines the fluctuating force and thus
 generates a trajectory in the variable $c$.\\
 \underline{\large{\it{Scission criteria}}}\\
 ``Scission" implies a transition from a continuous nuclear
 configuration (that becomes unstable for a number of reasons)
  to a configuration in which the nuclear system consists of
  separated shapes.
 The scission configuration is determined by the intersection
 points of the stochastic Langevin trajectories of the fissioning
 system, with the scission surface in the coordinate subspace.
 For an arbitrary dimensional model the crucial problem is how
 to define the scission surface. In fact, at the present time
 there is no unambiguous criterion of the scission condition.
 The condition of zero neck radius can be considered as one (the
 simplest) of the scission conditions.  However, this definition is
 unsatisfactory because
  the description of nucleus
 based on liquid-drop model looses significance when the neck
 radius becomes comparable with the distance between nucleons.
Hence, it is often supposed \cite{Brack,Popov,Brosa} that the
scission occurs at a critical deformation with a relatively thick
neck. Physical arguments lead to determine the scission surface as
the locus of points at which the following equation is satisfied:
\begin{equation}
{\left(\frac{{\partial}^2V}{\partial{h}^2}\right)}_{c=const}=0
\end{equation}
\noindent This means that stability against variations in the neck
thickness is lost. Such a criterion of scission can be called the
criteria of instability of the nucleus with respect to variations
in the thickness of its neck \cite{Adeev8,Brack,Adeev9,Kosenko1}.
It should be noted that this scission condition corresponds to the
shapes of the fissioning nucleus with a finite neck radius, with
$0.3R_0$ on the average \cite{Brack,Popov,Bao}, where $R_0$ is the
radius of the spherical compound nucleus and is given by
$1.16A^{1/3}$, A being the mass number of the fissioning nucleus.
Another acceptable and physically sensible criterion is based on
the equality of the Coulomb repulsion and the nuclear attraction
forces between future fragments. It was shown in \cite{Davies2}
that this scission condition leads to scission configurations for
the actinide nuclei with approximately the same neck radius
equaling $0.3R_0$.  In Ref. \cite{Nadtochy}, a probabilistic
criterion is proposed for the scission of a fissile nucleus, where
the probability is estimated by considering scission as a
fluctuation. The effect of the probabilistic criterion of
nuclear-scission on fission-process observables, such as the
moments of the mass-energy distribution of fission fragments and
prescission neutron multiplicity is demonstrated and it is shown
that the Strutinsky criterion\cite{Popov}, according to which
nuclear scission occurs at a finite neck radius of $0.3R_0$, is a
good approximation to the probabilistic scission criterion in
Langevin dynamical calculations employing reduced wall friction,
with the reduction factor being less than $0.5$.\\ These arguments
led us to choose the scission criterion for our work as those
configurations of the fissioning nucleus where the nuclear shapes
have finite neck radius of $0.3R_0$. However, it is worthwhile to
notice that results of fission dynamics of hot nuclei obtained
with this scission criterion and those with the criterion of zero
neck radius do not vary much.
 \subsection{Fission Rate}
 A Langevin
 trajectory will be considered as undergone fission if it reaches the scission
  point $c_{sci}$ (corresponding to the scission criteria defined above)
   in course of its time evolution. The calculations
 are repeated for a large number (typically 100,000 or more) of trajectories
 and
 the number of fission events are recorded as a function of time.
 At each iteration step, we calculate the probability of the system
 remaining as
 compound nucleus, $P_{C.N.}$, i.e, number of samples with $c < c_{scis}$
 divided by the total number of samples, and then calculate the
 fission rate as follows,
 \begin{equation}
 r(t) = -{1 \over P_{C.N.}}{dP_{C.N.} \over dt}\label{3g}
 \end{equation}
 As the rate calculated at each time step is still fluctuating, a time
 averaging is made over time $\Delta t$ as
 \begin{equation}
 r(t) = {1 \over \Delta t}\int_{t-{\Delta t}/2}^{t+{\Delta t}/2}
 r(t) dt\\
      = {1 \over \Delta t }\ln(P_{C.N.}(t-{\Delta t}/2)/P_{C.N.}(t+{\Delta
      t}/2))\label{3h}\\
 \end{equation}

\noindent The fission width $\Gamma_{f}(t)$ is given by $\hbar
r(t)$.

 The procedure described in this section to compute the fission
  rate is implemented in
 a computer code called FISSWDTH developed by us and is used for the
 numerical calculation presented in the next section. A brief
 description of FISSWDTH is given in Appendix F.

\section{Results}
\begin{figure}[b!]
\centerline{\psfig{figure=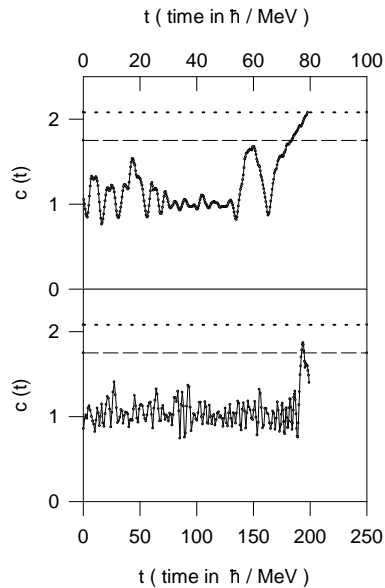,width=5cm}} \caption{The
upper panel shows a typical Langevin trajectory
 reaching the scission point (dotted line). The lower panel shows
 a trajectory which returns to the potential pocket after crossing
 the saddle point(dashed line).}\label{c1}
\end{figure}

\begin{figure}[t!]
\centerline{\psfig{figure=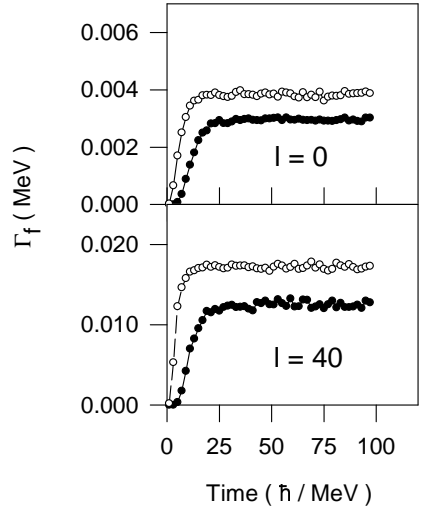,width=5.5cm}} \caption{Time
developments of fission widths for compound nuclear spins of $0$
and $40$ (in units of $\hbar$) at a temperature of $T=2.6$ MeV.
Open circles correspond to trajectories for which the saddle point
crossing is considered as fission. Solid circles represent
trajectories which reach the scission point.}\label{c2}
\end{figure}

A typical Langevin trajectory of $^{200}Pb$ nucleus which has
reached the scission point and has
 ended up as a fission event is
 shown in Fig. \ref{c1} (upper panel). Another trajectory, the kind of
 which is
 less frequent, is shown in the lower panel of the same figure. The Langevin
 trajectory in this case crosses the saddle point and after spending some time
 beyond the saddle point drifts back into the potential pocket again. Such
 trajectories may or may not finally reach the scission point within the
 observation time and corresponds to a to-and-fro motion
 across the saddle and
 essentially portrays the stochastic nature of the dynamics.
 This point is further illustrated in Fig. \ref{c2} where
 time development of the
 fission rates are plotted.
  Two different criteria are used to define a fission
 event here. The filled circles correspond to fission events defined as those
 trajectories reaching the scission point whereas the open circles correspond
 to those crossing the saddle point. The fission rate is very small for both
 the cases at the beginning when the compound nucleus is just formed and
 the Langevin dynamics
 has just been turned on.
Subsequently the fission rate grows with time and
 after a certain equilibration time it reaches a stationary value which
 corresponds to a steady flow across the barrier.
 The fission rate defined at the saddle point reaches the stationary value
 earlier than that defined at the scission point. The time difference
 between them gives the average time of descent from the saddle to
 the scission.
  This observation was also
 made in earlier works \cite{Wada1}. The main purpose of the present
 discussion is to investigate the role of backstreaming in the fission
 process. It is observed in Fig. \ref{c2} that the stationary
 fission rate at
 saddle point is higher than that at the scission point. The difference between
 these two stationary rates can be regarded as due to backstreaming.
 The backstreaming is thus small compared to the steady outward flow though
 it is not negligible. This also shows that crossing the saddle point is
 not an adequate criteria for fission in stochastic calculations and can lead
 to an overestimation of the fission rate.
\begin{figure}[htb]
\centerline{\psfig{figure=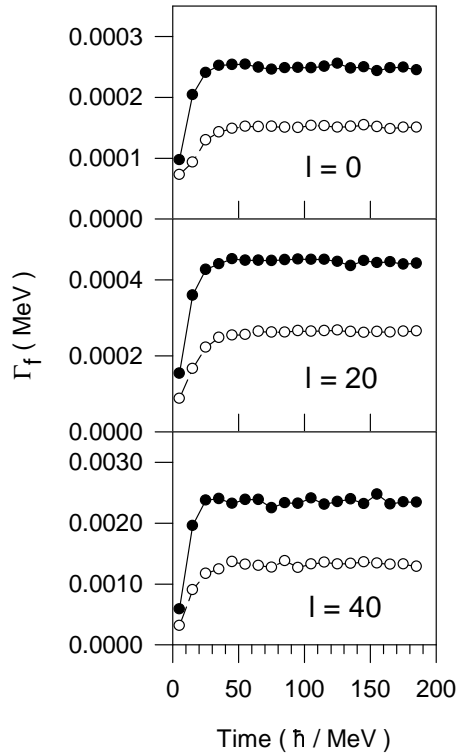,height=10cm,width=6cm}}
\caption{ Time developments of fission widths  calculated with
chaos-weighted wall-and-window friction(solid circles) and
wall-and-window friction(open circles) frictions for different
compound nuclear spins $l$ (in units of $\hbar$) at a temperature
of $T=1.83$ MeV.
  } \label{c3}
\end{figure}
\par
 The fission rates calculated with chaos weighted wall-and-
 window friction is  next compared with those obtained with wall-and-window
 friction. The calculations were performed for a wide range of spin and
 temperature of the compound nucleus.
 Fig. \ref{c3} shows the fission widths at three spins
 of the compound nucleus
 $^{200}Pb$. The effect of suppression in the chaos weighted wall
 friction
 shows up as an enhancement by about a factor of 2 of the stationary
 fission rates. Similar
 enhancement of the stationary fission rate calculated with chaos weighted
 wall-and-window friction in comparison with that obtained with wall-and-window
 friction are also observed for a wide range of compound nuclear spin
 and temperature. The enhancement factor (of about 2) remains almost the
 same when different choices of $c_{win}$ are used in the window
 friction.
\begin{figure}[htb]
\centerline{\psfig{figure=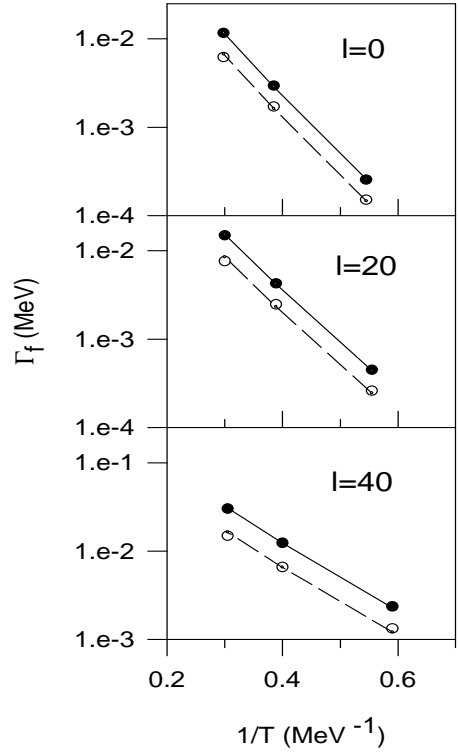,height=10cm,width=6cm}}
\caption{  Temperature dependence of stationary fission widths
calculated with chaos-weighted wall-and-window friction (solid
circles) and wall-and-window friction(open circles) frictions for
different com[pound nuclear spins $l$ (in units of $\hbar$). The
lines are fitted as explained in the text.
  } \label{c4}
\end{figure}

 We next systematically extracted the stationary fission widths at different
 temperatures for a given spin of the compound nucleus. This was done by
 taking the average of the fission rates in the plateau region.
 These fission rates are essentially the Kramers' limit of the Langevin
 equation under consideration and we expect the stationary fission
 widths $\Gamma_{f}$ to depend upon the
 temperature  $T$ as $\Gamma_{f}(l,T)= A_{l} \exp (-b_{f}/T)$ for a given
 spin ($l$)
 of the compound nucleus where $b_{f}$ is the height of the fission barrier
 in the free energy profile and $A_{l}$ is a parameter. Such a dependence
 of stationary fission widths on temperature
 was indeed found and is shown in Fig. \ref{c4}.
 The parameter $A_{l}$ can now be extracted by fitting the calculated fission
 widths with the above expression.
\begin{figure}[h!]
\centerline{\psfig{figure=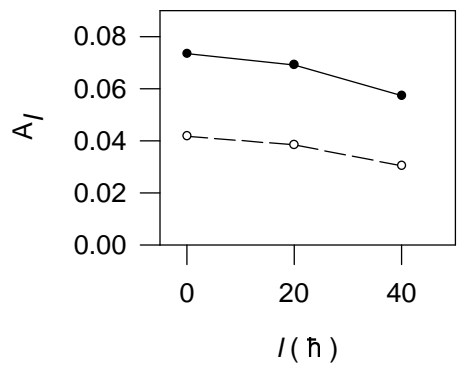,width=6cm}}
\caption{Variation of the parameter $A_l$ with compound nuclear
spin $l$.} \label{c5}
\end{figure}
\par
  Subsequently we looked into
 the dependence of the parameter $A_{l}$ on $l$, a few typical plots of which
 are shown in Fig. \ref{c5}. The variation of $A_l$ with the spin
 $l$ is plotted both for chaos-weighted wall friction (solid
 circles) and wall friction(open circles). The nature of variation
 is similar for both types of friction, while the magnitudes
 differ almost by a factor of 2 which is expected from the
 enhancement observed in the stationary fission rates with CWWF
 (Fig. \ref{c3}).
 Using these values of $A_{l}$, one can now obtain this parameter value
 for any arbitrary spin by interpolation. Even with a limited number of
 calculated values, the interpolated values will be quite reliable since
 $A_{l}$ depends on $l$ rather weakly as can be seen in Fig. \ref{c5}.
 Consequently
 it becomes possible to extract the fission width of a compound nucleus of
 any given temperature and spin from a set of a limited number of
 calculated widths. This fact will be very useful in statistical model
 calculations where fission widths are required at numerous values of
 temperature and spin which are encountered during evolution of
 a compound nucleus.
 Therefore in such cases where analytical expressions
 for fission widths cannot be obtained, the above systematics can generate
 fission widths from a limited set of calculations.
 \begin{figure}[h!]
\centerline{\psfig{figure=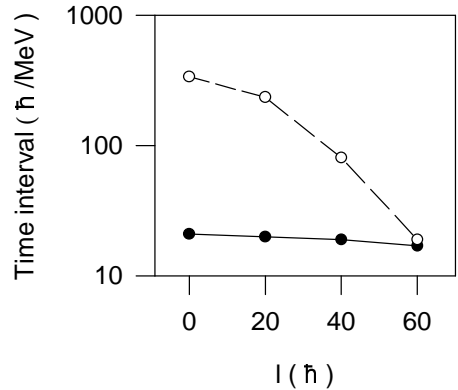,width=6cm}}
\caption{Dependence of the equilibration time $\tau_{eq}$(solid
circles) and the fission lifetime $\tau_{f}$(open circles) on
compound nuclear spin $l$.} \label{c6}
\end{figure}
\par
  Next we will demonstrate the importance of dynamical models at higher angular momentum.
 Two time scales are of physical significance in the Langevin description
 of dynamics of fission. One is the equilibration time $\tau_{eq}$, the time
 required to attain a steady flow across the barrier. The other is the
  stationary fission life time $\tau_{f}= \hbar / \Gamma_{f}$.
 Fig. \ref{c6} shows these
 time intervals for different values of spin of the compound nucleus
 $^{200}Pb$. At very small values of spin, the fission life time $\tau_{f}$ is many times
 longer than the equilibration time and one can neglect the equilibration time.
  This means that a statistical theory
 for compound nuclear decay is applicable in such cases. On the other hand,
 $\tau_{eq}$ and $\tau_{f}$ become comparable at  higher values of the
 compound nuclear spin and this corresponds to a dynamics dominated
 decay of the compound nucleus. Statistical models are not meaningful
 in these cases and dynamical descriptions such as Langevin equation
 become essential for fission of a compound nucleus. The results
 described in this section is reported in  Ref. \cite{Gargi1}.\par
 Before concluding this chapter, we would like to point out that
 our approximation of restricting the calculation to one-dimension
 is a reasonable one. This is so because it has been found in
 earlier works\cite{Wada1} that the fission rate in two-dimensional and
 one-dimensional cases differ by not more than $15\%$ while the
 stationary fission rate predicted by CWWF and WF vary by more than
 $100\%$. Moreover, we have also checked that the prescission
 neutron multiplicity and fission probability change by less than
 $5\%$ when the input fission rates are changed by $15\%$.
 Therefore we estimate that the uncertainty associated with our
 calculation is rather small allowing us to compare our results
 with experimental data, as we shall do in the next chapter.

 \section{Summary}

 In the preceding sections, we have presented a systematic study of fission
 widths using the Langevin equation. Among the various physical inputs
 required for solving the Langevin equation, we paid particular attention
 to the dissipative force for which we chose the wall-and-window
 one-body friction. We
 used a modified form of wall friction, the chaos weighted wall friction, in
 our calculation. The chaos weighted wall friction took into account
 the nonintegrabilty of
 single particle motion in the nucleus and it resulted in a strong suppression
 of friction strength for near spherical shapes of the nucleus. The fission
 widths calculated with chaos weighted wall friction turned out to be about
 twice the widths calculated with the normal wall friction. The
 chaos weighted wall friction thus enhances the fission rate substantially
 compared to that obtained with normal wall friction.

 We further made a parametric representation of the calculated fission widths
 in terms of the temperature and spin of the compound nucleus. It was found
 that this parameterized form can be well determined from the
 fission widths calculated over a grid of spin and temperature values
 of limited size.
 This fact would make it possible to perform statistical model
 calculation of the decay of a highly excited compound nucleus where the
 fission
 widths are to be determined from a dynamical model such as the Langevin
 equation. When the friction form factor has a strong shape dependence as in
 the chaos weighted wall friction, the corresponding fission widths cannot be
 obtained in an analytic form. In such cases, the frequently required values
 of the fission width in a statistical model calculation can be made
 economically accessible through the parameterized representation of the
 fission width which has to be obtained in a separate calculation similar to
 the present one. In this calculation of fission rates, we restricted our
 investigations to the $^{200}Pb$ nucleus as a representative
 example. This nucleus has been experimentally formed at a number
 of excitation energies \cite{Newton1,Forster,Fabris}. The fission probability and
 the prescission neutron and $\gamma$ multiplicity were measured
 in these experiments. These quantities can be calculated in a
 statistical model which requires the fission width as well as the
 neutron and $\gamma$ emission width as inputs to the
 calculation. In particular, the input fission width plays a
 critical role in order to reproduce the experimentally determined
 prescission neutron and $\gamma$ multiplicities at high
 excitation energies(typically a few tens of MeV or higher) in
 statistical model calculations. While the neutron and $\gamma$
 widths can be obtained from standard Weisskopf formula
 \cite{Frob},
 a dynamical theory is required to calculate the fission width of
 the hot compound nucleus. The work  reported in this chapter is a step in this
 direction and the application of these parametrized fission widths
 (described in this chapter) in dynamical calculations to
 reproduce experimental data is reported in the next chapter.

\chapter{Prescission neutron multiplicity and fission probability
from Langevin dynamics of nuclear fission}
 The  Langevin  equation
has  been used extensively in the recent years
\cite{Abe2,Frob,Nadtochy1,Richert2} in order to explain the
prescission  neutron  multiplicity  and  fission probability of
highly  excited  (typically a few tens of MeV and above) compound
nuclei  formed  in  heavy-ion  induced fusion reactions. It was
shown by Fr\"obrich and his coworkers\cite{Frob3} that the wall
friction cannot reproduce simultaneously experimental data for
excitation functions of prescission neutron multiplicity and
fission probability. It was found that different values of the
reduced friction parameter $\beta$ ($=\eta/m$) ranging from
$\beta=3 \times 10^{21} s^{-1}$ to $\beta= 20 \times 10^{21}
s^{-1}$ was required to fit experimental data for different
systems(compound nucleus). This is not a satisfactory situation
because it is essential that a physically meaningful reduced
friction coefficient $\beta$ should be a universal parameter for
different systems. This realization led to introduce shape
dependence in the friction coefficient and this shape dependent
empirical friction parameter of Fr\"obrich successfully explained
experimental data for different systems for prescission neutron
multiplicity and fission probability simultaneously \cite{Frob3}.
The application of the wall-plus-window dissipation to detailed
and systematic analysis of the data from fusion-fission reactions
have resulted in deduction of the value of $k_s$ (coefficient
which reduces the magnitude of wall friction) for light as well as
heavy fissioning systems\cite{Vanin2}. The conclusions reached are
as follows: (i) $k_s$ =$0.5$ for lighter fissioning systems, (ii)
$k_s$ = $0.1-0.2$ for the heaviest ones in order to get a good fit
of the parameters of the fission fragment mass energy
distribution, and (iii) a good quantitative description of the
prescission neutron multiplicities and angular anisotropy could be
achieved for heavier systems for $k_s$ = $0.5-1.0$. This was an
indication that the factor $k_s$ by which the wall friction needs
to be reduced might depend on the collective coordinate and
excitation energy. Shape dependent nuclear friction coefficient is
also extracted in Ref. \cite{Shaw1,Shaw2} in order to match giant
dipole resonance $\gamma$-ray spectra. It has already been
mentioned that an empirical suppression in the wall friction
coefficient similar to the one demanded by the experimental data
is achieved in the chaos weighted wall friction(CWWF)
microscopically, based on physical arguments. In the previous
chapter, the fission widths have been calculated using the CWWF in
Langevin dynamics and are found to be less by a factor of 2 from
those calculated by the standard wall friction. This reduction in
fission width is expected to influence the fission probability and
prescisssion neutron multiplicity favourably and this encouraged
us to do a full dynamical calculation coupled with particle
evaporation. The main motivation of the work presented in this
chapter is to verify to what extent the chaos-weighted wall
friction can account for the experimental prescission neutron
multiplicity and fission probability data.\par The Langevin
equation will be solved by coupling it with neutron and $\gamma$
evaporation at each step of its time evolution. Following the work
of Fr\"obrich {\it et al.} \cite{Mavil},  a combined dynamical and
statistical model will be used for our calculation in which a
switching over to a statistical model description will be made
when the fission process reaches the stationary regime. The
prescission  neutron multiplicity and fission probability will be
obtained  by sampling over  a  large number  of Langevin
trajectories. We  shall perform calculations at  a number of
excitation energies for each of the compound nuclei $^{178}$W,
$^{188}$Pt,$^{200}$Pb,$^{213}$Fr, $^{224}$Th, and $^{251}$Es. A
detailed comparison of the calculated values with the experimental
data will be presented. We have mainly considered heavier mass
nuclei ($A \geq 150$) since for the lighter mass nuclei, fission
width is much less than the corresponding neutron widths and hence
the prescission neutron multiplicities in those cases are not
sensitive to the choice of nuclear friction. It is for the heavier
mass nuclei that the neutron and fission widths become comparable
and their competition strongly dictates the final observables.\par
It is worthwhile here to  point  out  a special feature of the
present work. We do not have  any adjustable  parameter  in  our
entire calculation. All the input parameters except the friction
coefficients are fixed by standard nuclear  models.  The  chaos
weighted  wall  friction(CWWF) which is used for the nuclear
friction coefficient is obtained following a specific  procedure
\cite{Pal1}   which explicitly considers particle  dynamics in
phase  space  in order to calculate the chaoticity factor $\mu$.
There  is  no free parameter in this calculation of friction. In
fact, our main aim in this work is to calculate observable
quantities  using  the theoretically predicted friction   and
compare   them  with experimental values in order to draw
conclusions  regarding  the validity  of the theoretical model of
nuclear friction. As it would turn out,  our calculation  would
not  only  confirm  the theoretical model  of CWWF, it would also
provide physical justification for the empirical values of
friction used in other works \cite{Frob3,Shaw1}. The present  work
is thus expected to contribute significantly to our understanding
of the dissipative mechanism in nuclear fission.\par The different
steps of the combined dynamical and statistical model calculation
will be briefly described in the next section. The  calculated
prescission neutron multiplicities and fission probabilities will
be  compared with the experimental values in sec. 4.2.  A summary
of the results will be presented in the last section.
\section{Combined dynamical and statistical model }
\subsection{Introduction}
After the formation of a  fully equilibrated compound system in a
heavy-ion fusion reaction, the decay of the compound nucleus can
follow two different routes. On the first route the nucleus
undergoes fission, i.e, it predominantly separates into two heavy
fragments (binary fission) which is called a fusion-fission
process. During fission the intermediate system evaporates light
particles $(n,p,\alpha)$ and $\gamma$-quanta until scission when
the neck radius is shrinking to zero. These are called prescission
particles. After scission the heavy fragments are still excited
and continue to evaporate light particles and $\gamma$-quanta.
These are called post scission particles. It is possible to
distinguish experimentally between pre- and post-scission
particles.  Along the second decay route the nucleus does not
undergo fission and the excitation of the compound nucleus is
removed solely by the evaporation of light particles and
$\gamma$-rays. The evaporation of light particles of a particular
kind stops when the excitation energy has dropped to a value below
the corresponding binding energy. The deexcitation of the system
thus ends with the formation of the so-called evaporation
residues. For $\gamma$-quanta the emission process lasts until
zero excitation energy and the lowest possible spin value are
reached.
 During the formation process of the compound system, some
light particles can be emitted, which are of increasing importance
with increasing bombarding energy. These particles are called
pre-equilibrium particles and since our dynamical model starts
from the formation of a equilibrated compound nucleus, we do not
take into account these pre-equilibrium particles and hence the
experimental data should to be accordingly corrected for these
pre-equilibrium emission before any comparison is to be made.
Another possibility which may happen is that the intermediate
complex formed in the collision is not a fully equilibrated
compound system and the process is called fast-fission or
quasi-fission process. In our model we always assume that an
equilibrated compound system is formed and hence we will not deal
with fast fission or quasi-fission processes.
\subsection{Initial conditions}
 In our
calculation, we first specify the entrance channel through which a
compound nucleus is formed. In the reaction process, the compound
nucleus can be formed with different values of the angular
momentum. For starting a trajectory, an orbital angular momentum
value ($l$) is sampled from a fusion spin distribution as the
proper weight function. This spin distribution is usually
calculated with the surface friction model\cite{Frob5}. This
calculation also fixes the fusion cross section thus guaranteeing
the correct normalization of the fission and the evaporation
residue cross sections within the accuracy of the surface friction
model. Assuming complete fusion of the target with the projectile,
and if both the nuclei are assumed to be spherical, the spin
distribution of the compound nucleus calculated with the surface
friction model is usually found to follow the following analytical
form,
\begin{eqnarray}
\frac{d \sigma(l)}{dl} = \frac{\pi}{k^2} \frac{(2l+1)}{1+\exp
\frac{(l-l_{c})}{\delta l}} \label{4a}
\end{eqnarray}
\noindent where $k$ is the wavenumber of the relative motion. This
is used as the angular momentum weight function with which the
Langevin calculations for fission are started. The situation is
more complicated if one or both initial nuclei are deformed. In
principle one should then consider all possible relative
orientations of the nuclei and follow their relative trajectories
from an infinite distance up to fusion. We will avoid such
cumbersome calculations and restrict ourselves to the simplified
procedure by assuming that both target and projectile are
spherical in shape. The parameters $l_{c}$ and $\delta l$ should
be obtained by fitting the experimental fusion cross sections. It
is however found that these parameters for different systems as
calculated by the surface friction model, follow an approximate
scaling \cite{Frob} and hence it is not necessary to perform new
surface friction model calculations for the spin distributions of
each system. We shall, therefore, use the scaled values of these
parameters. The quantity $l_c$ scales as
\begin{equation}
l_c= \sqrt{A_P \times A_T/A_{CN}} \times ({A_P}^{1/3} +
{A_T}^{1/3}) \times (0.33+0.205 \times
\sqrt{E_{cm}-V_c}),\label{4b}
\end{equation}
\noindent when $0<E_{cm}-V_c<120MeV$; and when $E_{cm}-V_c>120
MeV$, the term in the last brackets is put equal to 2.5. $E_{cm}$
is the center of mass energy while $V_c$ is the Coulomb barrier.
For $V_c$, a simple Coulomb ansatz is used which is given by the
following relation,
\begin{equation}
V_c=(5/3)\times c_3 \times Z_PZ_T/({A_P}^{1/3} + {A_T}^{1/3}+1.6)
\label{4c}
\end{equation}
\noindent with $c_3 = 0.7053 MeV$. The diffuseness $\delta_l$ is
found to scale as
\begin{eqnarray}
\delta l &=& {(A_PA_T)}^{3/2}\times 10^{-5} \times [1.5+0.02\times
(E_{cm}-V_c-10)]\hspace{0.3cm}\mbox{for}\hspace{0.1cm}E_{cm}>V_c+10,\nonumber\\
         &=& {(A_PA_T)}^{3/2}\times 10^{-5} \times [1.5-0.04\times
(E_{cm}-V_c-10)]\hspace{0.3cm}\mbox{for}\hspace{0.1cm}E_{cm}<V_c+10.\label{4d}
\end{eqnarray}
\noindent The initial spin  of the compound nucleus will be
obtained by sampling the above spin distribution function.\par The
 Langevin equation and the different inputs for the
Langevin dynamics is described in chapter 2. In order to integrate
the Langevin equations, we need to fix the initial conditions from
which the evolution of the compound system starts. It is also
assumed that a fully equilibrated compound nucleus is formed at a
certain instant and this point of time is fixed as the origin of
our dynamical trajectory calculation. The initial distribution of
the coordinates and momenta $(c_0,p_0)$ is assumed to be close to
equilibrium and hence their initial values are chosen from
sampling random numbers following the Maxwell-Boltzmann
distribution. The energy available in the center-of-mass frame
$E_{cm}$ can be obtained from the beam energy of the projectile
and the target and projectile mass. The excitation energy
($E^{*}$) of the compound nucleus is obtained from $E_{cm}$ in the
following manner.
\begin{eqnarray}
E^*&=&E_{cm}-Q_{fus}\nonumber\\
Q_{fus}&=&\Delta_{CN}-(\Delta_{target}+\Delta_{projectile}).\label{4e}
\end{eqnarray}
\noindent where $Q_{fus}$ is the $Q$-value of the fusion reaction
which forms the compound nucleus(CN), and $\Delta$ is the mass
defect of the respective nuclei.  The target and projectile being
in their ground states, shell corrections are to be incorporated
to get the proper mass defects\cite{NWC}. The compound nucleus
however is formed in an excited state where shell corrections are
expected to disappear and hence not included in the mass defect of
the compound nucleus. The mass defect of the  hot compound nuclei
is taken from Ref. \cite{Myers2}. The total energy conservation of
the following form,
\begin{equation}
E^{*}=E_{int}+V(c)+p^{2}/2m    \label{4f}
\end{equation}
\noindent gives the intrinsic excitation energy $E_{int}$ and the
corresponding  nuclear  temperature $T=(E_{int}/a)^{1/2}$ at each
time step of integration. The centrifugal potential  is  included
in the potential energy $V(c)$ in the above equation.
\subsection{Particle emission}
The process of light particle emission from a compound nucleus is
governed by the emission rate $\Gamma_{\nu}^{\alpha}$ at which a
particle of type $\nu$ (neutrons, protons and $\alpha$-particles)
is emitted at an energy in the range
 $[e_{\alpha}-\frac{1}{2}\Delta
e_{\alpha},e_{\alpha}+\frac{1}{2}\Delta e_{\alpha}]$ before the
compound nucleus eventually undergoes fission. Several theoretical
approaches have been proposed in order to describe the emission
from a deformed, highly excited and rotating
nucleus\cite{Strum,Richert2,Frob6,Ohta1,Richert1}.\par
\begin{figure}[b!]
\centerline{\psfig{figure=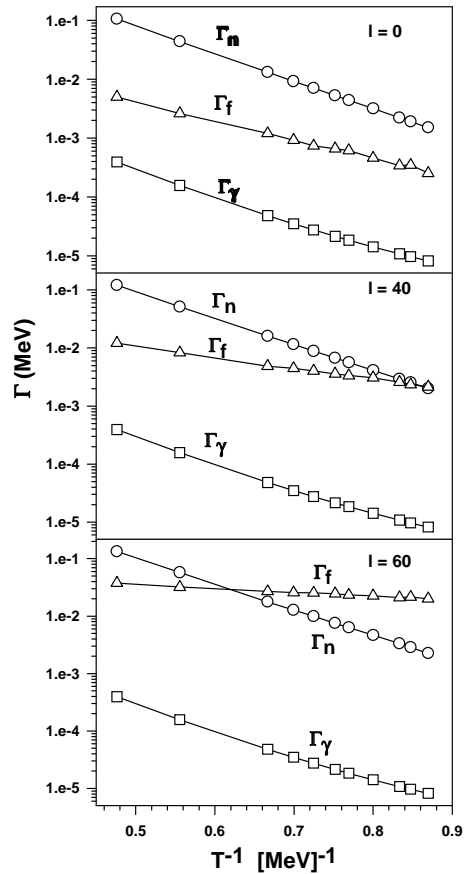,width=6cm}} \caption{Comparison
of neutron($\Gamma_n$), gamma($\Gamma_{\gamma}$) and fission
($\Gamma_f$)widths plotted as function of $T^{-1}$ for three
different angular momentum $l$.} \label{d1}
\end{figure}

 According to
Weisskopf's conventional evaporation theory \cite{Weiss}, the
partial decay width for emission of a light particle of type $\nu$
 is given
by
\begin{equation}
\Gamma_{\nu} = (2s_{\nu}+1){m_{\nu} \over
\pi^{2}\hbar^{2}\rho_{c}(E_{int})}\int_{0}^{E_{int}-B_{\nu}-\Delta
E_{rot}} d\varepsilon_{\nu}\rho_{R}(E_{int}-B_{\nu}-\Delta
E_{rot}-\varepsilon_{\nu})\varepsilon_{\nu}\sigma_{inv}
(\varepsilon_{\nu})\label{4g}
\end{equation}
 where $s_{\nu}$ is the spin of the emitted particle $\nu$,
 $m_{\nu}$ is the reduced mass with respect to the residual
 nucleus. $E_{int}$ is the  intrinsic excitation energy  of the parent nucleus,
  $B_{\nu}$ is the binding
 energy of the emitted particle calculated by the liquid drop
 model \cite{Swiat1}, $\varepsilon_{\nu}$ is the energy of the emitted
 particle and $\Delta E_{rot}$ is the change of the rotational energy
  due to the angular momentum carried away by the rotating particle. The procedure for
calculation of the binding energy
 $B_{\nu}$ of the emitted particle is given in Appendix E.
 The level densities of the compound and residual nuclei
 are denoted by $\rho_{c}(E_{int})$ and
 $\rho_{R}(E_{int}-B_{\nu}-\Delta E_{rot}-\varepsilon_{\nu})$ and is given by the
 following formula for the level density of a
 nucleus\cite{Bohr2},
 \begin{equation}
 \rho(E_{int},A,I)=(2I+1)\left[{\hbar^2 \over 2J_{0}}\right]^{3/2}{\sqrt{a}
 \over 12} {exp(2\sqrt{aE_{int}}) \over E_{int}^{2}}\label{4h}
 \end{equation}
 where $J_{0}$ is the moment of inertia and $I$ is the angular momentum of the rotating system.
  The level density
 parameter for particle emission is taken to be deformation independent and is given by
 $a=A/10 MeV^{-1}$.
The inverse cross sections are given by Ref. \cite{Blann},
\begin{eqnarray}
\sigma_{inv}(\varepsilon_{\nu})&=& \pi
R_{\nu}^{2}(1-V_{\nu}/\varepsilon_{\nu})\hspace{1cm} (\mbox{for}
\hspace{1cm} \varepsilon_{\nu}
>
V_{\nu})\nonumber\\
              &=& 0 \hspace{3.6cm}(\mbox{for} \hspace{1cm} \varepsilon_{\nu} <
              V_{\nu})\label{4i}
\end{eqnarray}
with
\begin{equation}
R_{\nu}=1.21[(A-A_{\nu})^{1/3}+A_{\nu}^{1/3}]+(3.4/\varepsilon_{\nu}^{1/2})\delta_{\nu
,n},\label{4j}
\end{equation}
where $A_{\nu}$ is the mass number of the emitted particle $\nu$
$(\equiv n,p,d,\alpha)$. The Coulomb barrier is zero for neutron
whereas for the charged particles the barrier is given by
\begin{equation}
V_{\nu}= [(Z-Z_{\nu})Z_{\nu}K_{\nu}]/(R_{\nu}+1.6)\label{4k}
\end{equation}
with $K_{\nu}=1.32$ for $\alpha$ and deuteron and 1.15 for proton.
For the emission of giant dipole $\gamma$-quanta we take the
formula given by Lynn\cite{Lynn}
\begin{equation}
\Gamma_{\gamma}={3 \over \rho_{c}(E^{*})}\int_{0}^{E_{int}-\Delta
E_{rot}}d\varepsilon\rho_{R}(E_{int}-\Delta
E_{rot}-\varepsilon)f(\varepsilon)\label{4l}
\end{equation}
with
\begin{equation}
f(\varepsilon)= {4 \over 3\pi}{1+\kappa \over mc^{2}}{e^{2} \over
\hbar c}{NZ \over A} {\Gamma_{G}\varepsilon^{4} \over
(\Gamma_{G}\varepsilon)^{2}+(\varepsilon^{2}-E_{G}^{2})^{2}}\label{4m}
\end{equation}
with $\kappa=0.75$, and $E_{G}$ and $\Gamma_{G}$ are the position
and width of the giant dipole resonance. Fig. \ref{d1} shows the
plot of comparison of neutron, gamma and fission widths plotted on
a logarithmical scale as function of $T^{-1}$ for three different
values of angular momentum $l$. The fission width ($\Gamma_f$)
shown here is calculated by following the procedure described in
chapter 3 using CWWF. The competition between neutron and fission
width is the main determining factor in deciding the fate of the
compound nucleus. At low angular momentum, fission width
$\Gamma_{f}$ is much less than the neutron width $\Gamma_{n}$ but
with rise of angular momentum the two widths become comparable.
 These widths depend upon the temperature, spin and
the mass number of the compound nucleus and hence are  to be
evaluated at each interval of time evolution of the fissioning
nucleus.\par Once the emission widths are known, it is required to
establish the emission algorithm which decides at each time step,
$[t,t+\tau]$ along each of the trajectories, whether a particle is
 being emitted from the compound nucleus.
 This is done by first
calculating the ratio $x=\tau / \tau_{tot}$  where $\tau_{tot}=
\hbar  / \Gamma_{tot}$, $\Gamma_{tot}=\sum_{\nu}\Gamma_{\nu}$ and
$\nu = n,p,\alpha,\gamma$. The probability for emitting any light
particle or $\gamma$ is given, for a small enough time step
$\tau$, by
\begin{equation}
P(\tau)=1-e^{- \tau/\tau_{tot}}\approx x. \label{4n}
\end{equation}
\noindent We shall then choose a random number $r_1$ by sampling
from a uniformly distributed set between 0 and 1. If we find $r_1
< x$, it will  be interpreted as emission  of either a light
particle or a $\gamma$ during that interval. If the time step
$\tau$ is chosen sufficiently small, the probability of emitting a
particle will be small. In this way we guarantee that in each time
interval at most one particle is emitted and we avoid to consider
the emission of more than one particle in each time interval. In
the case that a particle is emitted, the type of the emitted
particle is next decided by a Monte Carlo selection with the
weights $\Gamma_{\nu}/\Gamma_{tot}$ (partial widths). This
procedure simulates the law of radioactive decay for  the emitted
particles. The energy of the emitted particle is then obtained by
another Monte Carlo sampling of its energy spectrum by choosing
another random number following a probability distribution given
by the energy distribution laws in Eqs. \ref{4g} $\&$ \ref{4l}.
The intrinsic excitation energy, mass and spin of the compound
nucleus are recalculated after each emission and also the
potential energy landscape of the parent nucleus is replaced by
that of the daughter nucleus. The spin of the compound nucleus is
reduced only in an approximate way by assuming that each neutron,
proton or  a $\gamma$ carries  away $1\hbar$ while the $\alpha$
particle carries away $2\hbar$ of angular momentum. It is assumed
that the deformation of the nucleus
 is not changed due to particle emission. It is evident that each
  emission of
 a light particle carries away excitation energy and angular
  momentum and
 thereby increases the height of the fission barrier of the
  residual nucleus
 which, in turn, renders the fission event less and less probable.
 \subsection{Dynamical model}
 For each choice of the initial conditions, one generates a separate
 trajectory which is followed in time dynamically by solving the Langevin
equation numerically.
 The
 procedure of numerical integration of the Langevin equation which is
 described in details in the
 previous chapter is followed here.
 We will assume in the present work that
fission would proceed along the valley of the potential landscape
in $(c,h)$ coordinates, though we shall consider the one
dimensional Langevin equation in elongation coordinate $c$ alone
in order to simplify the computation. This approximation is good
enough for the analysis of prescission particle multiplicities and
fission probability. The emission of a particle (neutron, proton
or $\alpha$) or a photon and the nature of emission is checked at
each time interval.
 Emitted particles and their energies are registered
 along a trajectory.
 Each Langevin trajectory can either lead to fission if it
 overcomes the fission barrier
 and reaches
the scission point ($c_{sci}$  is defined in sec. 3.1.3 of the
previous chapter) in course of its time evolution. Alternately it
will be counted as an evaporation residue event if the intrinsic
excitation energy becomes smaller than the fission barrier ($B_f$)
and the binding energies  of  neutron ($B_n$), proton ($B_p$) and
alpha ($B_{\alpha}$), i.e., $E_{int} <
min(B_f,B_n,B_p,B_{\alpha})$. The calculation proceeds until the
compound nucleus undergoes fission or ends up as an evaporation
residue. The above scheme can however take an extremely long
computer time particularly  for those compound nuclei whose
fission probability is small.  We shall therefore follow  a
combined  dynamical  and statistical model, first proposed  by
Mavlitov  {\it  et al.} \cite{Mavil}, in the present calculation.
In this model, we  shall first  follow the time evolution of a
compound nucleus according to the Langevin equations as described
above for  a  sufficiently long period denoted by $\tau_{eq}$
($\tau_{eq}$ is taken as $300 MeV/\hbar$ in our model) during
which a steady flow across the fission barrier is established.
Beyond this period, a statistical model for compound nucleus decay
is expected to be a equally valid and more economical in terms of
computation. We shall therefore switch over to a statistical model
description after the fission process reaches the stationary
regime if the compound nucleus does not reach the scission
configuration within the time $\tau_{eq}$.
\subsection{Statistical model}
When entering the statistical branch we calculate the decay widths
$\Gamma_{\nu}$ for particle emission in the same way as described
before. We shall, however, require the fission width along with
the particle and $\gamma$ widths  in  the statistical  branch of
the calculation. This fission width should be the stationary limit
of the fission rate as determined  by the Langevin equation.
Though analytic solutions for  fission  rates can  be  obtained in
special  cases \cite{Kramers,Weid2,Frob4} assuming a constant
friction, this is not the case with  the chaos-weighted wall
friction(CWWF) which is not constant and is strongly shape
dependent. The fission widths for such shape dependent friction
can only be calculated by solving the Langevin equation
numerically as described in the previous chapter. Thus it becomes
necessary to find a suitable parametric form of the numerically
obtained stationary fission widths using the CWWF (and also WF) in
order to use them in the statistical branch of our calculation.
The details of this procedure is given in chapter 3 and also in
Ref. \cite{Gargi1}, following which we shall calculate all the
required fission widths for the present work.\par Once the fission
widths are known, we use a standard Monte Carlo cascade procedure
where the kind of decay at each time step is selected with the
weights $\Gamma_i/\Gamma_{tot}$ with ($i=fission,n,
p,\alpha,\gamma$) and
 $\Gamma_{tot}=\sum_{i}\Gamma_{i}$. This procedure allows for
 multiple emissions of light particles and higher chance fission.
  The time step
  $\tau$ is redefined after each step in the statistical branch
  as $\tau = \tau_{decay}/10000$, where $\tau_{decay}= \hbar/\Gamma_{tot}$.
  This procedure ensures economy in
  terms of computation time. The Monte-Carlo procedure chooses
  the fission route at a certain interval and the trajectory is
  then counted as a fission event. If the Monte-Carlo procedure
  does not select the fission channel at a certain interval but
  selects a particle/$\gamma$ emission,
   we again
  recalculate the intrinsic energy
 and angular momentum, and continue the cascade until the
 intrinsic energy is $E_{int} <$ min $(B_n,B_p,B_{\alpha},B_f)$.
   In this case we
 count the event as evaporation residue event. The combined
 dynamical plus statistical calculation is implemented in the
 fortran code ``DYSTNF" (developed as part of the thesis work)
  which is described in Appendix F. The flow
 chart of the calculation procedure of this combined dynamical
 plus statistical model which describes the logical sequence of
  the actual calculations is described schematically in Appendix G.
\subsection{Calculation}
 Following the above procedure, the number of  emitted
neutrons, protons, alphas and photons  is recorded for each
fission event. This calculation is repeated  for a  large   number
of Langevin trajectories followed by the statistical model and the
average number of neutrons emitted in the fission  events will
give  the required prescission neutron multiplicity. The
prescission neutron multiplicity is then given by
\begin{equation}
\langle \nu_{pre} \rangle =\frac{N_{\nu}}{N_{fiss}}\label{4o}
\end{equation}
\noindent where $N_{\nu}$ is the total number of neutrons emitted
for those events which have ended in fission and $N_{fiss}$ is the
total number of fission events.
 The fission probability  will be obtained as the
fraction of the trajectories which have undergone fission. The
fission cross section is given by the product of fission
probability ($ p_f=\frac{N_{fiss}}{N_{fus}}$) and the fusion cross
section $(\sigma_{fus})$, i.e.,
\begin{equation}
\sigma_{fiss}=\sigma_{fus}\cdot \frac{N_{fiss}}{N_{fus}}\label{4p}
\end{equation}
\noindent where $N_{fus}$ is the total number of fused
trajectories with which we have repeated the whole calculation.
\section{Results}
\begin{figure}[h!]
\vspace{0.5cm}
\centerline{\psfig{figure=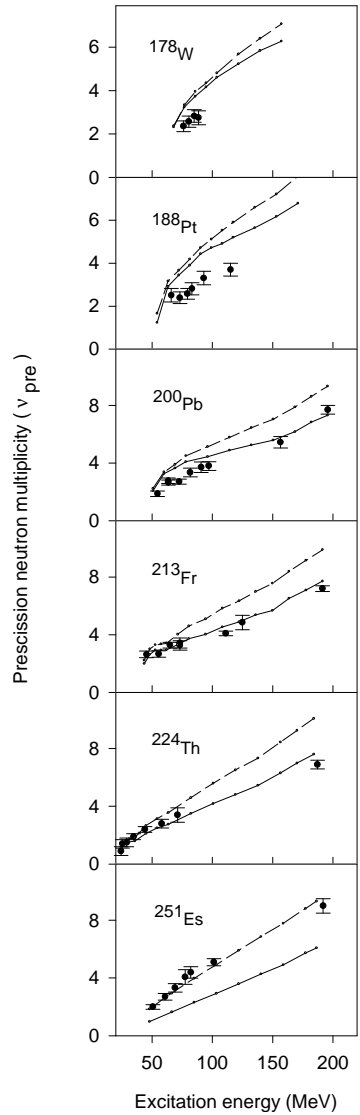,width=0.3\textwidth}}
\caption{\label{d2}Prescission      neutron      multiplicities
calculated with the CWWF  are shown as  points  connected by solid
lines whereas those calculated with the WF are shown as points
connected by dashed lines. The experimental  data for $^{178}$W,
$^{188}$Pt,  $^{200}$Pb, $^{213}$Fr, $^{224}$Th, and $^{251}$Es
are from Refs. [62],  [62,187],  [62,187,188], [62,187,188],
[187,189], and [62,187]  respectively.}
\end{figure}

 We have  calculated the prescission neutron multiplicity
  ($\langle \nu_{pre}\rangle$) and the
fission probability for a number of  compound  nuclei  formed  in
heavy-ion  induced  fusion  reactions. We have used both the
chaos-weighted wall friction(CWWF) and wall friction(WF) in our
calculation. Fig. \ref{d2} shows the results for prescission
neutron multiplicity along with the experimental data. A  number
of systematic features can be observed from these results. First,
the $\langle \nu_{pre}\rangle$ values calculated with the CWWF and
WF are very close  at smaller excitation energies, though at
higher excitation energies,  the WF predictions  are larger than
those obtained with the CWWF. This aspect is present in  the decay
of all  the compound nuclei which we  consider here  and can be
qualitatively understood as follows. The magnitude  of  the CWWF
being smaller than that of the WF, fission rate with the CWWF is
higher than that obtained with the WF.
It has been shown in the previous chapter  that the stationary
fission width with the CWWF  is about twice of that with the WF
\cite{Gargi1}. However at  a  low excitation energy where a
compound  nucleus is formed with a low value of spin, the fission
barrier is high and fission widths calculated with both CWWF and
WF  turn out to be many times smaller than the neutron width. This
can be seen from Fig. \ref{d1}, where the fission width $\Gamma_f$
is calculated using CWWF. The particle multiplicities is decided
by the ratio of particle width $\Gamma_{p}$ and the total width
$(\Gamma_{f}+\Gamma_{p})$. At lower energies, the fission width
$\Gamma_{f}$ being much less than the particle width $\Gamma_{p}$,
the particle multiplicity is practically independent of the
fission width and hence is insensitive to the particular type (WF
or CWWF) of nuclear friction used in calculation of the fission
width.  The neutrons, therefore, have enough time to be emitted
long before a compound nucleus undergoes fission irrespective of
its dynamics being controlled by either the CWWF or  the  WF. This
explains the observation of $\langle \nu_{pre}\rangle$ values
calculated with CWWF and WF being close at lower excitation
energies. Thus the prescission neutron multiplicities are rather
insensitive to fission time scales at lower excitation energies.
On  the  other hand,  a compound nucleus is formed with a larger
spin at higher excitation energies resulting  in  a reduction  of
the  fission barrier and hence an increase in the fission width.
The fission time scales and the neutron lifetimes start becoming
comparable at higher excitation energies (refer Fig. \ref{d1}),
and the dependence of $\langle \nu_{pre}\rangle$  on fission width
becomes sensitive. The fission width calculated with CWWF being
about twice than that with WF,  time  available for evaporation of
the neutrons is much less for the former type of friction  and
hence less neutrons are predicted from calculations with the CWWF
than those with  the WF. The prescission neutron multiplicity thus
becomes capable of discriminating between different  models of
nuclear friction at higher excitation energies of the compound
nucleus.\par A  similar explanation also holds for the systematic
variation of the calculated prescission neutron multiplicities
with respect to the mass number of the compound nucleus.  We  find
that  the  WF prediction  for prescission neutrons starts getting
distinct from that of the CWWF at smaller values of the excitation
energy  with increasing mass number of the compound nucleus. Since
the fission barrier decreases with the increasing mass of a
compound nucleus, the   fission  time  scales  and  the  neutron
lifetimes  become comparable  for  heavier  compound  nuclei  at
lower  excitation energies. This results in a fewer neutrons from
calculations with the CWWF than those with the WF as one considers
heavier compound nuclei. \par A  number  of interesting points can
be noted while comparing the calculated values  with  the
experimental data. For the compound nucleus   $^{178}$W,   the
available    experimental points \cite{Newton1}  are  at  low
excitation  energies  and therefore, cannot distinguish between
the calculated values using the  CWWF and  WF,  which are almost
identical. The calculated values slightly overestimate the
prescission neutron multiplicity compared to  the experimental
data.  A  more extensive  set of experimental values  for
prescission  neutron multiplicity  are available for   the
compound  nuclei $^{188}$Pt, $^{200}$Pb, $^{213}$Fr and $^{224}$Th
\cite{Newton1,Hinde6,Hinde7} covering a wider range of excitation
energy in which the calculated  values with  the CWWF and WF
differ. Clearly, the CWWF predicted values give excellent
agreement with the  experimental data  for  these compound nuclei
whereas  the  WF  predictions are considerably higher. However,
similar conclusions  cannot  be drawn  for  the heavier nucleus
$^{251}$Es.  It appears that the WF predictions are closer to the
experimental data \cite{Hinde1,Newton1,Rossner}
 whereas the CWWF predictions are somewhat lower. We shall return
to  this  point later for a detailed discussion. For the present,
we  shall   consider   the   results   of   fission   probability
calculations.\par
\begin{figure}[h!]
\vspace{0.5cm} \centerline
{\psfig{figure=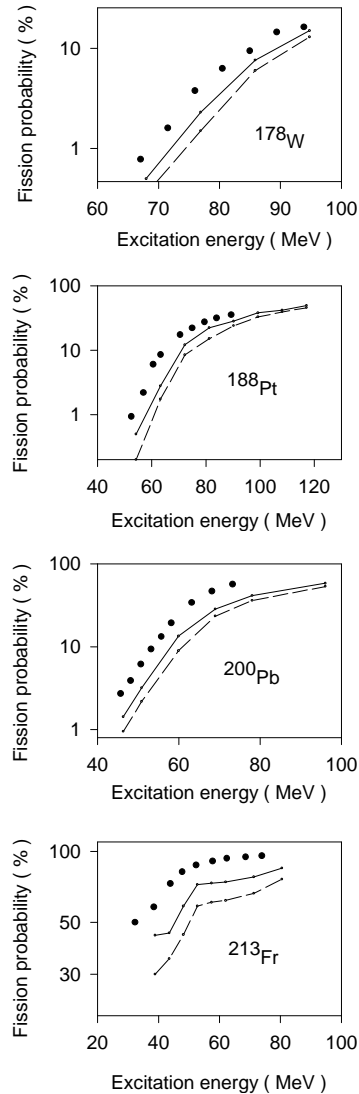,width=0.3\textwidth}}
\caption{\label{d3}Fission probabilities  calculated with the CWWF
 are shown as points connected  by  solid  lines  whereas those
calculated  with  the WF    are  shown  as points connected by
dashed lines. The experimental data  for  $^{178}$W,
$^{188}$Pt,$^{200}$Pb,  and $^{213}$Fr are from Refs. [190],
[190], [175], and [60], respectively.}
\end{figure}

\begin{figure}[b!]
\centerline {\psfig{figure=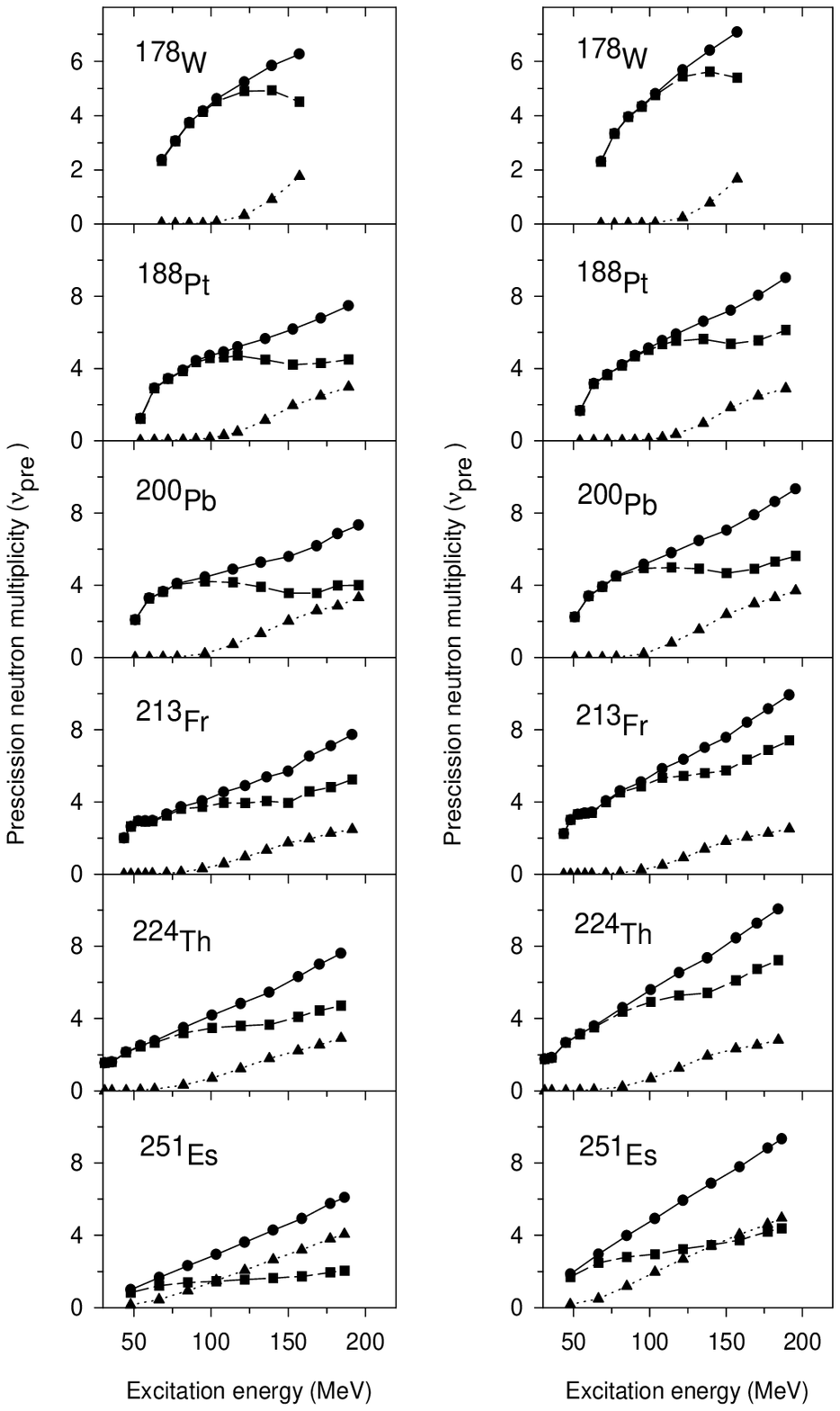,width=0.5\textwidth}}
 \caption{\label{d4}Neutrons
emitted  during  the presaddle and postsaddle (saddle to scission)
stages of fission. Figures in the left panel show values
calculated with the CWWF   whereas those  in  right panel are
obtained with the WF . In each plot,  the  solid circles, the
solid  squares  and  the  solid triangles represent the total
number of prescission neutrons, the number  of presaddle neutrons
and  the  number  of  postsaddle neutrons, respectively.}
\end{figure}

 The calculated and experimental values of
fission probability are shown  in  Fig. \ref{d3} for four compound
nuclei.   While the fission probability  for $^{251}$Es  is
almost $100 \%$, for $^{224}Th$, we shall consider the
complimentary cross-section, the evaporation residue cross section
separately in the next chapter. Hence they are excluded from the
present discussion. The calculated values of fission probability
complements  the picture of fission dynamics which was obtained
while  discussing the prescission neutron data. The fission
probability  is found to be more sensitive  to the choice of
friction at lower excitation energies than at higher excitations.
The CWWF predicted fission probabilities are  larger than  those
from  the  WF predictions. Moreover,  the  CWWF predictions are
consistently closer  to  the experimental  values of   fission
probability than those from the WF predictions.
In order to gain further insight into the dynamics of fission, we
have  also  calculated  the  presaddle  and postsaddle (saddle to
scission)  contributions  to  the  multiplicity  of   prescission
neutrons.  Figure \ref{d4} shows the results obtained with both
the CWWF and WF.
  For  all  the  cases,  starting  from
almost  zero  multiplicity  at  small  excitation  energies,  the
postsaddle contribution increases at higher excitation  energies.
It is further observed that the postsaddle neutron multiplicities
calculated with the CWWF and WF  are almost same for all the
compound  nuclei  over  the  range  of  excitation  energies
considered here. This would be due to the fact that the number of
postsaddle  neutrons depends on the time scale of descent from the
saddle  to  the  scission.  This,  in  turn, will depend upon the
strength of the friction between the saddle and the scission and
we  have  already  seen  in  Fig. \ref{b2} that the CWWF and WF
are indeed close at large deformations. We shall next compare the
presaddle contributions calculated with the CWWF and WF  for each
of  the  nuclei  under consideration. We immediately notice that
the  WF  predictions are consistently larger than those from the
CWWF at  higher excitation energies. This gives rise to the
enhancement of the WF prediction for total prescission
multiplicity  compared  to  that from   the CWWF prediction, which
we have already noticed in Fig. \ref{d2} and have discussed
earlier.  Since  the  CWWF predicted neutron multiplicities agree
with the experimental values for the nuclei $^{178}$W, $^{188}$Pt,
$^{200}$Pb, $^{213}$Fr, and $^{224}$Th, we conclude  that the
chaos-weighted wall friction provides the right kind of friction
to describe the presaddle dynamics  of  nuclear fission.
\begin{figure}[t!]
\vspace{0.5cm}
 \centerline
{\psfig{figure=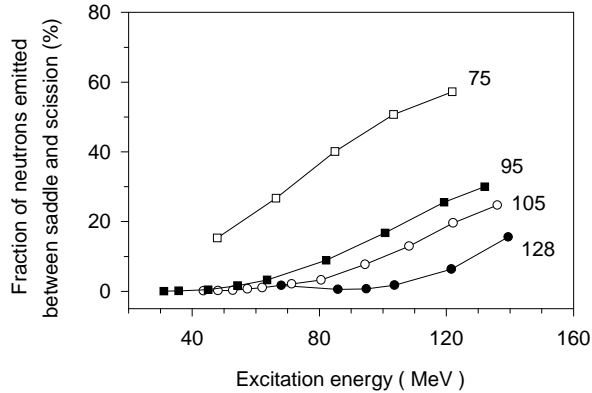,width=0.5\textwidth}}
 \caption{\label{d5} Fraction
of neutrons emitted between saddle and  scission  is  shown  as  a
function of excitation energy for different compound nuclei. The
the open square, the solid square, the open circle and the solid
circle  represent  the  calculated values  for  $^{251}$Es,
$^{224}$Th,  $^{213}$Fr, and $^{178}$W, respectively. The critical
excitation energy (in units  of  MeV), as  defined  in  the  text,
is  indicated  for  each  nucleus. }
\end{figure}
\par While  comparing  the  relative  importance  of the presaddle
and postsaddle neutrons, we further note that the postsaddle
neutrons are more frequently emitted from  heavier  compound
nuclei.  For $^{251}$Es,  most  of  the  prescission neutrons
predicted by the CWWF are accounted for by the postsaddle
neutrons. The underlying physical picture can be described as
follows.  When  a  compound nucleus  is  formed  in  a heavy-ion
induced fusion reaction, its spin distribution is assumed to be
given by Eq. \ref{4a}. If the compound nucleus is formed with a
spin  at  which  there  is  no fission  barrier,  its  transition
to the scission point will be essentially  considered  as
postsaddle  dynamics.  In  order  to simplify  our discussion, let
us assume that most of the compound nuclei at a given excitation
energy  are  formed  with  the  spin $l_{c}$  of  Eq. \ref{4a} and
let $l_{b}$ be the limiting spin value at which the fission
barrier vanishes. We can then  find  a critical  excitation
energy,  $E_{crit}$,  above  which  $l_{c}$ becomes greater than
$l_{b}$ and most of the fission dynamics  at excitations  above
this  critical  value  can  be  considered as comprising of only
postsaddle trajectories. In  Fig.  \ref{d5}, we have  plotted  the
fraction  of  neutrons  emitted  in  the postsaddle stage as a
function of the  excitation  energy  for  a number  of  compound
nuclei.
  The critical excitation energy for
each  nucleus  is  also given in this plot. We have used the CWWF
predicted neutron multiplicities for this plot where we find that
the critical excitation energy decreases  with  increase  in  the
compound  nuclear mass. Thus the dominance of postsaddle neutrons
sets in at lower excitation energies for heavier nuclei which, in
turn, gives rise to the increase in the  fraction  of  postsaddle
neutrons with increasing mass of the compound nucleus. \par Though
the above discussion clearly establishes the importance of
postsaddle neutrons for a very heavy compound nucleus, the number
of  postsaddle  neutrons  calculated with the CWWF still falls
short of making the total prescission multiplicity equal to the
experimental values for $^{251}$Es. We consider the  apparent
better  agreement  between  the  WF predicted prescission neutron
multiplicity and the experimental data for $^{251}$Es as shown in
Fig. \ref{d2} as a mere coincidence and  we  do  not  find  any
physical  justification for abandoning the chaos-weighted factor
in one-body friction for such heavy nuclei. Instead, we feel that
the  mechanism  of neutron  emission  in  the  postsaddle  stage
requires  a  closer scrutiny  essentially  because  the  nucleus
becomes strongly deformed beyond the saddle  point.  The  neutron
decay  width  of such a strongly deformed nucleus could be quite
different from that of the  equilibrated  near-spherical  nucleus
which   we use in   our   calculation.   In  particular,  the
neutron-to-proton ratio is expected to  be  higher  in  the  neck
region  than  that in  the  nuclear bulk and this can cause more
neutrons to  be emitted.  Further,  dynamical  effects  such  as
inclusion  of the neck degree of freedom in the Langevin equation
can influence the time scale of the postsaddle dynamics and hence
the  number  of emitted  neutrons.  Such possibilities should be
examined in future for a better understanding of  the  postsaddle
dynamics of nuclear fission. The results presented in this section
is published in Ref. \cite{Gargi2}.
 \section{Summary}
We have applied a theoretical model of one-body nuclear friction,
namely   the   chaos-weighted   wall   friction,  to  a  dynamical
description of compound nuclear decay where fission  is  governed
by the Langevin equation coupled with the statistical evaporation
of light particles and photons. We have used both the normal wall
friction  and  its modified form with the chaos-weighted factor in
our calculation in order  to  find  its  effect  on  the  fission
probabilities and prescission neutron multiplicities for a number
of  compound  nuclei.  The  strength  of  the chaos-weighted wall
friction(CWWF)  being  much  smaller  than  that  of  the  wall
friction, the fission probabilities calculated with the CWWF are
found  to be larger than those predicted with the WF. On the other
hand, the  prescission neutron multiplicities predicted with the
CWWF  turn out to be smaller  than  those  using the WF. Both the
prescission  neutron  multiplicity  and fission probability
calculated  with  the CWWF for the compound nuclei $^{178}$W,
$^{188}$Pt,  $^{200}$Pb,  $^{213}$Fr, and $^{224}$Th agree  much
better  with  the experimental data compared to the predictions of
the WF.\par We  have subsequently investigated the role  of
presaddle and postsaddle neutrons  at different excitation
energies   for different compound nuclei. It has been shown that
the majority of the prescission neutrons are emitted in the
postsaddle stage for a very heavy nucleus like $^{251}$Es. The
CWWF, however, cannot  produce  enough neutrons  to match the
experimental prescission multiplicities for such a nucleus. It is,
therefore, possible that in  the  postsaddle region, either the
fission dynamics  gets considerably slowed down or the neutrons
are more easily emitted. These aspects require further studies
before  we draw conclusions regarding the postsaddle dynamics of
nuclear fission.\par The presaddle neutrons are however found  to
account for most of the prescission neutrons for lighter nuclei at
lower excitation energies.  On the  basis of  the  comparison  of
the calculated prescission multiplicities with experimental data
as given in the preceding section, we can conclude that the
chaos-weighted  wall friction can adequately describe the fission
dynamics in the presaddle region.

\chapter{Evaporation   residue  cross-sections  as  a  probe  for  nuclear
dissipation}

Experimental studies of the prescission multiplicities of neutrons
\cite{Hinde6}, $\gamma$ rays \cite{Dioszegi}, and charged
particles \cite{Lestone2} have shown that the fission process is
strongly hindered relative to expectations based on the
statistical model description of the process, as we have already
discussed in the previous chapters. However, it is not possible to
infer from these experiments whether the emission of the particles
occur mainly before or after the traversal of the saddle point as
the system proceeds toward scission. This kind of information
could be useful to discriminate between various dissipation models
which are strongly dependent on the deformation and shape symmetry
of the system. The evaporation probability for hot nuclei formed
in heavy-ion fusion reactions can be useful for such purposes
which is sensitive only to the dissipation strength inside the
fission barrier. As the hot system cools down by the emission of
neutrons and charged particles there is a finite chance to undergo
fission after each evaporation step. If the fission branch is
suppressed due to dissipation there is therefore a enhanced
probability for survival which manifests itself as an evaporation
residue cross section which is larger than expected from
statistical model predictions. It turns out that the evaporation
residue(ER) cross-section depends strongly on the strength of the
nuclear dissipation whenever it is a very small fraction of the
total fusion cross section\cite{Back1}. The fate of a compound
nucleus, i.e., whether it will  undergo fission or survive as an
evaporation residue is decided mainly within the saddle point.
Hence the measurement of evaporation residue formation probability
is expected to be a sensitive probe for nuclear friction and may
therefore provide the desired separation between presaddle and
post-saddle dissipation. It is concluded in Ref. \cite{Frob2} that
evaporation residue cross sections give restrictions for possible
$\eta$( friction parameter) values, and seem to be even more
sensitive probes for friction than $\gamma$-rays which is
generally considered as a good probe for investigating dynamics in
fission.\par In  a recent work, Di\'oszegi {\it et al.}
\cite{dioszegi} have analyzed the $\gamma$ as  well  as  neutron
multiplicities  and evaporation residue cross-section  of
$^{224}$Th and have concluded that the experimental data can be
fitted equally well with either a temperature or a
deformation-dependent nuclear dissipation. Interestingly, the
deformation-dependence of the above dissipation corresponds to a
lower value of the strength of the dissipation inside the saddle
and a higher value outside  the saddle,  similar to  the
phenomenological dissipation  of Ref. \cite{Frob3}.
 It is worthwhile to note here that our model for nuclear
 friction i.e., the  shape-dependent  chaos-weighted wall friction
  (CWWF)  has  features
  similar  to  the
empirical dissipations discussed above. In  the  present  work, we
shall employ the CWWF to  calculate the evaporation   residue
excitation function for  the $^{224}$Th nucleus.  Our  main
motivation here will be to put the CWWF to a further test and
verify to what extent it can account  for  the experimental
evaporation residue data which is a very sensitive probe for
nuclear dissipation. Calculation  will be performed at a number of
excitation energies for $^{224}$Th formed in the
$^{16}$O+$^{208}$Pb system. We  have chosen  this  system
essentially  because of the availability of experimental data on
both  evaporation  residue  and  prescission neutron  multiplicity
covering  the  same  range  of  excitation energies and the fact
that earlier analysis  of  the  evaporation residue excitation
function have already indicated the need for a dynamical     model
for     fission     of    this    nucleus
\cite{Brinkmann,Rossner,Morton}.
\section{Calculation}
  The  various inputs required
 for the model and the different steps involved in the calculation
 of evaporation residue
probability are exactly same as those involved in the calculation
of fission probability as described in chapters 2 and 4. The same
notations and procedure will be followed in the present
calculation and hence is not repeated here.
 A  Langevin trajectory will be considered as undergone fission
if it reaches the scission point ($c_{sci}$) in course of  its
time evolution.  Alternately  it  will  be  counted  as an
evaporation residue event if the intrinsic excitation energy
becomes  smaller than  the  fission  barrier  and  the binding
energies of neutron, proton and alpha. The calculation proceeds
until the compound nucleus undergoes   fission  or  becomes  an
evaporation residue. This calculation  is  repeated  for  a large
number of Langevin trajectories and the evaporation residue
formation probability is obtained  as the fraction of the
trajectories which have ended up as evaporation residues. The
evaporation residue cross-section is subsequently obtained by
multiplying the experimental  value  for fusion  cross-section  in
the  entrance channel  with the formation probability of the
evaporation residue. Similarly,  the average number of particles
(neutrons, protons or alphas) emitted in the fission events will
give the required prescission particle multiplicities. The
calculated evaporation residue excitation function and prescission
neutron multiplicities will be compared with the experimental
values in the next section.
\section{Results}
 The results which are presented in this section is reported in
Ref. \cite{Gargi3}. We have calculated the prescission neutron
multiplicity($\nu_{pre}$) and the evaporation residue(ER)
cross-section for the compound nucleus $^{224}$Th  when  it is
formed  in  the fusion of an incident $^{16}$O  nucleus  with a
$^{208}$Pb target nucleus. The calculation is done at a number of
incident energies in the range of $80$ MeV to $140$ MeV using both
the WF and the CWWF. \par
\begin{figure}[h!]
 \centerline {\psfig{figure=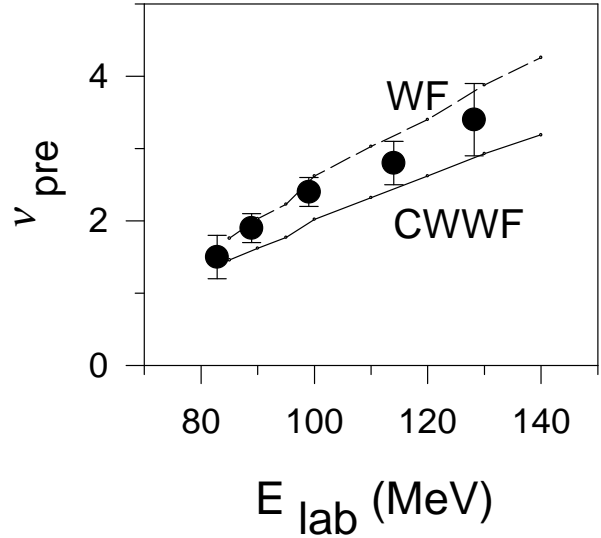,width=0.5\textwidth}}
\caption{\label{e2}Prescission neutron       multiplicity
($\nu_{pre}$) excitation  function calculated  with  WF  (dashed
line)   and   CWWF   (full   line) frictions  for  the  reaction
$^{16}$O+$^{208}$Pb. The experimental  points  (dots)  are  also
shown.}
\end{figure}

 Fig.  \ref{e2}  shows  the  calculated
prescission  neutron multiplicity along with  the  experimental
data  \cite{Rossner}. Both the wall friction(WF) and
chaos-weighted wall friction(CWWF) predictions for $\nu_{pre}$ are
quite close to the experimental values and this  shows that
neutron multiplicity is not very sensitive to the dissipation in
fission in  the energy range under consideration. It must be
pointed out, however, that the CWWF predictions for neutron
multiplicity  are closer  to  experimental data  compared to those
from WF at much higher excitations of the compound
nucleus\cite{Gargi2}. \par
\begin{figure}[b!]
\vspace{0.4cm} \centerline {\psfig{figure=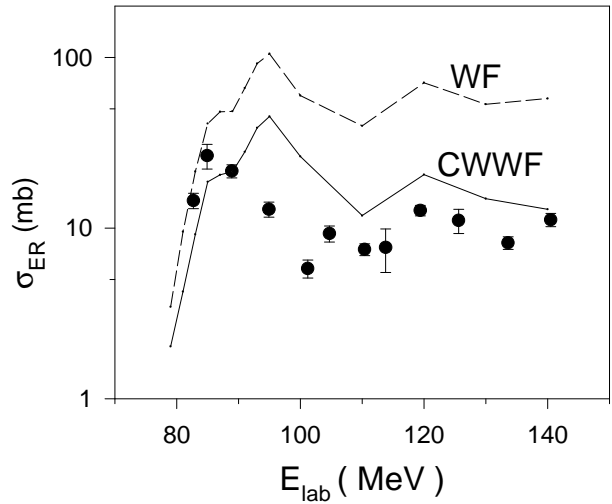,width=8cm}}
\caption{\label{e3}Evaporation residue cross-section excitation
function  calculated  with  WF (dashed line) and CWWF (full line)
frictions for the reaction as in fig.2. The  experimental  points
(dots) are also shown.}
\end{figure}

 We  shall  next  consider  the results
of the evaporation residue calculation. Fig. \ref{e3}  shows the
evaporation  residue cross-section($\sigma_{ER}$) excitation
functions calculated using both the  WF  and CWWF.
 The
experimental values\cite{Brinkmann} of evaporation residue
cross-section are also shown in this figure.
We first note that the calculated evaporation residue
cross-section is very sensitive to  the dissipation in the fission
degree of freedom. The  WF predictions are  a  few times
(typically 2-5) larger than those obtained with the  CWWF.  Next
we make the important observation that the CWWF predicted
excitation function is  much closer  to the experimental values
than that obtained with the WF. This observation clearly shows
that  the chaos-weighted  factor in CWWF changes its strength in
the right direction. We must take note of the fact, however, that
the CWWF still considerably overestimates the ER cross-section.
Since the present dynamical calculation considers only  one
(elongation) fission degree  of freedom,  it is expected that
inclusion of the neck degree of freedom will increase the fission
probability \cite{Wada1} further and  hence reduce   the ER
cross-section. We  plan to extend our work in this direction in
future. We further observe that while  a peak appears  in  the
experimental excitation function at  about 85 MeV, the same is
shifted by 10 MeV in our calculated results. We do not  have  any
explanation for  this discrepancy except pointing out that there
is no free parameter in our calculation and  thus  no  parameter
tuning  has been attempted in order to fit experimental data. A
similar  shift has also been observed  in an  earlier
work\cite{dioszegi}.\par The structure of the evaporation residue
excitation function can also reveal certain interesting features.
Since  the  calculated values of  the evaporation residue
cross-section are obtained as the product of the fusion
cross-section and  the  probability  of evaporation residue
formation,  the  initial  rise  of  the  ER cross-section with
beam energy  essentially  reflects  the  steep rise of fusion
cross-section in this energy region \cite{Morton}. The high energy
part of the ER cross section excitation function is due to charged
particle emission and this part even seems to rise slowly with
excitation energy\cite{Frob2}. The reason for that is that after
the emission of some charged particles the daughter nuclei becomes
less fissile and survive with higher probability. This mechanism
works more successful for higher excitation energy because of the
reduced role of the Coulomb barrier. This establishes the
significant role of charged particle emission in the survival
probability of a nucleus. An experimental finding like this is
reported in Ref. \cite{Ninov}.
 At
higher beam energies, the ER cross-section becomes approximately
stable which results from a delicate balance between the
increasing trend of the fusion cross-section and the decreasing
trend of the probability of ER formation.  Had the  ER formation
probability decreased at a rate higher than those obtained  in the
present calculation, the resulting   ER cross-section  would have
decreased  at higher compound nuclear excitations.  In  fact, such
an observation was made   in Ref. \cite{Brinkmann}  where the ER
\begin{figure}[h!]
\vspace{1.0cm}
 \centerline {\psfig{figure=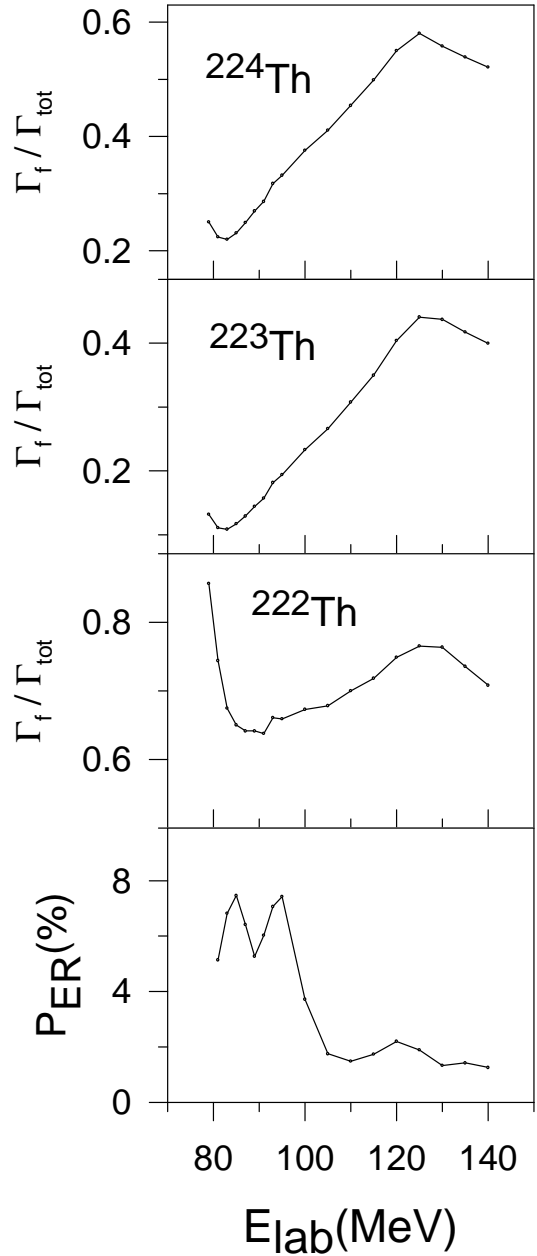,width=7cm}}
\caption{\label{e4}The top three  panels  show  the  fission
partial widths for $^{224}$Th, $^{223}$Th  and  $^{222}$Th  (see
text).  The  total width $\Gamma _{tot}$ includes the neutron,
proton, alpha and $\gamma$ evaporation widths in addition to  the
fission  width. The bottom panel displays the excitation function
of the evaporation residue formation probability for the reaction
as in Fig. 5.1}
\end{figure}
cross-section obtained from standard statistical model calculation
was found to decrease very steeply beyond 100 MeV of beam energy.
In order to explore  this point  a  little  further, we  have
calculated  the excitation function of the partial width for
fission. Since fission can take place at any stage during neutron
(or any other light  particle) evaporation, the partial widths are
calculated for $^{222}$Th and $^{223}$Th  as well at excitation
energies reduced by the neutron separation energy after each
neutron emission.  The  compound nuclear  spin  was taken  as
$l_{c}$ from Eq. \ref{4b} (section 4.1.2) while only
chaos-weighted wall friction was considered for this
calculation.\par The calculated excitation functions of the
fission partial widths are shown  in  Fig. \ref{e4}.
 The
calculated values of the ER formation probability  ($P_{ER}$)  are
also  displayed  in  this figure. Each partial width excitation
function is found to have a minimum  around  90  MeV  of  beam
energy  after which it starts increasing till this trend is
arrested  and  reversed  at  higher excitations. Recalling the
fact that the above results on partial widths are only indicative
while $P_{ER}$ is obtained from a full dynamical  calculation,  it
is of interest to note that a bump in the excitation function of
$P_{ER}$ also  appears  in  the  above ($\sim$ 90 MeV) energy
range. Subsequently, the value of $P_{ER}$ drops  rather  sharply
before  it  settles  to a steady value at higher excitations. This
feature is also complementary to that of the excitation functions
of the partial  widths  of  fission.  We thus  demonstrate  in a
schematic manner how the structure in the excitation function of
the ER cross-section  is  related  to  the competition between
fission and other decay channels at different stages of fission.
\section{Summary}
We have applied a theoretical model of one-body nuclear friction,
namely   the   chaos-weighted   wall   friction,  to  a  dynamical
description of compound nuclear decay where fission  is  governed
by the Langevin equation coupled with the statistical evaporation
of  light  particles. We have used both the standard wall friction
and its modified form with the chaos-weighted factor in order  to
calculate  the  prescission  neutron multiplicity and evaporation
residue excitation functions for the $^{224}$Th  nucleus.  Though
the  number of the prescission neutrons calculated with either
wall friction or chaos-weighted wall friction are found to be very
close to each other in the energy range considered, the
evaporation residue cross-section is found  to depend very
strongly on the choice of nuclear friction. The evaporation
residue cross-section calculated with the  CWWF  gives a much
better agreement with the experimental data compared to the WF
predictions. This result demonstrates that the consequences  of
chaos  in particle motion give rise to a strong suppression of the
strength of  the wall friction  for  compact shapes   of   the
compound   nucleus which, in  turn,  brings theoretically
calculated  evaporation residue   cross-sections considerably
closer  to  the experimental values. Thus the chaos considerations
may  provide a plausible  explanation  for  the shape-dependence
of  the strength of nuclear friction which was found
\cite{Frob3,dioszegi} to  be  necessary  in  order  to  fit
experimental data.

\chapter
{ Effect of transients in nuclear fission }
 Induced nuclear fission
had been viewed by H. A. Kramers as a diffusion process of the
fission degree of freedom over the fission barrier long before the
successful developments of transport theories for description of
heavy ion reactions \cite{Kramers}. In the eighties, forty years
after Kramers, Weidenm\"uller \textit{et al.} made a detailed
study of nuclear fission using the Fokker-Planck equation within
the framework of this diffusion model \cite{Weid2}. The work of
Weidenm\"uller \textit{et al.} first revealed that it requires a
certain interval of time to develop a steady probability flow at
the saddle point across the fission barrier. During this time
interval, also referred to as the transient time $(\tau)$, the
probability flow at the saddle point increases from zero to its
stationary value. This stationary probability flow also defines
the stationary fission width ($\Gamma_{0}$) and the associated
fission life time ($\tau_ {f}= \hbar/\Gamma_{0}$). More
specifically, the fission-decay width $\Gamma_f(t)$ is inhibited
at its earliest times and thus at the beginning of the process,
during a delay of the order of the transient time $\tau$, the
fission-decay width differs from its asymptotic value
$\Gamma_{f}^{K}$($\simeq \Gamma_0$), originally derived by Kramers
by solving the stationary Fokker-Planck
equation\cite{Kramers}.\par The transient time which arise from
the relaxation of the collective degrees of freedom has
significant effects on the
 prescission particle and $\gamma$
multiplicities and on the future evolution of the nucleus, in
particular its fission probability. The crucial quantities which
govern the time evolution of the  probability current across the
barrier are: the excitation energy of the system, the height of
the fission barrier and the nuclear friction coefficient.
Depending on the values of these quantities, completely different
situations may be encountered. One possibility is a slow
attainment of a quasistationary regime of probability flow over
the transient time $\tau$. On the other extreme, the entire
distribution may pass the barrier in a single swoop and the whole
fission process becomes a transient\cite{Grange2}. These different
situations are clearly not taken into account in the ``statistical
model'' of Bohr and Wheeler, which assumes from the outset the
existence of a quasistationary regime at the saddle point.
 In particular, when the entire distribution comes close to pass the barrier
in a single swoop it is not possible to define a fission width in
the usual sense. Thus the standard treatment of the cascade
de-excitation and associated cooling of a compound nucleus via
particle emission and fission needs to be modified accordingly.
 \par Weidenm\"uller and his coworkers generalized the
quasistationary approach of Kramers to a time dependent one and
derived for the first time a time dependent fission width solving
the Fokker-planck equation analytically as well as numerically
within the framework of a simplified model. From their numerical
calculations, the authors of Ref. \cite{Bhatt} extracted the
following information for the transient time, defined as the time
until the fission width $\Gamma_f(t)$ reaches $90\%$ of its
asymptotic value:
\begin{eqnarray}
\tau &=& \frac{1}{\beta}\ln \left( \frac{10B_f}{T}\right)
\hspace{1cm}
         \mbox{for} \hspace{1cm} \beta<2{\omega}_{g} \nonumber\\
     &=& \frac{\beta}{2{\omega_g}^2}\ln\left(\frac{10B_f}{T}\right)
     \hspace{0.5cm} \mbox{for} \hspace{1cm} \beta>2{\omega}_g
 \end{eqnarray}
\noindent where $B_f$ is the fission-barrier height, $T$ is the
nuclear temperature, $\omega_g$ is the effective oscillator
frequency (harmonic oscillator osculating the nuclear potential at
the first minima) at the ground state, and $\beta$($=\eta/m$) is
the effective reduced dissipation coefficient which rules the
relaxation of the collective degrees of freedom towards thermal
equilibrium.
 They distinguished two regimes for the motion of
the collective variable in the first minimum of the potential,
which is characterized  by a specific value $\beta_0$
$(=2\omega_g)$ of the  reduced nuclear friction coefficient
$\beta$ : for $ \beta < \beta_0$ the motion of the collective
variable is underdamped while for $\beta
> \beta_0$ it is overdamped. Thus a semi-quantitative
estimate of the transient time $\tau$ was obtained and it was
found to increase with decreasing values of $\beta$ in the
underdamped case while it increases linearly with $\beta$ in the
overdamped case.
 In both the cases they derived
simple analytical formula for the time-dependent fission width.
   \par The dominant role of transients on lifetime of
induced fission at high excitation energies($\geq$ 100 MeV) is
emphasized in Ref. \cite{Bhatt}. The detailed study and effects of
transients in the case of overdamped motion can be found in Ref.
\cite{Zhang}. In cases where the system is highly excited and the
potential minimum is very shallow or non existent(no fission
barrier), the entire fission process is predominantly or
completely a transient phenomenon. The transient time for such
cases is redefined and a simple analytical expression is given in
Ref. \cite{Grange2}. It appears essentially as the time for the
onset of the exponential growth of fluctuations i.e. the time when
the system becomes globally unstable and breaks apart. The
solution of Fokker-Planck equation to arrive at an analytical
expression for the time-dependent fission width is significant
since incorporation of this $\Gamma_f(t)$ in a statistical model
evaporation code is equivalent to a dynamical study of nuclear
fission by Langevin or Fokker-Planck equation \cite{Butsch}. Due
to the high computing time required by the Langein or the
Fokker-Planck approaches, this equivalent procedure of using a
cascade code (where fission is treated as one of the decay
channels and time dependence of fission width is explicitly taken
into account) is often preferred to interpret experimental data.
This realization motivated improved deduction of $\Gamma_f(t)$
from the analytical solution of the Fokker-Planck equation.  One
of the work worth mentioning is the meticulous investigation by B.
Jurado {\it et al.} of the evolution of the probability
distribution of the system in phase space all along its dynamical
path which resulted in extracting the main features of the
relaxation process towards equilibrium \cite{Jurado1}.
Characteristic features of the evolution of the amplitude of the
probability distribution and the velocity profile at the fission
barrier were derived. Making use of these results, they have
developed an easily calculable approximation of the time dependent
fission-decay width that is based on realistic physical
assumptions, taking the initial conditions into account properly.
This new analytical formulation of $\Gamma_f(t)$ was able to
reproduce rather closely the trend of the exact numerical solution
in the under- as well as in the over-damped regime\cite{Jurado2}.
\section{Experimental signatures}
The tools most frequently applied to measure fission time scales
are the neutron clock\cite{Ross} and the gamma clock\cite{Paul}.
They have yielded the majority of the available information on the
time interval a heavy nuclear system needs to cross the scission
point. However, the mean scission time is an integral value,
including the transient time, the inverse of the stationary decay
rate(the statistical decay time) and an additional dynamic
saddle-to-scission time. Thus it does not give direct access to
the transient time that is connected to the equilibration process
of the compound nucleus. The total fission or evaporation residue
cross section have been used to investigate dissipation at low
deformation, but they are not sufficient to determine transient
effects in an unambiguous way. The challenge to observe transient
effects is increased by the fact that they show up only in a
restricted energy range. The calculations of Ref. \cite{Jurado3}
have shown that that fission is affected by transient effects only
for excitation energies at saddle within the interval $150MeV <
E^*_{saddle}<350MeV$. At excitation energies below 150 MeV, the
statistical decay times for fission is appreciably longer than
typical dynamical time scales, making the dynamical observables
rather insensitive to the transient time. This point may explain
why in several experiments performed at rather low excitation
energy no transient effects at all were observed\cite{Hui}. In
fact, the observation of transient effects requires a reaction
mechanism that forms excited nuclei with an initial population in
deformation space far from equilibrium and an experimental
signature that is specifically sensitive to the delayed population
of transition states.\par In peripheral relativistic heavy-ion
collisions using $^{238}U$ at 1 $A$ GeV, fission studies  in
inverse kinematics were carried out at GSI\cite{Jurado4}. They
studied projectile-fragmentation - fission reactions and
introduced two experimental signatures to observe transient
effects in fission. The total fission cross sections of $^{238}U$
projectiles at 1 $A$ GeV were studied as a function of the target
mass and also the partial fission cross sections and the partial
charge distributions of the fission fragments were investigated
for the reaction of $^{238}U$ on ${(CH_2)}_n$ target. The first
signature exploited to measure transient effect was given by the
partial fission cross-sections, i.e. the fission cross sections as
a function of $Z_1 + Z_2$. At high excitation energies particle
decay times become smaller than the transient time and the nucleus
can emit particles while fission is suppressed. Therefore, for the
lightest fissioning nuclei (lowest values of $Z_1 + Z_2$)
transient effects will lead to a considerable reduction of the
fission probability. The second signature is based on the charge
distribution of the fission fragments that result from a given
fissioning element. The width of the charge distribution of the
fission fragments is a measure of the saddle point
temperature\cite{Jurado4} and thus for the lower values of $Z_1 +
Z_2$ where the initial excitation energy is large and fission is
suppressed with respect to evaporation, the nucleus will evaporate
more particles on its way to fission. Therefore, transient effects
will reduce the temperature of the system at saddle and
consequently the width of the charge distributions\cite{Ben1}.
These experimental observables were compared with an extended
version of the abrasion-ablation Monte-Carlo code ABRABLA to
deduce quantitative results on transient effects. The results
demonstrated the suppression of fission at high excitation
energies and thus established the importance of transients.

\section{ Transients in our model}
 The studies of transients as described earlier in this chapter
  were carried out under a number
of simplifying assumptions. The potential chosen was not
realistic(simplified) and the inertia and friction parameter were
taken as constants so that analytical solution of the
Fokker-Planck equation is possible as well as the numerical
solution becomes easier. However it has been established by
extensive experimental data that friction coefficient is not 
constant but strongly shape dependent and hence numerical study of
the fission process becomes inevitable. This motivated us to use
our model for friction namely the chaos-weighted wall friction in
Langevin dynamics and study the effects of transients in nuclear
fission in a much more realistic framework with a Yukawa plus
exponential double folding potential.\par In  the  present work,
we would examine certain issues related to the time dependence of
fission  widths  and  its effect  on  the multiplicity  of  the
prescission neutrons. First, we would study the effect of lowering
the fission barrier on the time dependence of the rate of fission.
The motivation for this study is to find the transition from a
diffusive process in the presence of  a fission  barrier  to  a
transient  dominated picture  when  there is no fission barrier.
We would indeed find that the diffusive nature of fission
continues  to  some  extent even  for  cases which  have  no
fission barrier. The underlying physical picture that would emerge
for fission in the absence  of a  fission barrier  would be as
follows. Consider an ensemble of fission trajectories which have
started together sliding down the potential (with no fission
barrier) towards the  scission  point. However,  the  random force
acting  on  the  trajectories  will introduce a dispersion in
their arrival  time   at  the  scission point.  In  other words,
the trajectories will cross the scission point at different
instants and a flow will thus  be  established at  the  scission
point.  However, the effect of this dispersion will be reduced
when the conservative force becomes much stronger than the random
force. This would happen at  very  large  angular momentum  of the
fissioning nucleus due to the strong centrifugal force. Therefore,
the single swoop picture  for  fission  becomes more  appropriate
at very large values of spin of the nucleus. We would establish
the above scenario in the first part of our work. \par It is
already stated that the study of fission dynamics using Langevin
or Fokker-Planck equation requires huge amount of computation time
and can be avoided by adopting an alternative and easier approach
which is to perform  a statistical  model calculation by modifying
a cascade code  in which  fission  is treated  as one of the decay
channels and the time dependence of the fission width is
explicitly taken into account \cite{Butsch}. In such calculations,
the input fission widths and the  transient times are   usually
taken from the analytical  expressions deduced from the solution
of the Fokker-Planck equation under simplifying assumptions
\cite{Kramers, Weid2, Grange3, Jurado2}. These analytical
expressions for $\Gamma_f(t)$  were all obtained under the
approximation of constant, shape-independent nuclear friction.
However, since friction was shown to be strongly shape dependent
by extensive experimental work and theoretical analysis, numerical
solution of the dynamical equation is essential. In our work, the
fission widths will be obtained numerically by solving the
Langevin equation using the chaos-weighted wall
friction\cite{Gargi1} (as described in chapter 3) and the
transient time will be obtained by fitting the numerically
calculated fission widths with an analytical expression. Thus the
expression for $\Gamma_f(t)$ to be used in the statistical
calculation has inputs from the numerically solved Langevin
dynamics and hence is a much realistic description of the actual
process. In the next part of our work, we would perform such
statistical model calculations for prescission neutron
multiplicity using the time dependent fission widths  as well as
the single swoop description of fission. This would be done with
the aim of finding how  well the statistical model calculations
with and without the single swoop assumption agree with  each
other.  We  would subsequently calculate  the prescission  neutron
multiplicity in a dynamical model of fission and compare the
results  with those obtained from the statistical calculations
with time dependent fission widths. Though one would expect the
results from the statistical and the dynamical calculations to be
the same, there could be some differences and we would ascertain
the magnitude of such differences from our calculation. \par As we
have shown in chapter 4, the prescission neutron multiplicity and
fission probability calculated from Langevin dynamics using the
chaos-weighted  wall friction  were found  to agree fairly well
with the experimental data for a number of heavy compound nuclei
($A \sim 200$) over  a wide range of excitation energies
\cite{Gargi2}.  This observation as well as the fact that CWWF
does not contain any free parameter to fit experimental data
motivated us to use this modified form of one-body friction to
pursue our study of transients in this work. In the 3rd chapter, a
systematic study of fission widths using this friction is already
reported. That study was confined to cases with fission barriers
whereas we would concentrate upon fission in the absence of a
barrier in the present chapter. The details of our model including
the nuclear shape, potential, inertia and friction is given in the
2nd chapter. The next section will contain the numerical results
of our study while the last section  will present a summary of
this chapter. \vspace{-0.4cm}
\section{Results}
\subsection{Fission widths from Langevin equation}
The time-dependent behaviour of fission widths under different
physical conditions is  being studied here using the Langevin
equation. Starting with a given total excitation energy ($E^{*}$)
and angular momentum ($l$) of the compound nucleus, the energy
conservation in the following form,
\begin{equation}
E^{*}=E_{int}+V(c)+p^{2}/2m \label{(5)}
\end{equation}
\noindent gives the intrinsic excitation energy $E_{int}$ and the
corresponding  nuclear  temperature $T=(E_{int}/a)^{1/2}$ at each
integration step. A Langevin trajectory  will  be  considered  as
having  undergone  fission  if  it  reaches  the  scission  point
($c_{sci}$) in the course of its time evolution. The calculations
are repeated for a large number (typically 100 000  or  more)  of
trajectories  and  the  number of fission events is recorded as a
function of time. Subsequently the time dependent fission   rates
can be easily evaluated as described in chapter 3.\par
\begin{figure}[h!]
\vspace{0.6cm}
\centerline{\psfig{figure=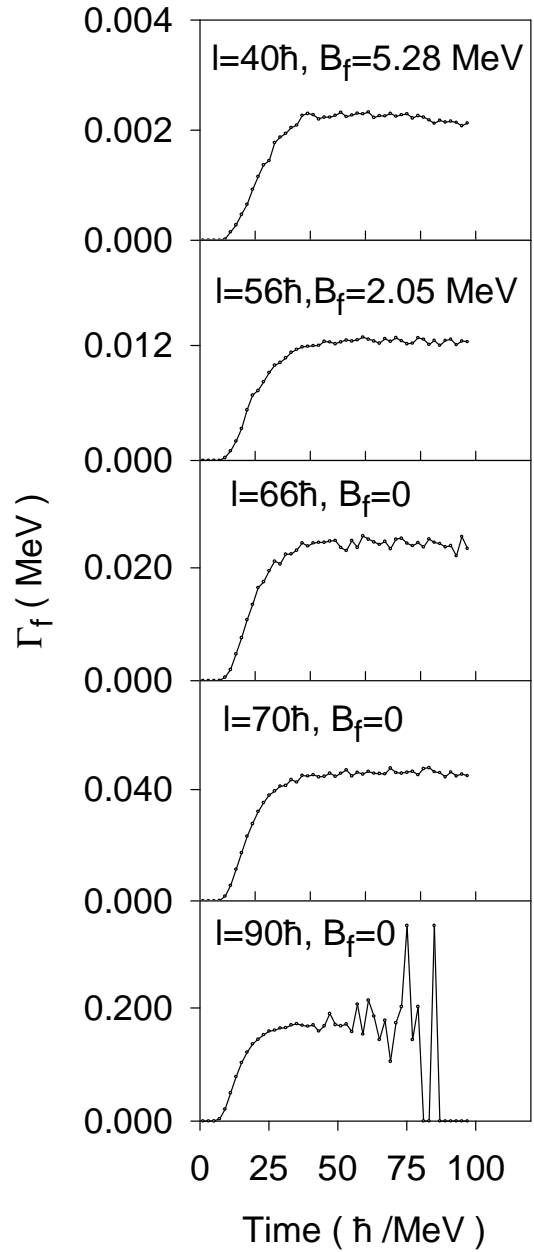,width=0.45\textwidth}}
\caption{\label{fig1} Time development   of   fission   widths
calculated  for  the compound nucleus $^{200}$Pb at a temperature
of    2  MeV  for different nuclear spins $l$. The corresponding
values  of  the fission barriers  $B_{f}$  are   also   given.}
\end{figure}

 We  have
chosen  the  $^{200}$Pb  compound nucleus for our study which has
been  experimentally  formed  at  different  excitation energies
in  a  number  of  heavy  ion  induced fusion reactions
\cite{Newton1,Hinde6,Hinde7}. The results which will be described
in this section is reported in Ref. \cite{Gargi4}.
   Fig.  \ref{fig1}   shows the
calculated time  dependent  fission widths at different spins of
the compound nucleus  for  a  given  temperature.  A  number of
interesting observations  can be made from this figure. The time
dependence of the fission width of the compound  nucleus with  a
spin  of 40$\hbar$  (and with a fission barrier) is typical
 of a
diffusive flow across the fission barrier which has been  studied
extensively in chapter 3\cite{Gargi1}. The fission width is found
to remain practically zero till a certain interval of time
($t_{0}$) which essentially  corresponds to  the interval after
which  the fission trajectories  start arriving at the scission
point. The fission width  subsequently increases with time till it
reaches its stationary value ($\Gamma_{0}$). The following
parametric form will be used for the time dependent fission width
in order to enable us to use it in our later calculations,
\begin{eqnarray}
\Gamma  (t)  =  \Gamma_{0}   [1-   exp(-(t-t_{0})/\tau)] \Theta
(t-t_{0}) \label{(6)}
\end{eqnarray}
\noindent  where  $\tau$ is a measure of the transient time after
which  the  stationary flow is established and $\Theta(t)$ is the
step function. The intervals $t_{0}$ and $\tau$ are  obtained  by
fitting the calculated fission widths with the above
expression.\par It is observed from  Fig. \ref{fig1} that the
nature of the time dependence of the fission width  remains almost
same even though the fission barrier  decreases  and subsequently
vanishes  with increasing spin.   At  very  large values  of spin,
however, fluctuations appear at the later stages of time
evolution.  These fluctuations are statistical  in nature  because
the  number of nuclei which have not yet undergone fission
decreases  very  fast with increasing  time  for higher  values of
spin and therefore introduces large statistical errors in the
measured numbers.  The magnitude  of the fluctuations can thus be
reduced by considering a larger number of fission trajectories. In
our  calculation,  we have  taken particular  care by using larger
ensembles at higher values of nuclear spin in order to enable us
to check  whether  a stationary value of the fission width is
attained at all.\par The  above  observation  is of particular
interest since it shows that the diffusive nature of  fission
persists  even  for  cases which  have  no  fission  barrier. This
diffusive  nature  is a consequence  of  the  random  force acting
on   the   fission trajectories  as we have discussed earlier. As
a compound nucleus is formed having no potential pocket in the
fission  channel,  it starts  rolling  down  the potential towards
the scission point. However, the random force acting on  these
fission  trajectories introduces  a spread in their arrival time
at the scission point. The spread in the arrival time of the
fission trajectories  gives rise to a finite fission width as we
find in Fig. \ref{fig1}.\vspace{0.5cm}
\begin{figure}[h!]
 \centerline{\psfig{figure=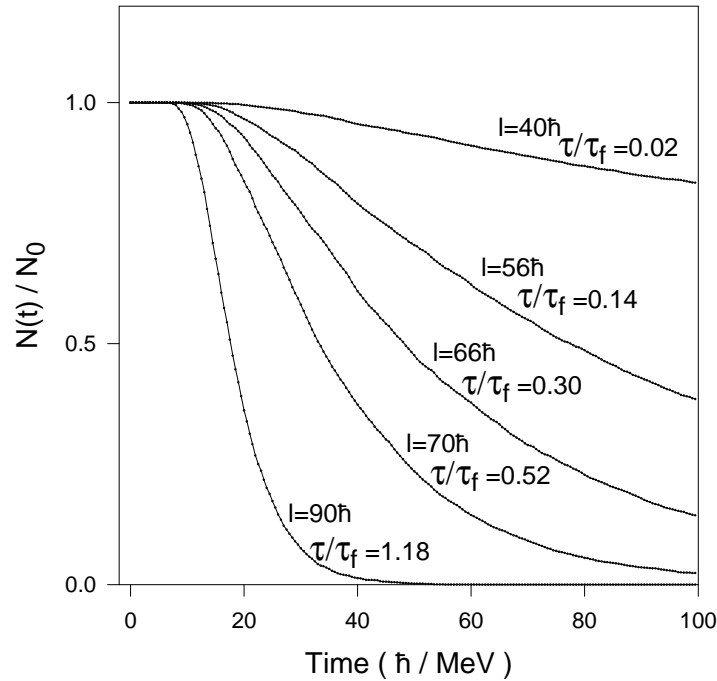,width=0.6\textwidth}}
\caption{\label{fig2}Survival probability of the compound nucleus
$^{200}$Pb against fission at a temperature of 2 MeV for different
nuclear  spins  $l$. The corresponding values of the ratio of the
transient  time to the fission life time ($\tau  / \tau_{f}$) are
also given.}
\end{figure}
\par
\begin{figure}[h!]
\centerline{\psfig{figure=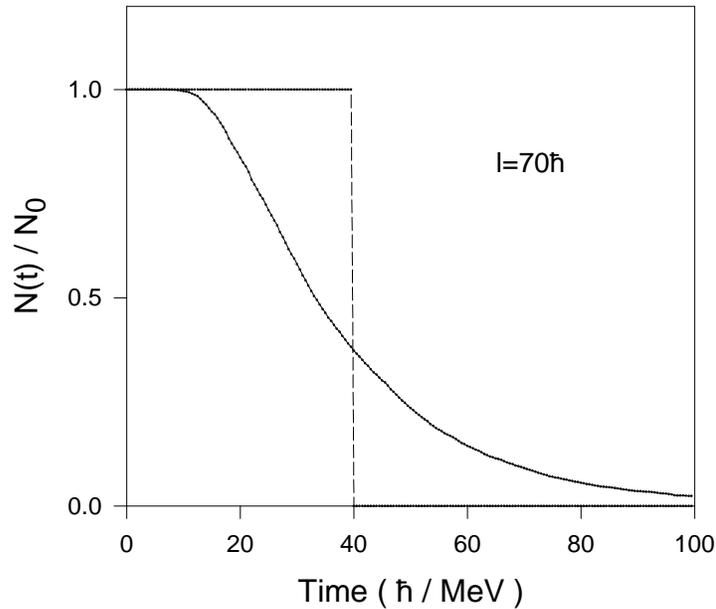,width=0.6\textwidth}}
\caption{\label{fig3}Survival probability of the compound nucleus
$^{200}$Pb  against  fission at a temperature of 2 MeV calculated
with (solid line) and without (dashed line) the random  force  in
the Langevin equation.}
\end{figure}

In  order  to  further  investigate the above diffusive nature of
fission, the fraction of the number of compound nuclei which have
survived  fission  is  shown  as  a  function  of  time  in  Fig.
\ref{fig2}.  The same compound nuclei as in Fig. \ref{fig1} has
been considered for this figure. Here we find a gradual shift in
the decay rate with increasing  spin of the compound nucleus.
Specifically, the exponential decay of  the number  of compound
nuclei  having a fission barrier (with spins 40 and 56$\hbar$) is
found to continue for those without fission barriers (with  spins
66,  70 and 90$\hbar$).  Subsequently the fraction of the
surviving  compound  nuclei have been calculated from the Langevin
dynamics by switching  off  the random force. Fig. \ref{fig3}
shows this decay in which all the nuclei have the same life time
which is simply the swooping  down  time ($\tau_{s}$) from  the
initial to the  scission  configuration. The spread in the life
time of the trajectories around this value when the random  force
is switched on  can also be seen in this figure. It may also be
noted that for very large values of the compound nuclear  spin,
the decay is very fast and consequently, the above spread is very
small. For such cases, fission is dominated by the transients and
can be approximated by a single swoop process. \vspace{0.6cm}
\par
  The  relevance of the different time scales in order to
distinguish between the  roles  of  stationary flow and transients
in  fission will now be investigated.  When  the  fission life
time ($\tau_{f}= \hbar/ \Gamma_{0}$) is much longer than the
transient time $\tau$, most of the fission  events  take  place
after  the establishment of a stationary  flow. Evidently, this
holds for nuclei with a barrier in the fission channel. However,
it is also possible to have $\tau_{f} > \tau$  for  cases  which
have  no fission barrier.
\begin{figure}[h!]
\vspace{0.5cm}
\centerline{\psfig{figure=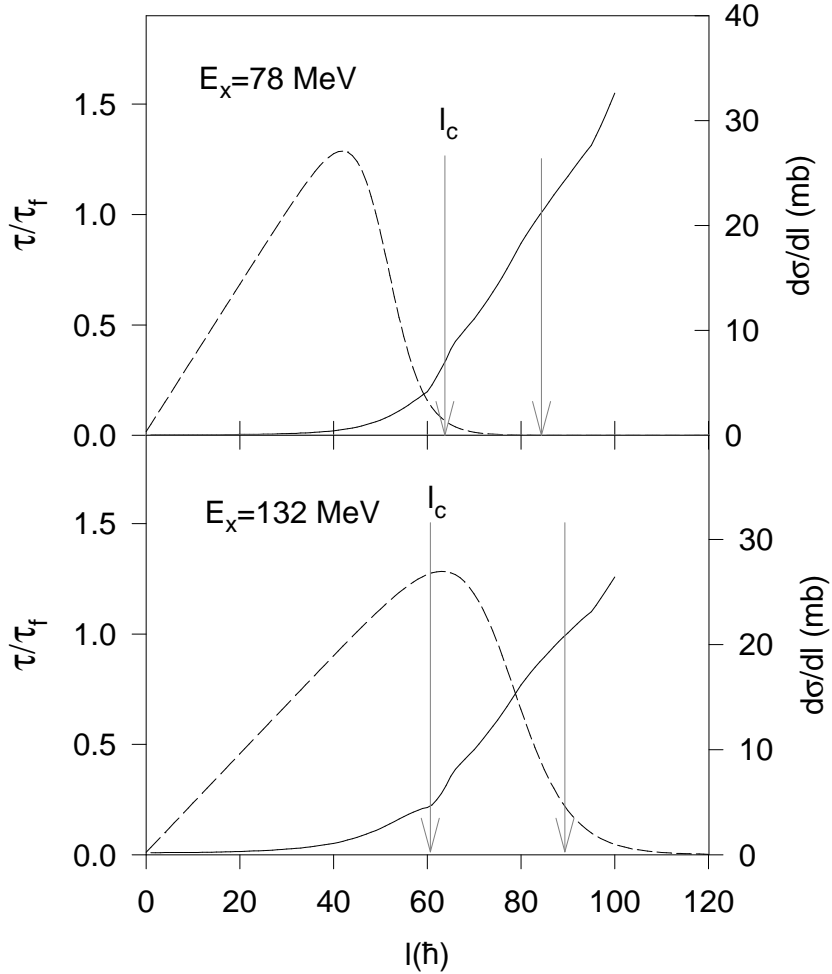,width=0.7\textwidth}}
\caption{\label{fig4}The ratio of the  transient  time  to the
fission life time ($\tau / \tau_{f}$) as a function of  the  spin
$l$ of the compound nucleus $^{200}$Pb at two excitation energies
(solid  lines). The transition  region is  indicated by the two
arrows. The arrow at the  critical angular  momentum  ($l_{c}$)
marks  the  beginning of  the transition region. The next arrow
corresponds to $\tau / \tau_{f} =1$ and indicates the end of  the
transition  region. The partial cross  sections  for  compound
nucleus formation are also shown (dashed lines).}
\end{figure}
 This is illustrated in fig. \ref{fig4}
where the ratio $\tau  / \tau_{f}$ is plotted as a function of the
spin of the nucleus. Beyond the critical angular  momentum
$(l_{c})$  at which  the fission barrier vanishes, we find a
window of angular momentum where $\tau_{f}$ is indeed  greater
than  $\tau$.  This window represents  the  transition region over
which the fission dynamics changes  from  a  steady  flow  to
transients.  Fission becomes transient   dominated   for   spin
values   at  which $\tau>\tau_{f}$. A single swoop description of
fission  can  be applied  for such cases. However, a single swoop
picture would be rather inaccurate in the transition region where
a  steady  flow still  persists.  In  the  next  subsection, the
consequences of using the single swoop description of fission in
statistical  model calculations in terms of the multiplicities of
prescission neutrons will be explored. It would  be of  interest
for our later discussions to locate the transition region with
reference to the spin  distributions  of  the compound nuclei
formed in heavy ion induced  fusion  reactions. The spin
distribution of the compound nucleus $^{200}$Pb obtained in the
fusion  of $^{19}F$+$^{181}$Ta at two excitation energies is
therefore plotted. It is observed that the transition region lies
beyond the range of the spin distribution when the compound
nucleus is excited to 78 MeV, whereas  it is well within the range
of the spin values populated at an excitation of 132 MeV.  One
would thus expect that the number  of prescisssion neutrons would
be affected more at higher excitation energies when the single
swoop picture is used in  the transition region.

 \subsection{Prescission  neutrons from dynamical and statistical
model calculation}
 A comparison of the  prescission neutron multiplicity from
  the Langevin dynamics
of fission as well as  from  a  statistical model calculation
(where  time-dependent  fission widths will be used) is studied in
details in this section. The details of the dynamical model along
with statistical evaporation of neutron and giant dipole $\gamma$
is described in the 4th chapter. The same procedure is followed in
the present calculation.  A Langevin trajectory will be considered
as undergone fission if  it reaches the  scission point in  course
of its time evolution. Alternately it will be counted as  an
evaporation residue  event if  the   intrinsic excitation energy
becomes smaller than either the fission barrier or  the binding
energy of a neutron. The calculation proceeds until the compound
nucleus undergoes fission or  ends  up  as  an evaporation
residue.  The number of emitted neutrons and photons is recorded
for each fission event. This calculation is  repeated for  a large
number of  Langevin trajectories and the average number of
neutrons emitted in the fission events  will  give  the required
prescission neutron multiplicity.\par The statistical model
calculation of prescission neutron emission proceeds in a similar
manner where a time-dependent fission width is  used to decide
whether the compound nucleus undergoes fission in  each interval
of  time  evolution. The intrinsic excitation energy at each step
is given by the total excitation energy minus the rotational
energy since no kinetic energy is associated  with the fission
degree of freedom in the statistical  model  and  the compound
nucleus   is   assumed  to  be  in  its  ground  state
configuration   (zero   potential   energy). Two prescriptions for
the  time-dependent  fission  widths  will be used in  our
calculation. In the first one, we shall use the parametric form of
the  width given by  Eq. \ref{(6)}  for  all  spin  values
including those for which there  is  no  fission barrier. The
parameters $\Gamma_{0}$, $t_{0}$  and  $\tau$  are  obtained  by
fitting the numerically calculated time-dependent widths. In the
other statistical model calculation, the  above parametric form
will be used only for those spin values which have fission
barriers. For higher spin values for which there  is  no  fission
barrier including those in the transition region, the swooping
down picture will be applied. For  these  cases, the swooping down
time $\tau_{s}$  is evaluated numerically as explained earlier. In
this statistical model calculation,  neutron  and $\gamma$
evaporation can take place during this period $\tau_{s}$ while the
nucleus will be considered as undergone fission at the end of this
interval.
\begin{figure}[h!]
\vspace{0.5cm}
\centerline{\psfig{figure=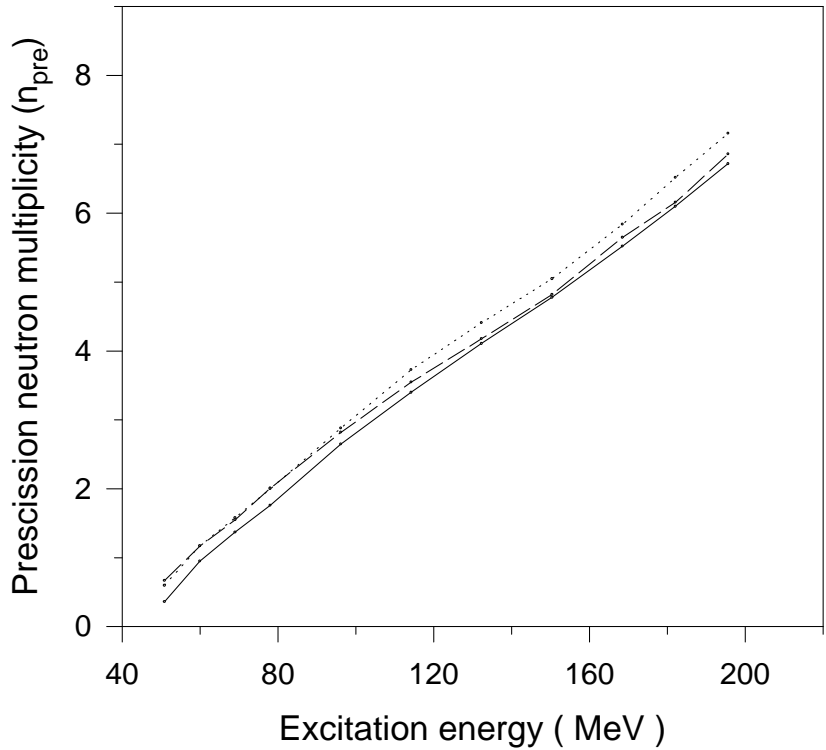,width=0.7\textwidth}}
\caption{\label{fig5}Prescission neutron      multiplicities
calculated from the statistical  model with  (dotted  line)  and
without  (dashed line) the single swoop approximation (see text).
Results from the dynamical model (solid line) are also shown.}
\end{figure}
\par Fig.  \ref{fig5}  shows  the  calculated
prescission  neutron multiplicity at different excitation energies
of  the  compound nucleus $^{200}$Pb  formed in the $^{19}$F +
$^{181}$Ta reaction. Results shown in this figure are obtained
from the dynamical  and statistical  model  calculations which are
continued for a period of 300$\hbar/MeV$. This time period is not
sufficient for all the nuclei in the ensemble either to reach the
fission  fate  or  to become  evaporation  residues. Pushing  the
Langevin calculation much beyond the above time period becomes
prohibitive in terms of computer time. The above time duration  is
however  much  longer than  the  transient times and hence are
adequate for our purpose of comparing the dynamical and
statistical results.
\par
It is observed from Fig. \ref{fig5}  that  the  neutron
multiplicity calculated from the statistical model  using  the
time-dependent fission widths with  and  without  swooping down
assumption are almost same at lower  excitation  energies  though
they  differ marginally  at higher excitation energies. Such a
difference was anticipated in the earlier subsection  since  the
swooping  down assumption is invoked more frequently for compound
nuclei at high excitation energies which are mostly formed with
large values of spin and consequently  with  no  fission  barrier.
\begin{figure}[h!]
\vspace{0.5cm}
\centerline{\psfig{figure=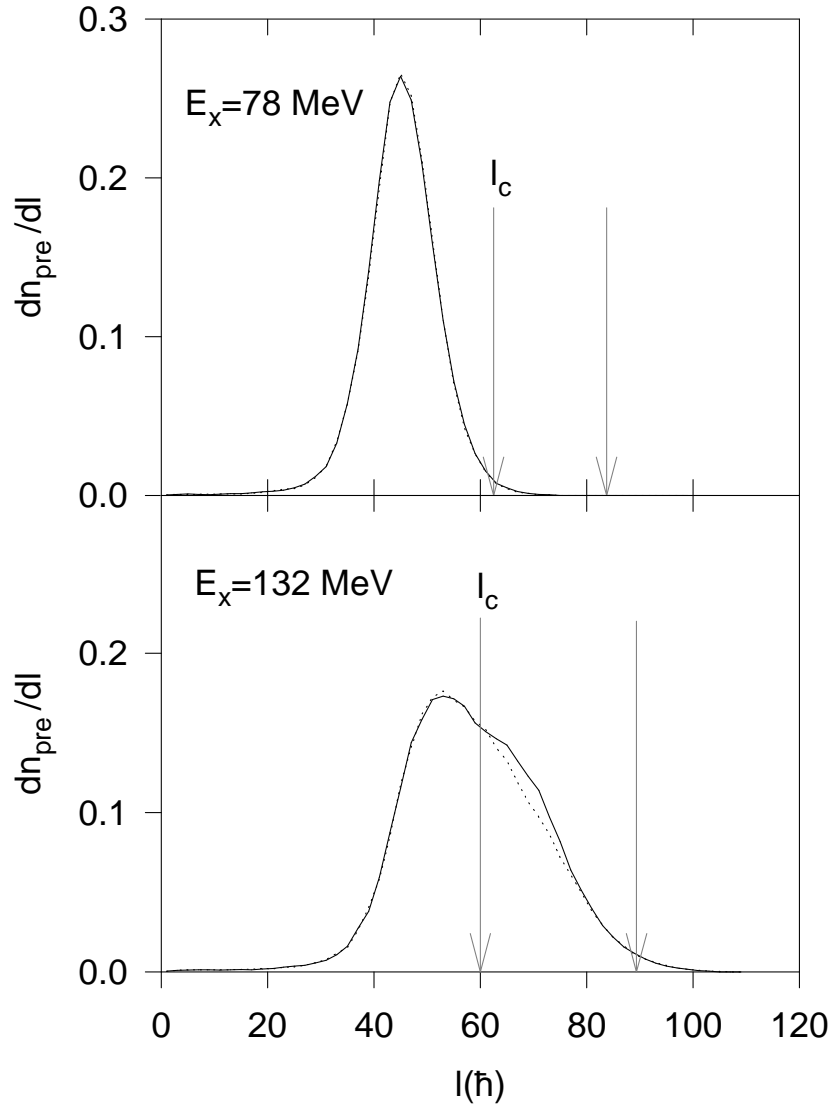,width=0.7\textwidth}}
\caption{\label{fig6} Differential prescission neutron
multiplicities  calculated  with the single swoop approximation at
two excitation energies  (solid lines). The corresponding
distributions  without the single swoop approximation are shown by
the  dotted  lines.  The  transition regions are also indicated as
in fig.4.}
\end{figure}
In  order  to explore this   point   further,   the   differential
neutron multiplicities  are   obtained   from   the   statistical
model calculations with as well as without the single swoop
description and   are   shown   in   Fig. \ref{fig6}.  The  two
calculated distributions  at an excitation energy of 132 MeV are
found to be different beyond $l_{c}$ though they merge again  at
the  higher end   of  the  transition  region.  This  difference
essentially reflects the approximate nature of the single  swoop
description in   the  transition  region.  However,  the magnitude
of  this difference is found to be rather small ($\sim$ a few
$\%$). At  a lower  excitation  of  78  MeV,  the two
distributions are almost identical as one would expect since they
have very little overlap with  the  transition  region.  The
significance  of  the  above observations is of interest since  it
shows  that  for  compound nuclei without a fission barrier,
considering a sharp valued life time  (the  swooping down time
$\tau_{s}$) instead of a life time with a dispersion does not make
any  appreciable  effect  in  the number of emitted neutrons
before fission. It is next observed in Fig. \ref{fig5} that the
neutron multiplicity from the statistical (both  calculations) and
dynamical models are also very close to each other though the
statistical models marginally  overestimate the  neutron
multiplicity  compared  to  the  dynamical model. A possible
explanation for this observation would be the fact  that the
compound  nuclear  temperature  in  the statistical model is
higher than that in the dynamical model since a part of the total
excitation energy is locked up as kinetic energy of  the  fission
mode   in   the  dynamical  model.  This  reduces  the  intrinsic
excitation energy and hence  the  temperature  in  the  dynamical
model resulting in a smaller number of evaporated neutrons.
\par
It is already mentioned that a full dynamical calculation can take
an extremely long computer  time  particularly  for  those
compound nuclei whose fission  probability  is small. Hence the
combined  dynamical  and  statistical  model, first proposed by
Mavlitov {\it et al.} is followed \cite{Mavil}, in  order to
perform a full calculation.  This approach has been described in
details in the previous chapter. In this model, the time evolution
of a compound nucleus is followed according  to the Langevin
equations for a sufficiently long period  (during which a steady
flow across the fission barrier is established) and  then switch
over to a statistical model description after the fission process
reaches the stationary regime. It is possible to continue this
calculation for a sufficiently long time such that every compound
nucleus  can be accounted  for  either  as an evaporation residue
or having undergone fission.
\begin{figure}[ht!]
\vspace{0.5cm}
\centerline{\psfig{figure=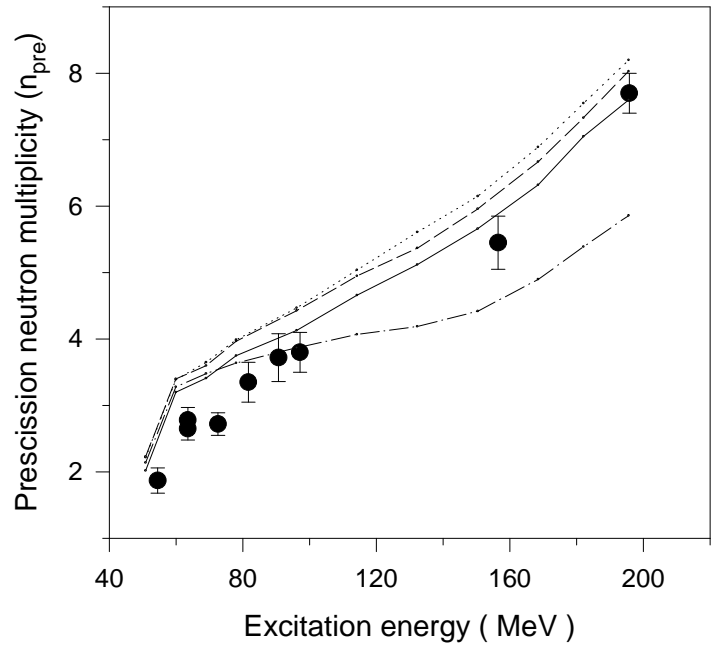,width=0.6\textwidth}}
\caption{\label{fig7} Prescission    neutron   multiplicities
calculated   from   the statistical model with (dotted line) and
without  (dashed  line) the  single swoop approximation and also
from the dynamical model (solid line) along with the experimental
data.  The results of a statistical  calculation  using  the
stationary  values  of  the fission widths are also shown
(dash-dotted line). }
\end{figure}
\par
The  prescission  neutron  multiplicity calculated with the above
combined  dynamical  and  statistical  model  is  shown  in  Fig.
\ref{fig7}  along  with  the full statistical model calculations.
The  statistical  model  calculations  are  made  with as well as
without the swooping down assumption in  the  time-dependence  of
the  fission  widths.  The  experimental values are also shown in
this figure. The observations made in this figure are similar  to
those  in  Fig. \ref{fig5},  namely,  the statistical calculations
slightly overestimate the neutron multiplicity  compared  to  the
dynamical    (plus   statistical)   calculation.   However, the
statistical and dynamical results are quite close to  each  other
and  are  also  close  to  the  experimental  values. This result
therefore   shows   that   the   statistical   calculation   with
time-dependent   fission   width   can  represent  the  dynamical
calculation with reasonable accuracy. The results of a statistical
calculation  is also shown in this figure where the fission widths
are assumed to be  independent  of  time  and  are given by their
stationary values. This calculation substantially underestimates
the  neutron multiplicity  and  illustrates  the importance of
transients at higher excitation energies.
 \section{Summary }
We  have  presented  in  the  above  a  numerical  study  of  the
transients in the fission of  highly  excited  nuclei  and  their
effect  on  the  number  of neutrons emitted prior to fission. To
this end, we first investigated the  time-dependence  of  fission
widths using the Langevin dynamics of fission. We have shown that
the  fission  width  reaches a stationary value after a transient
period even for those nuclei which have no  fission  barrier.  We
have discussed the role of the random force acting on the fission
trajectories in introducing a dispersion in their arrival time at
the  scission  point  and thereby giving rise to a finite rate of
fission for such cases. We have also shown that  this  stationary
fission  rate  for very large values of spin of the nucleus loses
significance  since  the  stationary  fission  life  time  itself
becomes  much  smaller  than  the  transient time for such cases.
Therefore, fission  of  nuclei  rotating  with  a  large  angular
momentum  can  be  considered  to  proceed in a single swoop. Our
study demonstrates a gradual transition from  a  diffusive  to  a
single  swoop  picture  of  fission  with  increasing spin of the
compound nucleus.
\par
We have subsequently examined the effect of the transients on the
multiplicity  of  the  prescission  neutrons emitted in heavy ion
induced  fusion-fission  reactions.  We  used  both the diffusive
description  and  the  swooping  down   picture   separately   in
statistical  model calculations and found close agreement between
the two calculated neutron numbers  at  low  excitation  energies
whereas  they  differed  marginally at higher excitations. It was
also  shown  that   the   differential   neutron   multiplicities
calculated  with  and  without the single swoop assumption differ
only in  the  transition  region  though  the  magnitude  of  the
difference is small. We therefore conclude that  the single swoop
description  of  fission  can  be  used  in   statistical   model
calculations without making any  significant  error  in the final
observables.\par We  finally  compared    the number of neutrons
calculated from a dynamical model with that obtained from a
statistical  model  in which  time-dependent  fission widths are
used. We found that the statistical model marginally overestimates
the  neutron  numbers than those from the  dynamical  calculation.
We  explained  this difference  in  terms  of  the  temperature
which is lower in the dynamical model than the statistical
calculation. The temperature turns out to be  smaller  in  the
dynamical  model  because  the excitation  energy  is shared
between the collective fission mode and the thermal mode in the
dynamical calculation in contrast  to the  statistical
calculation where the full excitation energy is assumed to be
available in the thermal mode. However, in most  of the  fission
events  in  the  dynamical calculation, the kinetic energy builds
up to values which are a little above  the  fission barrier
before  it  proceeds to fission. Since the values of the fission
barrier (typically a few MeV or less)  are  much  smaller than
the  excitation  energies  (a  few  tens  of  MeV  or more)
considered  here,  the  temperature   differences   between   the
statistical  and  dynamical calculations remain small for most of
the cases. Consequently the difference  between  the  prescission
neutron   multiplicities   calculated   from  the  dynamical  and
statistical models become small,  as  we  have  observed  in  our
calculation.

\chapter{Summary, discussions and future outlook}

\section{Summary and discussions}
 A detailed study of fission dynamics of highly excited nuclei
formed in heavy-ion collisions is presented in this thesis with a
view to extract knowledge about the dissipative properties of hot
fissioning nuclei. Chaos-weighted wall friction(CWWF) which is a
microscopic model of nuclear friction, is  incorporated in a
Langevin dynamical model, and different observables namely
prescission neutron multiplicity, fission probability and
evaporation residue cross-sections  are calculated. CWWF takes
into account
 the nonintegrabilty of
 single particle motion in the nucleus and it resulted in a strong suppression
 of friction strength for near spherical shapes of the
 nucleus. A general introduction to the subject and application
 of Langevin dynamics in fission with description of different
  input parameters are given in chapter 1 and 2 respectively.
 \par In chapter 3 a systematic study of the fission
 widths is made for different excitation energies and spins of a
 compound nucleus, using both CWWF and WF in the Langevin
 equation. The fission widths calculated with CWWF turned out to
 be about twice the widths calculated with the normal wall
 friction(WF). A parametric representation of the calculated fission
 widths in terms of the temperature and spin of the compound
 nucleus was made so that they can be used in subsequent statistical
  model
 calculation of the decay of a highly excited compound nucleus
 which requires dynamically calculated fission widths as inputs.
 The substantial enhancement in the fission widths with CWWF
 was expected to influence fission probability and neutron
 multiplicity  and this motivated the use of these
 parameterized widths in the next part of our work.\par
 In chapter 4 of this thesis, the chaos-weighted wall friction is
 applied to a combined dynamical plus statistical description of
 compound nuclear decay where fission is governed by the Langevin
 equation coupled with statistical evaporation of light particles
 and photons. The calculation was done using
 both CWWF and WF and the results show that both the prescission
 neutron  multiplicity as well as the
fission probability  calculated  with  the chaos-weighted wall
friction for the compound nuclei $^{178}$W,  $^{188}$Pt,
$^{200}$Pb, $^{213}$Fr, and $^{224}$Th  agree  much  better  with
the experimental data compared to the predictions of the normal
wall friction. The separate contributions of presaddle and
postsaddle neutrons at different excitation energies were
investigated in order to gain further insight into the dynamics of
fission. The postsaddle neutron multiplicities calculated with the
CWWF and WF  are almost same for all the  compound nuclei over the
range  of  excitation  energies considered. This is due to the
fact that the number of post saddle neutrons depend on the
strength of friction between saddle and scission, and CWWF and WF
are indeed close to each other in this regime of large
deformation. The wall friction predictions for the presaddle
contribution to neutron multiplicity are consistently higher than
those from CWWF at higher excitation energies and this gives rise
to the enhancement of the WF prediction for total prescission
neutron multiplicity as compared to that from CWWF predictions.
Since CWWF predicted neutron multiplicities  better agree with the
experimental values for all the above nuclei, we conclude that the
chaos-weighted wall friction provides the right kind of friction
to describe the presaddle  dynamics  of  nuclear fission.
\par It was also noted that the majority of
the  prescission neutrons are emitted in the postsaddle stage for
a very heavy nucleus like $^{251}$Es. The chaos-weighted wall
friction, however, cannot produce  enough  neutrons  to  match the
experimental prescission multiplicities for such a nucleus. It is,
therefore, possible that in  the  postsaddle  region,  either  the
fission dynamics  gets considerably slowed down or the neutrons
are more easily emitted. The neutron widths from a highly deformed
nucleus could be quite different from that of the equilibrated
near-spherical nucleus which we use in our calculation. Also,
dynamical effects, like inclusion of the neck degree of freedom in
the Langevin equation can influence the time scale of post saddle
dynamics and hence the number of neutrons.\par In chapter 5, we
have used the evaporation residue cross-sections  as probe for
nuclear dissipation. We have used both the standard wall friction
and its modified form with the chaos-weighted factor in order  to
calculate  the  prescission neutron multiplicity and evaporation
residue excitation functions for the $^{224}$Th  nucleus. It is
found that WF and CWWF predictions for neutron multiplicity are
very close to each other in the energy range considered, and hence
is not a sensitive probe for nuclear friction in this energy
regime. It is further noted from the results that the calculated
evaporation residue cross section is very sensitive to the
dissipation in the fission degree of freedom, the  WF predictions
being few times (typically 2-5) larger than those obtained with
the  CWWF. The  most important observation is  that the CWWF
predicted excitation function is  much closer  to the experimental
values than that obtained with the wall friction, which clearly
shows that the chaos-weighted  factor in CWWF changes its strength
in the right direction. Thus the chaos considerations may provide
a plausible explanation  for  the shape-dependence of the strength
of nuclear friction which was found \cite{Frob3,dioszegi} to  be
necessary in  order  to  fit experimental data.\par In chapter 6,
a numerical study of the transients in the fission of highly
excited nuclei was presented using our dissipative dynamical model
of nuclear fission and also the effect of the transients on the
number of prescission neutrons was investigated. The detailed
study of the time-dependent fission widths demonstrated a gradual
transition from a diffusive to a single swoop picture of fission
with increasing spin of the compound nucleus. It was found that
the fission width reaches a stationary value after a transient
period even for those nuclei which have no fission barrier. The
stationary fission life time for such nuclei is much smaller than
the transient time and hence fission can be considered to proceed
in a single swoop for nuclei rotating with large angular momentum.
For nuclei with no fission barrier, the diffusive picture and the
swooping down assumption were used separately in statistical model
 calculations and close agreement was found on the calculated
 prescission neutron numbers which justified the use of swooping
 down description of fission in statistical model calculation
 without making any significant error in the final
 observables.\par
 Finally, it was noted that statistical model calculation
 with time dependent fission widths marginally overestimates the
 prescission neutron multiplicity than a dynamical calculation.
 This is mainly because the intrinsic excitation energy turns out to be smaller in
 the dynamical model since total excitation energy is shared between
 collective and thermal degrees whereas in statistical description
 the full excitation energy is available in the thermal mode. The number of neutrons emitted
  depends  sensitively on the compound nuclear temperature which is directly
   proportional to the intrinsic excitation energy and thus the dynamical model calculation
   ends up
    with a marginally smaller number of neutrons.  It
 is to be noted that this temperature difference between the two
 types of calculation is small for most cases and hence the
 difference between the neutron numbers as observed from the
 calculations is marginal and thus statistical model calculation
 with time dependent fission widths can represent a dynamical
 model calculation with reasonable accuracy.
 \section{Future Outlook}
 The dynamical model of fission of hot nuclei using the chaos weighted
 wall friction can be extended to include the asymmetry degree of freedom
 in the dynamics in order to study the fission fragment mass
 energy distributions. The available experimental data of the
 prescission neutron multiplicities as a function of the fragment
 mass asymmetry and kinetic energy can be compared with the
 results of this dynamical model which includes asymmetry parameter
 as a collective coordinate. The comparison of such exclusive
  experimental data with the theoretical results is in fact a crucial
  test for the stochastic approach to fission dynamics
   based on the Langevin equations.
  This will also help in analyzing and
 elucidating correlations between the prescission neutron
 multiplicities and fission fragment mass energy distribution and
  to study the fission fragment angular anisotropy. It is
 claimed in \cite{Ross} that fission fragment angular anisotropy
 is strongly affected by evaporation of presaddle neutrons and
 hence is a strong probe of dissipation, in particular for compact
 configurations. Three dimensional Langevin calculations of the
 fission fragment mass energy distributions have  been carried
 out\cite{Vanin1,Nadtochy1} using wall friction with a adjustable
 reduction coefficient. Our model of nuclear friction (CWWF) should be used
 for such  three-dimensional  dynamical calculations and
  analysis of the relevant experimental
 data.
  This  calculation will be of great significance
 since it can predict the production
 cross section of exotic nuclei as asymmetric fission fragments.
 The chance of formation of exotic nuclei by this method will
 depend sensitively on nuclear dissipation and CWWF  is expected to
  play a significant role. The production
 cross-section of fission fragments predicted by the fission dynamics
  calculation is also expected to have significant bearing on transmutation
   of nuclear waste.  \par
 The neck degree of freedom designated by the coordinate $h$ in
 our shape parametrization, will be included in our dynamical
 model in near future in order to compare the results of the
 already reported one dimensional calculation with the two
 dimensional one. Fission rates obtained from the two dimensional calculation
 is expected to improve upon that of the one dimension though the
 difference will not be very significant. On the whole,
 the effect of incorporation of this
 additional collective coordinate in the dynamical model
 is expected to influence the prescission neutron multiplicity
 and fission probability in a favourable direction. The inclusion
 of an additional degree of freedom will increase the accuracy of our
 dynamical model and hence will be a better test of our model of
 friction. The production cross section of evaporation residues
    which can be better predicted from this multi-dimensional
     fission dynamics model will help in search
 of  superheavy elements
  formed as evaporation residues.\par
 It was concluded from the analysis of our results of postsaddle,
 presaddle and total prescission neutrons that chaos weighted wall
 formula is the right kind of friction to describe the presaddle
 dynamics of a hot rotating nucleus. It was also noted that for a
 heavy nucleus like $^{251}Es$, most of the prescission neutrons
 are contributed by the post saddle ones and the calculated number
  using CWWF number falls
 short of the experimental value by a considerable margin. In
 fact,
 it was seen that in order to match experimental data,
 the empirical friction of Frobrich {\it et al.}
  needs to be much stronger in the large deformation region(post
 saddle part) than that given by the chaos weighted wall formula.
 However,
 there is no physical justification of increasing the strength of the
 nuclear dissipation in the post saddle region beyond that of the
  wall friction.
 Instead, we feel that
the  mechanism  of  neutron  emission  in  the  postsaddle  stage
requires  a  closer  scrutiny  essentially  because  the  nucleus
becomes strongly deformed beyond the saddle  point.  The  neutron
decay  width  of  such a strongly deformed nucleus could be quite
different from that of the  equilibrated  near-spherical  nucleus
which   we   use   in   our   calculation.   In  particular,  the
neutron-to-proton ratio is expected to  be  higher  in  the  neck
region  than  that  in  the  nuclear bulk and this can cause more
neutrons to  be  emitted.  Further,  dynamical  effects  such  as
inclusion  of the neck degree of freedom in the Langevin equation
can influence the time scale of the postsaddle dynamics and hence
the  number  of  emitted  neutrons.  Such possibilities should be
examined in future for a better understanding of  the  postsaddle
dynamics of nuclear fission.\par
 In our model we have not used any explicit temperature dependence
 of the dissipation coefficient. However, the necessity for
 clarifying the role of the deformation and the temperature
 dependence is exemplified in a recent paper by Dioszegi {\it et
 al.}
 \cite{dioszegi} who were able to reproduce their data with a modified
 statistical model by applying either a strong temperature dependent
 friction form factor or with a deformation dependent form factor.
 A temperature dependence of nuclear friction is expected from
 general considerations since large phase space becomes accessible
 for particle-hole excitations at higher temperature. In fact in a
 microscopic calculation using linear response theory Hofmann {\it et
 al.} have obtained a temperature dependence of the form
 of $0.6T^2$. The role of the temperature dependence of the friction factor
 predicted by microscopic theory is still to be clarified by using it in
  Langevin calculations and confronting the results with experimental data.
    Our model can be extended to include a form factor in
 the friction coefficient which will represent the temperature
 dependence in order to examine its effect on the final
 observables. Questions such as whether both shape and temperature
 dependence should exist simultaneously or either one of them is
 sufficient should be addressed in future. Investigations regarding
 the nature of dependence on temperature is also to be carried out
 in order to arrive at a conclusive picture of the friction form factor
  with respect to its deformation and temperature dependence.\par
  Quantal corrections are expected to modify the fission rates from
   classical Langevin results up to quite high temperatures. A
  fission rate calculated with an influence functional path
  integral technique gives a $20\%$ enhancement as compared to
 a Kramers rate for fission of $^{224}Th$ at a temperature of
 1.57MeV\cite{Tillack1}. The inclusion of quantum effects
 like the shell and the pairing correlations on the potential energy landscape
 is expected to alter the barrier height which in turn will influence
 the fission rates and the fragment mass distributions as seen in Ref. \cite{Bartel1}.
   At lower temperatures, e.g. when dealing
 with Langevin models for superheavy element formation, quantum
 effects are more important.\par
 It has already been discussed in chapter 1 (section 1.2.3)
  that the Langevin
 equation needs to be generalized to allow for finite memory
 effects when the time scale of the fission degree of freedom
 becomes comparable to that of the intrinsic degrees of freedom.
 Thus one has to deal with a non-Markovian process when the
 collective motion is faster than it is assumed in our dissipative
 dynamical model. The chaos weighted wall friction can be extended
 in future to include
 the memory effects in order to examine its influence on the final
 observables. It is shown in a paper by Kolomietz et al.
 \cite{Shlomo} that the elastic forces produced by the memory
 integral in the friction kernel lead to a significant delay for
 the descent of the nucleus from the barrier. Numerical
 calculations for the nucleus $^{236}U$ show that due to memory
 effect the saddle-to scission time grows by a factor of about 3
 with respect to the corresponding saddle-to -scission time
 obtained in liquid drop model calculations with friction
 forces\cite{Shlomo}. This observation implies that incorporation
 of the memory effects in our calculation may increase the saddle
 to scission time and in turn the total prescission neutron
 multiplicity. In fact this effect can account for the empirical
 need of large increase in the strength of friction in the post
 saddle region. Thus the non Markovian dynamics may be a possible
 explanation for the extra neutrons in the post
 saddle region.\par The study of fission dynamics using the concept of
 ``Mean First Passage Time"(MFPT) has invoked interest in recent
  times\cite{Hof1,Hof3,Yord}. This time interval represents the average time
  it takes for the system to start at the potential minimum and to
   make its motion all the way out to scission. It included relaxation
   processes  around the first minimum as well as the sliding down
    from saddle to scission. The concept of the ``transient effect" is
     examined with respect to MFPT in \cite{Hof4,Hof5}. Our
  dissipative dynamical model of fission can be used to
  investigate this concept and its dependence on initial
  conditions. A comparative study of MFPT with the concept of
  transient time is to be made to reach at a more definite
  conclusion. In Ref. \cite{Jia}, the concept of ``Mean Last
   Passage Time"
  (MLPT) is proposed for the fission rate defined at the saddle
   point and it is concluded that this is a better concept than that
   of the mean first passage time(MFPT) since a dynamical effect
    of descent from the saddle point to the scission point
     has been induced in the MLPT. This idea can be checked
   using our dissipative dynamical model. \par
   In Ref. \cite{Charity0}, one-dimensional Langevin simulations
   are performed to emphasize the strong sensitivity of fission
   transients to the assumed initial shape distribution of the
   compound nuclei.  Fission delays or transient
   fission suppressions are found if the compound nucleus is
   initially spherical or near spherical, whereas a moderate
   initial fissionlike deformation can reduce the magnitude of this
   suppression (transient fission enhancement). It is argued that
   the initial conditions are determined by the fusion dynamics
   and thus fission transients are dependent on the entrance
   channel. The nature of the transients may change from
   suppression to an enhancement as the entrance-channel changes
   from asymmetric to symmetric. Transient fission will only be
    important when there is strong competition from evaporation of
    light particles and thus  calculations which invoke fission
    delays (transient effects) to explain the large number of
    prescission neutrons measured in experiments should be
    reexamined in the light of these considerations and our
    dissipative dynamical model of fission can be used for this
    purpose.\par
    In this work we have considered only those systems
where fission follows the formation of an equilibrated compound
system and the process is called fusion-fission. In our model, we
have not taken into account any delay effects in the formation
phase i.e., the previous pre-equilibrium stage is not considered
explicitly. This assumption is valid as long as the decay time of
the system is much longer than the equilibration time. However, at
sufficiently high excitation energies when the transient time is
comparable to or even greater than the stationary fission life
time, quasi-fission or fast-fission process needs to be
considered. The presence of quasi-fission process inhibits heavy
element formation and thus experimental studies of this process is
crucial for the search of superheavy
elemnts\cite{Berriman,Mukherjee}. There is an increasing amount of
data in which contributions of fast-fission or quasi-fission are
identified; i.e. there is a need for modelling these
processes.\par
  In our model, though we calculate the number of prescission
  protons, alphas and GDR $\gamma$'s, we do not compare them
  with experimental data because these numbers are rather small
  with large statistical uncertainties in the present work. In
  order to obtain the energy spectrum of the $\gamma$ multiplicity
  with a reasonable statistical accuracy, in particular, it is
  necessary to perform computation using a much larger ensemble of
  trajectories than the one used in the calculation presented in
  this thesis. This puts a severe demand on computer time making
  such computations impractical at present. However, an
  alternative approach would be to make use of the time-dependent
  fission widths in a full statistical calculation of the
  compound-nucleus decay. This calculation would be much faster
  than the present Langevin dynamical model calculation though the
  time-dependent fission widths would be required as input to this
  statistical model calculation. The results of the calculation
  are to be folded with the appropriate detector response function
  so that the calculated numbers can be compared to the experimental
  data. We plan to perform such calculations in future.\\

 \underline{\large{\bf{ Experimental scenario:}}}
 The majority of the experimental
 approaches dedicated to the study of  nuclear dissipation are based on
 nucleus-nucleus collisions at energies that range from 5 $A$ MeV
 to about 100 $A$ MeV. Among the experimental observables studied
 in this type of reactions the most common are the particle\cite{Ross}
  and
 the $\gamma$-ray\cite{Paul} multiplicities, the angular, mass and charge
 distributions of the fission fragments\cite{Schr}, and the
 fission and evaporation-residue cross sections. Except for the
 fission and evaporation-residue cross sections, all these
 observables give information on dissipation on the whole path
 from ground-state deformation to scission, but they do not allow
 exploring the deformation range from the ground state to the
 saddle point independently. Also, fusion-fission and
 quasi-fission reactions, which are mostly used, induce initial
 composite systems with large deformation, and therefore they do
 not offer suitable conditions for extracting the relevant
 information at small deformation. Contrary to fusion-fission and
 quasi-fission reactions, antiproton annihilation
 experiments\cite{Lott,Hof2,Kim, Schmid}, very peripheral transfer
 reactions\cite{Guinet} and spallation reactions\cite{Ben1} lead
 to fissioning nuclei with small deformation and small angular
 momentum, simplifying the theoretical description considerably.
 Fission induced by heavy ion collisions at relativistic energies
 offers ideal conditions for investigating dissipation at small
 deformation \cite{Jurado5}. Two new experimental signatures,
 namely the partial fission cross sections and the partial widths
 of the fission fragment charge distributions are introduced by
 these peripheral heavy-ion collisions\cite{Jurado4},
 in order to observe transient effects in fission. These
 observations exploit the influence of the excitation energy on the
 fission probability and on the fluctuations of the mass-asymmetry
 degree of freedom. They are based on the particle-emission clock;
 however the emission of particles is translated into a reduction
 of excitation energy before the system passes the fission barrier.
 These new signatures, being sensitive to the dissipation at small
 deformation, is expected to give new insights into still open
 questions on the strength of the nuclear dissipation coefficient
 and its variation with deformation and temperature. These
 investigations are planned to be extended to projectiles between
 uranium and lead in order to separately vary fissility and induced
 energy by using secondary beams, presently available at GSI.
 Further progress in this field is expected when advanced
 installations, e.g, in the planned GSI or RIA future projects, will
 become available. They will allow for more sophisticated fission
 studies by extending the isospin range of available secondary
 beams and by adding new capabilities for mass-identification and
light-particle-detection, aiming for kinematically complete
experiments with a measure of excitation energy in individual
events.
\par
Chaos-weighted wall friction, which is the friction model used in
this thesis does not have any adjustable parameter and has the
same order of magnitude for low deformation as the empirical
frictions which have successfully reproduced experimental data for
different observables\cite{Frob,Vanin2}.  In fact CWWF is the only
model of deformation dependent friction derived from physical
considerations which is closest to the phenomenological frictions
of Fr\"obrich for compact shapes \cite{Fro04}.  However, CWWF does
not increase strongly for large deformations as required by the
phenomenological friction\cite{Frob} in order to match the
prescission neutron multiplicity data for heavy systems like
$^{251}Es$ with a long saddle to scission path.  Therefore  CWWF
should be applied to wider variety of systems and for different
types of observables so that distinction between different models
could be made more confidently  in order to reach at a better
picture.
 A systematic analysis and explanation  of  all the available
  experimental data  should
be attempted using CWWF in a multi-dimensional Langevin dynamical
model.   This will be a crucial test for our theoretical model of
friction and will help to reach at a definite conclusion regarding
the friction form factor with respect to its deformation (and
temperature) dependence, and finally to arrive at an unified
picture for fission of hot nuclei.

\addcontentsline{toc}{chapter}{Appendix A: Evaluation of the
nuclear potential }
\chapter*{Appendix A}
\section*{Evaluation of the nuclear potential}
\setcounter{equation}{0}
\setcounter{figure}{0}
\def\theequation{A.\arabic{equation}}
\def\thefigure{A.\arabic{figure}}

The potential energy is obtained from the finite-range liquid drop
model, where we calculate the generalized nuclear energy by double
folding the uniform density within the nuclear surface with a
Yukawa-plus-exponential potential. The six dimensional double
folding integral for evaluation of the potential is as follows:
\begin{equation}
I= \int d^{3}r_{1}d^{3}r_{2} f(\vec{r_{1}})f(\vec{r_{2}}) v(\mid
\vec{r_{1}}-\vec{r_{2}} \mid) \label{ac1}
\end{equation}

\noindent where $f$ and $v$ gives the nuclear density and
potential respectively. The above integral is reduced to that of
lower dimensions by the method of Fourier transform. The Fourier
transform in $k$ space of the charge densities and the potential
are given by the following relations.

\begin{eqnarray}
f(\vec{r_{1}})&=& \frac{1}{{(2\pi)}^{3}}\int d^{3}k_{1}
e^{-i\vec{k_{1}}\cdot \vec{r_{1}}} \tilde{f}(\vec{k_{1}}) \nonumber \\
f(\vec{r_{2}})&=& \frac{1}{{(2\pi)}^{3}}\int d^{3}k_{2}
e^{-i\vec{k_{2}}\cdot \vec{r_{2}}}
\tilde{f}(\vec{k_{2}}) \nonumber \\
v(\mid \vec{r_{1}}-\vec{r_{2}} \mid)&=& \frac{1}{{(2\pi)}^{3}}\int
d^{3}k e^{-i\vec{k}\cdot
(\vec{r_{1}}-\vec{r_{2}})}\tilde{v}(\vec{k}) \nonumber
\end{eqnarray}

\noindent Substituting these fourier transforms in Eq.~
(\ref{ac1}), and using the following identities,
\begin{eqnarray}
\int d^{3}r_{1} e^{-i(k_{1}+k)\cdot r_{1}}&=& {(2\pi)}^{3}
\delta(k_{1}+k)\nonumber \\
\int d^{3}r_{2} e^{-i(k_{2}-k)\cdot r_{2}}&=& {(2\pi)}^{3}
\delta(k_{2}-k)\nonumber.
\end{eqnarray}
\noindent and exploiting the properties of the delta function,
the six dimensional integral is reduced to the following three
dimensional integral.
\begin{equation}
I= \frac{1}{{(2\pi)}^{3}}\int d^{3}k \tilde{f}(\vec{k})
\tilde{f}(-\vec{k})\tilde{v}(\vec{k}) \label{ac2}
\end{equation}
Since the charge distribution is symmetric i.e.,
$f(\vec{r})=f(-\vec{r})$, it can be shown that
$\tilde{f}(\vec{k})=\tilde{f}(\vec{-k})$ and hence the integration
takes the form
\begin{equation}
I=\frac{1}{{(2\pi)}^{3}}\int d^{3}k {(\tilde{f}(\vec{k}))}^{2}
\tilde{v}(\vec{k}). \label{ac3}
\end{equation}
\noindent where the inverse Fourier transform relations,

\[\tilde{v}(\vec{k})= \int d\vec{r} e^{i\vec{k} \cdot
\vec{r}}v(\vec{r}) \]

 \noindent and similarly for $\tilde{f}(k)$ are used. If potential is of the
coulomb form i.e., $v(\vec{r})=1/r$, then it can shown using
contour integration that $\tilde{v}(\vec{k})=\frac{4\pi}{k^{2}}$.
If the potential takes the exponential form i.e., $v(\vec{r})=
e^{-\mu r}$, then
$\tilde{v}(\vec{k})=\frac{8\pi\mu}{({{\mu}^{2}+k^{2}})^{2}}$. For
Yukawa type of potential, i.e., $v(\vec{r})= {e^{-\mu r}}/r$,
$\tilde{v}(\vec{k})=\frac{4\pi}{({\mu}^{2}+k^{2})}$.\\
\noindent $\tilde{f}(\vec{k})$ is evaluated in cylindrical
coordinate system. Due to axial symmetry in $f(\vec{r})$,
$\tilde{f}(\vec{k})$ will also have axial symmetry in $k$ space.
Assuming $\vec{k}$ to lie in $(y-z)$ plane, it can be shown that
$\vec{k} \cdot \vec{r} = \rho k_{\rho}\cos\phi + zk_{z}$. Hence in
cylindrical coordinate system,
\begin{equation}
\tilde{f}(k_{\rho},k_{z}) = \int e^{(i\rho k_{\rho}\cos\phi +
izk_{z)}}f(\rho,z)\rho d\rho dz d\phi \label{ac4}
\end{equation}
\noindent For uniform density, $f(\rho,z)$ = constant within the
defined surface.
\begin{equation}
\int_{0}^{2\pi} e^{(i\rho k_{\rho}\cos \phi)}d\phi = 2\pi
J_{0}(\rho k_{\rho})\nonumber
\end{equation}
\noindent where $J_{0}$ is the zeroth order Bessel function.
Using the above result,
\begin{equation}
\tilde{f}(k_{\rho},k_{z}) =2\int_{0}^{z_{max}}
cos(zk_{z})\frac{1}{k_{\rho}^{2}} I_{2}(k_{\rho} \rho(z))dz
\label{ac5}
\end{equation}
where
\begin{equation}
I_{2}(\beta) = \int_{0}^{\beta} I_{1}(x)x dx
\end{equation}
and
\begin{equation}
I_{1}(x)= 2\pi J_{0}(x)
\end{equation}
$I_{1}(x)$ is calculated for $x= 0$ to $x_{max}$, where $x=\rho
k_{\rho}$, and using these values of $I_{1}(x)$, $I_{2}(\beta)$ is
calculated for a wide range of of $\beta$ ranging from $0$ to
${\beta}_{max}$ ($\beta =\rho k_{\rho}$). The integral is
evaluated for different values and the required values are
extracted later by interpolating from the table. The function
$I_2(\beta)$ is thus required to be computed only once and can be
used as a standard input for any subsequent double folding
calculation. These values are used to evaluate
$\tilde{f}(\vec{k})$ in Eq.~(\ref{ac5}).  The integral in
Eq.~(\ref{ac2}) is finally evaluated in spherical polar
coordinates. The final form  of Eq.~(\ref{ac2})  is given by
\begin{equation}
I= \frac{1}{2{\pi}^{2}}\int_{0}^{\pi/2}\int_{0}^{k_{max}} k^{2} dk
\sin\theta d\theta\tilde{f}(k\sin\theta,k\cos\theta)\tilde{v}(k)
\end{equation}
where $k_{\rho}=k\sin\theta$ $\&$ $k_{z}=k\cos\theta$. The $k$
integration is done by dividing the range in two parts i.e, from
$0$ to $k_{1}$ and $k_{1}$ to $k_{2}$. The upper cut-off $k_{max}$
is chosen after ensuring a very good convergence of the integral.
Since the integrand for lower values of $k$ is very oscillating,
the integration here is done with very small step size, while for
the second part integration is performed with a bigger step size.
The stability of the potential calculation by this method is of
the order of 1 in $10^8$.

\newpage
\addcontentsline{toc}{chapter}{Appendix B: Generation of random
numbers }
\chapter*{Appendix B}
\section*{ Generation of random numbers  }
 \setcounter{equation}{0}
 \setcounter{figure}{0}
\def\theequation{B.\arabic{equation}}
\def\thefigure{B.\arabic{figure}}

Random number generation following a particular distribution occurs repeatedly
at different stages of our calculation. The solution of Langevin equation for
 fission dynamics requires the generation of Gaussian distributed random
number at each step of time evolution. Choosing of initial coordinates,
 momenta and spin of the compound nucleus also required the generation
 of random numbers following particular type of distribution function.

  The emission of particles
during the fission process as well as the energy of the emitted particles is
decided  by Monte-Carlo selection where random numbers are required
to be generated following uniform probability distribution.
The method for generation of random numbers following a particular
distribution function is described
here.

If the numbers $x_1, x_2,\dots, x_{n-1},x_{n}$, are the values of
one and the same random quantity $X$ under independent trials with
recurrent conditions following a particular distribution law, then
the sequence of random numbers $\{x_{n}\}$ is called a random
sequence with that particular distribution. To generate random
numbers by computers, it is convenient to consider the sequence of
random numbers uniformly distributed on the unit interval $0\leq x
\leq 1.$ The probability of generating a number between $x$ and $x
+ dx$, denoted by $p(x)dx$, for a sequence of random numbers with
a uniform and normalised ($\int_{-\infty}^{+\infty} p(x)dx =1$)
probability distribution, is given by
\begin{eqnarray}
p(x)dx &=& dx \hspace{1cm}\mbox{for}\hspace{.5cm} 0 < x < 1,\nonumber\\
       &=& 0  \hspace{1.2cm}  \mbox{otherwise}.
\end{eqnarray}
If the random sequence $\{x_{n}\}$ is uniformly distributed on the
interval $[0,1]$, then the linear transformation
\begin{equation}
y_{n}=A+(B-A)x_{n}\hspace{1cm} (n=1,2,3\dots) \\
\end{equation}
(A and B are given numbers) reduces to the random sequence
$\{y_{n}\}$ uniformly distributed on the interval $[A,B]$.\\
Now if we generate a uniform deviate $x$ and then take some
prescribed function of it, say $y(x)$, the probability
distribution of $y$, denoted by $p(y)dy$ is determined by the
fundamental transformation law of probabilities, which is simply
\begin{eqnarray}
\mid p(y)dy\mid = \mid p(x)dx\mid \nonumber \\
p(y) = p(x)\mid\frac{dx}{dy}\mid.
\end{eqnarray}
For a uniform deviate $p(x)=1$ for $0\leq x \leq 1$. Hence
\begin{equation}
p(y)dy = \mid\frac{dx}{dy}\mid dy.
\end{equation}
Having a random sequence $\{x_{n}\}$ uniformly distributed on the
interval $[0,1]$, we can construct a random sequence $\{y_{n}\}$
with a specified distribution, say one with $p(y)=f(y)$ for some
positive function $f$ whose integral is 1, using the above
transformation method. We need to solve the equation
\begin{equation}
\frac{dx}{dy}=f(y)
\end{equation}
to get y. The solution is
\begin{equation}
x= \int_{-\infty}^{y} f(y)dy = F(y)
\end{equation}
i.e., indefinite integral of $f(y)$. Hence the transformation
which takes a uniform deviate into one distributed as $f(y)$ is
therefore
\begin{equation}
y(x) = F^{-1}(x).
\end{equation}
where $F^{-1}$ is the inverse function to $F$. The inverse
function can be found analytically if feasible, otherwise computed
numerically by forming a table of the integral values and the
corresponding value is found from the table by interpolation.\par
\begin{figure}[h!]
\centerline{\psfig{figure=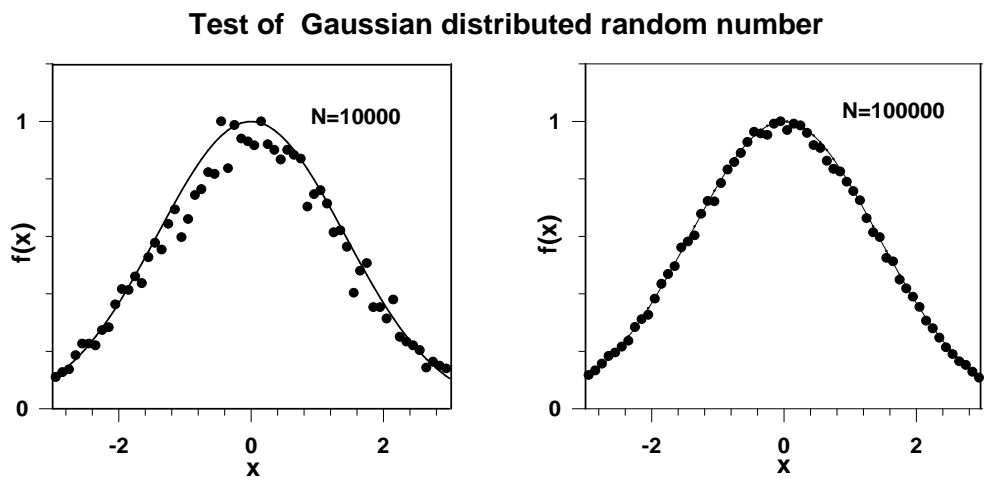,width=13cm}}
\caption{ The solid line in the figures correspond to the function $\exp{(-x^{2}/4)}$;
the filled circles correspond to the generated random numbers following
 the gaussian distributed function. }
\label{appa}
\end{figure}

A Gaussian distributed random number is numerically generated
following
 the method described above. Fig. (\ref{appa})
clearly implies that the quality of the random number improves considerably
as one increases the number of samplings $N$. The accuracy of the
 algorithm followed is also established
by the exactness of the numerically generated random number. This
method is used in all cases
 for generation of random numbers following any particular
distribution law. \vspace{1cm}

\addcontentsline{toc}{chapter}{Appendix C: Numerical integration
of the Langevin equation}
\chapter*{Appendix C}
\section*{Numerical integration
of the Langevin equation}
\setcounter{equation}{0}
\setcounter{figure}{0}
\def\theequation{C.\arabic{equation}}
\def\thefigure{C.\arabic{figure}}

The Langevin equation describing the motion of a free Brownian
particle can be solved analytically to get the mean values of the
position and momentum as a function of time $t$. As $t$ approaches
infinity, the Brownian particle is expected to be in equilibrium
with the heat bath. The average value of the kinetic energy of the
Brownian particle becomes equal to $\frac{1}{2}T$(the temperature
T is here in units of energy, i.e. we set the Boltzman constant
$k=1$).
 The one dimensional Langevin equation in $(p,q)$ space with $V=0$ and neglecting
the coordinate dependence of the inertia reads as follows:
\begin{eqnarray}
\frac{dp}{dt} &=&  -\eta \dot{q}+ g\Gamma(t), \nonumber\\
\frac{dq}{dt} &=& \frac{p}{m} ,
\end{eqnarray}

\begin{figure}[h]
\centerline{\psfig{figure=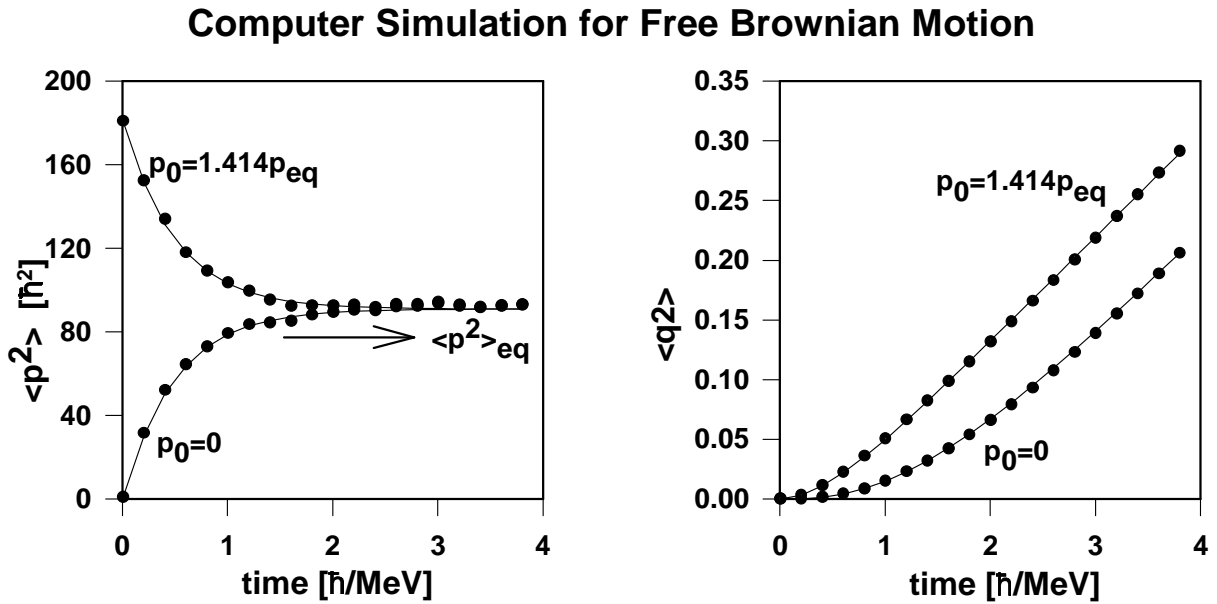,height=8cm,width=16cm}}
\caption{ $\langle p^{2}\rangle$ $\&$ $\langle q^{2} \rangle $ of
the free Brownian particle starting with two different initial
momenta ${p}_{0}$. Solid line represent the analytical solution
while the filled circles depict the numerical results.}
\label{appb}
\end{figure}

\noindent The analytical solution of the above equation is given
by\cite{Abe1}
\begin{eqnarray}
p &=& p_{0}\exp({-\frac{\eta}{m} t}) +
\int_{0}^{t}dt'\exp({-\frac{\eta}{m} (t-t')})\cdot
g\Gamma(t'), \\
q &=& q_{0} + \frac{p_{0}}{\eta}[1-\exp({-\frac{\eta}{m}t)}] +
\frac{1}{\eta}\int_{0}^{t}dt'[1-\exp(-{\frac{\eta}{m}(t-t')})]\cdot
g\cdot\Gamma(t')
\end{eqnarray}
\noindent where $p_{0}$ and $q_{0}$ are the respective initial
values. Averaging over all possible realizations of the random
force, the mean value of the momentum and coordinate are given by
\begin{eqnarray}
\langle p\rangle &=& p_{0}\exp({-\frac{\eta}{m}t}),\\
 \langle q\rangle &=&
 q_{0}+\frac{p_{0}}{\eta}[1-\exp({-\frac{\eta}{m}t})],
 \end{eqnarray}
 The mean values of the square of position and momentum
 are given as follows:
\begin{eqnarray}
\langle p^{2}\rangle &=& m\cdot
kT\cdot[1-\exp({-2\frac{\eta}{m}t})] +
p_{0}^{2}\exp({-2\frac{\eta}{m}t})\\
\langle (q-q_{0})^{2}\rangle &=& (p_{0}^{2}-3mkT)/{\eta}^{2} +
2kT/\eta \cdot t + 2\cdot (2mkT-p_{0}^{2})/{\eta}^{2}\cdot
\exp({-\frac{\eta}{m}\cdot t})\nonumber\\ && +
(p_{0}^{2}-mkT)/{\eta}^{2}\cdot\exp({-2\frac{\eta}{m}\cdot t}).
\end{eqnarray}
The above equation is also solved numerically by integrating it
directly by successive iterations(method explained in chapter 3).
The numerical results are compared with the analytical results for
the cases starting with $p_{0}=0$ and $p_{0}=\sqrt{2\cdot mT}$. \\
It is seen from Fig. (\ref{appb})  that the numerical results almost
coincide with the analytical solution at any steps. We also see
that the Brownian particle approaches  the thermal equilibrium
with the heat bath, i.e., $\langle p^{2} \rangle \rightarrow {\langle p^{2} \rangle}_{eq}
(=m\cdot T)$ irrespective of the initial momenta. The numerical results for $\langle q^{2}\rangle$ are also very well reproduced.
The errors are within $(1\sim2)\%$ which is comparable
with $1/\sqrt{N}$ where $N$ is the number of trajectories. These
results establish the accuracy and convergence of our algorithm
for solving the Langevin equation which is subsequently used in
fission dynamics.

\vskip 1cm

\newpage
\addcontentsline{toc}{chapter}{Appendix D: Units and Dimensions}
\chapter*{Appendix D}
\section*{ Units and Dimensions }
 \setcounter{equation}{0}
\def\theequation{D.\arabic{equation}}
\def\thefigure{D.\arabic{figure}}

\subsection*{Langevin dynamics}
The units and dimensions used for different dynamical variables in
the Langevin dynamics will be described here. The different input
quantities for the fission dynamics including the shape degrees of
freedom for the collective motion are described in details in
chapter 2.

(A) \underline{Shape degrees of freedom}:\\
  $c$ and $h$ are the shape variables for the fission degrees of
  freedom. Both $c$ $\&$ $h$ are dimensionless while $c_o(=cR)$
  ($R$ is radius)
  has the dimension of length and its unit is $fm$. The surface of
   the nucleus is
  defined by the following expression in cylindrical coordinates,
  \begin{equation}
  \rho^2(z) = \left(1 - \frac{z^2}{c_o^2}\right)(a_oc_o^2 + b_oz^2),
\label{ad1}
\end{equation}
\noindent where $a_o$ and $b_o$ are dimensionless (expressions
given in section 2.2), $\rho$ has the dimension of length and unit
is $fm$.

(B) \underline{Inertia}:\\
The inertia is given by the following equation, the different
quantities used being already explained in the subsection 2.3.4.
\begin{equation}
M_{ij}= \pi\rho_{m} \int_{z_{min}}^{z_{max}} P^{2}(A_{i}A_{j} +{1
\over 8}P^{2}A_{i}^{\prime}A_{j}^{\prime})dz \label{ad2},
\end{equation}
In the above expression, $\rho_m$ is the matter density, the unit
is $amu/{fm}^3$. $P$ is the value of $\rho$ on the nuclear surface
and hence has the dimension of $length$, unit is $fm$. The
mathematical definition of $A_i$ is given in section 2.3.4, from
which its dimension reduces to that of length, since in our choice
of units, $q$($c$ $\&$ $h$) is dimensionless. The unit of $z$
being $fm$, inertia has the unit of $amu\cdot {fm}^2$.

(C) \underline{Friction}:\\
The friction coefficient $\eta$ is given by the following
expression
\begin{equation}
 \eta=
{1 \over 2} \pi \rho_m {\bar v} \int_{z_{min}}^{z_{max}} { \left(
\frac{\partial \rho^2}{\partial c} \right)}^2  {\left[\rho^2 +
{\left({1 \over 2}\frac{\partial \rho^2} {\partial
z}\right)}^2\right]}^{-{1 \over 2}} dz, \label{ad3}
\end{equation}
The unit of ${\bar v}$ (average nucleon speed) is
${MeV}^{1/2}{amu}^{-1/2}$, the units of other quantities being
already, defined $\eta$ has the unit of
${amu}^{1/2}{MeV}^{1/2}fm$.

(D) \underline{Momentum}: \\
It is conventional to express time interval in units of
$\hbar/MeV$ in the calculations of fission dynamics and we will
also follow the same convention. The following relations will be
useful for the required conversion of units.
\begin{equation}
\hbar=6.4655 {MeV}^{1/2}{amu}^{1/2}fm=65.82 \times {10}^{-23} MeV
sec \label{ad4}
\end{equation}
\noindent and
\begin{equation}
{fm}^2 amu=\frac{{\hbar}^2}{MeV}\cdot\frac{1}{{(6.4655)}^2}
\label{ad5}
\end{equation}
It is convenient to express inertia and friction coefficients in
units of $\hbar$ and MeV. To convert $\mu(=1/m)$ (inverse of
inertia) given in units of ${amu}^{-1}{fm}^{-2}$ to that in
$\frac{MeV}{{\hbar}^2}$, one needs to multiply it by the
conversion factor ${(6.4655)}^2$ as seen is seen from Eq.
\ref{ad5}.  Similarly to express the friction coefficient $\eta$
in units of $\hbar$, one needs to divide $\eta$ given in units of
${amu}^{1/2}{MeV}^{1/2}fm$ by the factor $6.4655$ as is evident
from Eq. \ref{ad4}. The equation $\frac{dc}{dt}=\mu p$ gives for
momentum $p$ the units of
$(\frac{{\hbar}^2}{MeV}\cdot\frac{MeV}{\hbar})$, $c$ being
dimensionless. Hence momentum has units of $\hbar$.

(E) \underline{Energy}: \\
Energy is expressed in units of MeV and so is the temperature $T$.
It can be checked that the kinetic energy $\frac{p^2}{2}\mu$ has
units of ${\hbar}^2\times \frac{MeV}{{\hbar}^2}$ $\Rightarrow$
MeV.\\
 The units of different input quantities for the Langevin equation
  being defined, it can be checked whether the different terms of the
  equation has the correct dimensions and units.
The time evolution equation for the momentum $p$ is as follows
\begin{equation}
p(t+\Delta t)=p(t) +(-\frac{p^2}{2}\frac{\partial\mu}{\partial
c}-\frac{\partial F}{\partial c}-\eta\mu p)\Delta t + \sqrt{\Delta
t}\sqrt{\eta T}\omega_1(t). \label{ad6}
\end{equation}
All the quantities in the above Langevin equation is converted and
expressed in units involving $MeV$ and $\hbar$. The l.h.s has unit
of $\hbar$. It can be checked that all the
terms in r.h.s also has the same unit.\\
$\frac{p^2}{2}\frac{\partial\mu}{\partial c}\Delta t$
$\Rightarrow$
${\hbar}^2\cdot\frac{MeV}{{\hbar}^2}\cdot\frac{\hbar}{MeV}$
$\Rightarrow$ $\hbar$\\
$\frac{\partial F}{\partial c}\Delta t$ $\Rightarrow$
$MeV\cdot\frac{\hbar}{MeV}$ $\Rightarrow$ $\hbar$\\
$\eta\mu p\Delta t$ $\Rightarrow$
$\hbar\cdot\frac{MeV}{{\hbar}^2}\cdot\hbar\cdot\frac{\hbar}{MeV}$
$\Rightarrow$ $\hbar$\\
$\sqrt{(\Delta t)\eta T}{\omega}_1$ $\Rightarrow$
${(\frac{\hbar}{MeV}\cdot\hbar\cdot MeV)}^{1/2}$ $\Rightarrow$
$\hbar$
  ($\omega_1$ is a number and is hence dimensionless)\\
Thus it is verified that the system of units used  in the Langevin
dynamics is consistent and all the variables has the correct
dimensions.
\subsection*{\underline{Particle emission width}}
In Langevin dynamics, though it is conventional to express time in
units of $\hbar/MeV$, in the calculation of particle emission
widths, time is normally expressed in units of $fm/c$
($=3.33\times {10}^{-24} sec$). Since in our model it is required
to couple particle emission with the fission dynamics, appropriate
conversions for the corresponding units should  be made so that
the final dimensions are correct. The particle emission width is
given by the following formula
\begin{equation}
\Gamma_{\nu} = (2s_{\nu}+1){m_{\nu} \over
\pi^{2}\hbar^{2}\rho_{c}(E^{*})}\int_{0}^{E^{*}-B_{\nu}}
d\varepsilon_{\nu}\rho_{R}(E^{*}-B_{\nu}-\varepsilon_{\nu})\varepsilon_{\nu}\sigma_{inv}
(\varepsilon_{\nu})\label{ad7}
\end{equation}
The different quantities used in the above equation is explained
in section 4.2.4. It is important to note that
\begin{equation}
\hbar c=197.32 MeV fm\\ \label{ad8}
\end{equation}
\begin{equation}
 m_pc^2=938.9 MeV .\label{ad9}
 \end{equation}
 To make proper use of the units, the numerator and denominator of the
 r.h.s of Eq. \ref{ad7} is multiplied by $c^2$ and the final
 expression has the dimension of (substituting the conversions
 used in Eqs. \ref{ad8} $\&$ \ref{ad9})
 \begin{equation}
r.h.s \Rightarrow
\frac{d\varepsilon_{\nu}(MeV)\sigma_{inv}({fm}^2)m_{\nu}c^2(MeV)\varepsilon
_{\nu}(MeV)}{{(\hbar c)}^2({MeV}^2{fm}^2)} \Rightarrow
MeV.\label{ad10}
\end{equation}
 The decay width thus rightly is expressed in
units of $MeV$ and the corresponding decay time $\tau_{\nu}$ which
equals $\hbar/\Gamma_\nu$ is in units of time
($\frac{\hbar}{MeV}$)and can be converted to units of $fm/c$ using
Eq. \ref{ad8}.

\newpage
\addcontentsline{toc}{chapter}{Appendix E: Energetics}
\chapter*{Appendix E}
\section*{ Energetics}
 \setcounter{equation}{0}
\def\theequation{E.\arabic{equation}}
\def\thefigure{E.\arabic{figure}}

The energy conservation followed during the emission of a particle
from the compound nucleus is given by the following equation.
\begin{equation}
M_A+({B.E})_A+E_{A}^{*}+\frac{l_A(l_A+1){\hbar}^2}{2I_A}=
M_d+({B.E})_d+E_{d}^{*}+\frac{l_d(l_d+1){\hbar}^2}{2I_d}+M_p+({B.E})_p+E_p
\end{equation}
\noindent The subscript $A$ is for the parent compound nucleus
with mass number $A$, atomic number $Z$ and neutron number $N$,
whereas $d$ stands for the daughter nucleus after emission of a
particle (subscript $p$) from the compound nucleus. $M$ denotes
the mass which for the parent nucleus equals $Nm_n+Zm_p$ where
$m_p$ and $m_n$ are proton and neutron masses respectively. Hence,
by definition, $M_A$ will cancel with $M_d + M_p$. $B.E$ stands
for the binding energy which is calculated by the liquid drop mass
formula of Myers and Swiatecki which is given at the end of this
Appendix. If the emitted particle is not a composite particle,
i.e, if it is a neutron or a proton then the corresponding binding
energy is zero. $E_{A}^{*}$ and $E_{d}^{*}$ gives the excitation
energies of the parent and daughter nucleus respectively. $E_p$ is
the kinetic energy of the emitted particle which can vary from
zero to a maximum value fixed from the above energy balance
equation by setting  the excitation energy of the daughter nucleus
$E_{d}^{*}$ to a minimum possible value. This minimum excitation
energy of the daughter nucleus is determined from the
considerations of level density of the nucleus which should have
some finite value. Zero value of kinetic energy of the emitted
particle correspond to maximum excitation of the daughter nucleus.
$l(l+1){\hbar}^2/{2I}$ gives the rotational energy of a nucleus
with angular momentum $l\hbar$ and moment of inertia $I$. $l_A$
and $l_d$ gives the angular momenta of the parent and the daughter
nucleus where we usually take $l_d=l_A-1$ for the emitted neutron
or $\gamma$. \par The binding energy of of a nucleus with mass
number $A$, proton number $Z$ and neutron number $N$ is given by
the liquid drop model of Myers and Swiatecki\cite{Swiat1} which is
given by the following expression.
\begin{equation}
B.E=-c_1A+c_2A^{2/3}+\frac{c_3Z^2}{A^{1/3}}-\frac{c_4Z^2}{A}+\Delta
\label{ae2}
\end{equation}
where
\begin{eqnarray}
c_1&=&15.677\left[1-1.79{\left(\frac{A-2Z}{A}\right)}^2\right]\nonumber\\
c_2&=&18.56\left[1-1.79{\left(\frac{A-2Z}{A}\right)}^2\right]
\end{eqnarray}
\noindent $c_3=0.717$ and $c_4=1.2113$.  The first term on the
r.h.s of Eq. \ref{ae2},  i.e., $c_1A$ is the sum of the volume
energy term which is proportional to the mass number $A$ and the
volume-asymmetry energy term which is proportional to
${(A-2Z)}^2/A$. The second term $c_2A^{2/3}$ is the sum of the
surface energy term being proportional to $A^{2/3}$ and the
surface asymmetry energy term proportional to $I^2A^{2/3}$ where
$I$ equals ${(A-2Z)}/A$. The third term $c_3Z^2/{A^{1/3}}$ is the
direct sharp-surface Coulomb energy whereas $c_4Z^2/A$ gives the
surface-diffuseness correction to the direct Coulomb energy. The
shell correction is not included in the binding energy formula
since in fission dynamics of excited nuclei, both the parent and
the daughter nuclei are hot, and shell corrections need not be
considered. $\Delta$ gives the pairing energy correction and is
given by the following formulas.
\begin{eqnarray}
\Delta &=& -\frac{11}{\sqrt{A}} \hspace{1cm}\mbox{for even-even nuclei},\nonumber\\
       &=& 0  \hspace{1.9cm}  \mbox{for even-odd or odd-even nuclei}, \nonumber\\
       &=& +\frac{11}{\sqrt{A}} \hspace{1cm}\mbox{for odd-odd
       nuclei}.
\end{eqnarray}

\newpage
\addcontentsline{toc}{chapter}{Appendix F: Brief description of
the computer codes}
\chapter*{Appendix F}
\section*{ Brief description of the
computer codes}
 \setcounter{equation}{0}
\def\theequation{F.\arabic{equation}}
\def\thefigure{F.\arabic{figure}}

The different computer codes developed and used in the thesis will
be briefly mentioned here. The codes were all written in Fortran
language and the computational work was done using ES-40 server
with alpha CPU (21264A) having clock speed 667 MHz.
\subsection*{CHAOTICITY}
The code named by us as ``CHAOTICITY'' was obtained from Professor
J. Blocki of Institute for Nuclear Research, Swierk, Poland. This
code calculates the chaos factor of different nuclear shapes
starting from spherical to the scission configuration. The method
is based on calculating the Lyapunov exponent of a large number of
classical trajectories (typically 10000 or more), from which the
chaos factor is extracted. The procedure for calculation of
Lyapunov exponent of a trajectory by following it classically in
time is explained in section 2.4.3. As part of the thesis work,
the above code was modified suitably to incorporate the Brack
shape parametrization which is used in our model to represent a
hot fissioning nuclei. The initial code as obtained was for volume
sampling of trajectories. It was modified by us so that surface
sampling of trajectories is possible which is required for our
purpose. The chaos factor is calculated for a wide range of the
dynamical coordinates $c$ and $h$. Values of $c$ range from 0.6 to
2.1 in steps of 0.01, whereas for each value of $c$ , $h$ ranges
from -1.5 to 1.5 in steps of 0.1. A typical calculation of chaos
factor(for 1000 trajectories) for a particular value of $c$ and
$h$ takes about 2.22 minutes of computer time.

\subsection*{POTFOLD}
The code named as ``POTFOLD"(POTential FOLDing) was developed as
part of this thesis work. This code calculates the potential
energy of nuclei of different mass and atomic number as function
of the deformation coordinates $c$ and $h$.  ``POTFOLD" includes
both the nuclear and the coulomb part of the potential. The
potential landscape in $(c,h)$ coordinates is generated from the
finite-range liquid drop model where we calculate the generalized
nuclear energy by double folding the uniform density with
Yukawa-plus-exponential potential. The Coloumb energy is obtained
by double folding another Yukawa function with the density
distribution. The detailed description of the potential used can
be found in chapter 2 and the techniques used for simplifying and
solving the integrals is given in Appendix A. The potential is
calculated for the same range of $c$ and $h$ as the chaos factor
mentioned above. A typical calculation of potential for a nuclei
of mass number 224 for a particular value of $c$ and $h$ takes
about 8.8 seconds of computer time.
\subsection*{FISSWDTH}
The calculation of prescission neutron multiplicity and fission
probability proceeds through two stages. In the first part the
fission width of a nucleus is calculated dynamically which is
required for the next part of the calculation. This code named as
``FISSWDTH"(FISSion WiDTH) is developed to calculate the dynamical
fission rate or fission width of a nucleus for a particular
angular momentum and excitation energy, by numerically solving the
Langevin equation. This code considers only one mode of decay of
the compound nucleus i.e, fission, and does not include particle
or gamma emissions. The fission widths calculated from this code
for different values of energy and angular momentum is used for
the parametric representation of fission width which serve as
input for the statistical branch of the main code(to be described
next). The chaos factor and potential which are calculated using
the codes mentioned above are used as input for this code.  The
other inputs required for the calculation of fission width like
the friction and inertia are calculated within this code
``FISSWDTH". The different steps involved in the procedure for
calculation of fission width is described in section 3.2. A
Langevin trajectory is followed in time till it reaches the
fission fate and the calculation is repeated for a large number
(typically 100,000) of trajectories so that fluctuations in the
steady value is minimised
 (better averaging is possible) and a reliable value of fission
width can be extracted from its saturation value. A typical
calculation of fission width for thorium nucleus(A=224) for
angular momentum  $30\hbar$ and temp $2$ MeV takes about 4.72
hours of computer time.

\subsection*{DYSTCNF}
This code called ``DYSTCNF" is developed as part of the thesis
work to calculate the prescission neutron multiplicity and fission
probability of different nuclei. The name stands for `DYnamical
plus STatistical Code for Nuclear Fission'. As the name implies,
the code uses a combined dynamical plus statistical model where
light particle evaporation is coupled with Langevin dynamics
followed by a statistical branch based on Monte Carlo cascade
procedure. The different steps of the calculation is discussed
thoroughly in section 4.2. The mass and atomic number of target
and projectile, laboratory energy of the projectile and $Q$ value
of the reaction are given as inputs to this code. The calculation
in the code starts from the formation of a compound nucleus from
the fusion of target and projectile, the total excitation energy
of the fused nuclei being calculated from the projectile energy
and $Q$ value. The initial conditions for the dynamical
coordinates as well as the angular momenta are obtained by
sampling suitable distribution functions which are suitably
incorporated in different subprograms of the code. The potential
values being supplied by the code 'POTFOLD', a suitable
minimization routine finds the minimum in the two dimensional
(c,h) valley.  Different subprograms are developed in the main
code to calculate the free energy, inertia, level density
parameter, moment of inertia of the nucleus for different values
of the dynamical coordinates $c$ and $h$. Wall and Window friction
values for different nuclear shapes calculated in a subprogram is
suitably combined with the values of the chaos factor obtained
from the code 'CHAOTICITY' to generate the values of the
chaos-weighted wall friction values, the testing of which in
fission dynamics being the main emphasis of the thesis. The
formulas and mathematical expressions for different input
parameters calculated in this code are given in chapter 2 of the
thesis. The particle emission widths required for the evaporation
probability of the particles are calculated within this code while
the fission widths are obtained from the separately developed code
'FISSWDTH'. The neutron widths for different mass number(A),
angular momentum(l) and excitation energy($E^*$) of the parent
nucleus are calculated and stored by constructing a three
dimensional matrix wdthn(i,j,k) where the indices i, j and k stand
for mass number, angular momentum and excitation energy
respectively. The required values are obtained from this matrix by
interpolation. The energy of the emitted particles obtained by
sampling the energy spectrum of each type and the binding energy
of the emitted particles calculated from the mass of the parent
and daughter nuclei using the global liquid-drop parameters of
Myers and Swiatecki (Appendix E) are taken care of in different
subroutines of the main code. The different time steps like $\tau$
(time step for numerical integration of Langevin equations),
$\tau_{eq}$ (time from where the statistical part of the code
starts) and $t_{max}$(total time up to which a trajectory is
followed) are all given as inputs to this code. The calculation in
the code terminates with each trajectory reaching the scission
point or ending up as an evaporation residue. A few trajectories
may not reach either of these fates within the total specified
time $t_{max}$ (which is chosen to be sufficiently large) and the
calculation is stopped there for those trajectories. The whole
calculation is repeated for a large ensemble of trajectories and
the final outputs like the prescission neutron multiplicity,
fission probability or evaporation residue probability are
extracted by suitably averaging the numbers obtained from the
ensemble. A typical calculation for $^{224}{Th}$ for 20000
trajectories for projectile energy 100MeV takes about 60 hours  of
computer time.
 A
schematic sketch of the calculational procedure used in this code
is given by the flow chart in Appendix G.

\newpage
\addcontentsline{toc}{chapter}{Appendix G: Schematic sketch of the
calculational procedure}
\chapter*{Appendix G}
\section*{ Schematic sketch of the
calculational procedure }
 \setcounter{equation}{0}
 \setcounter{figure}{0}
\def\theequation{G.\arabic{equation}}
\def\thefigure{G.\arabic{figure}}

\begin{figure}[h!]
\centerline{\psfig{figure=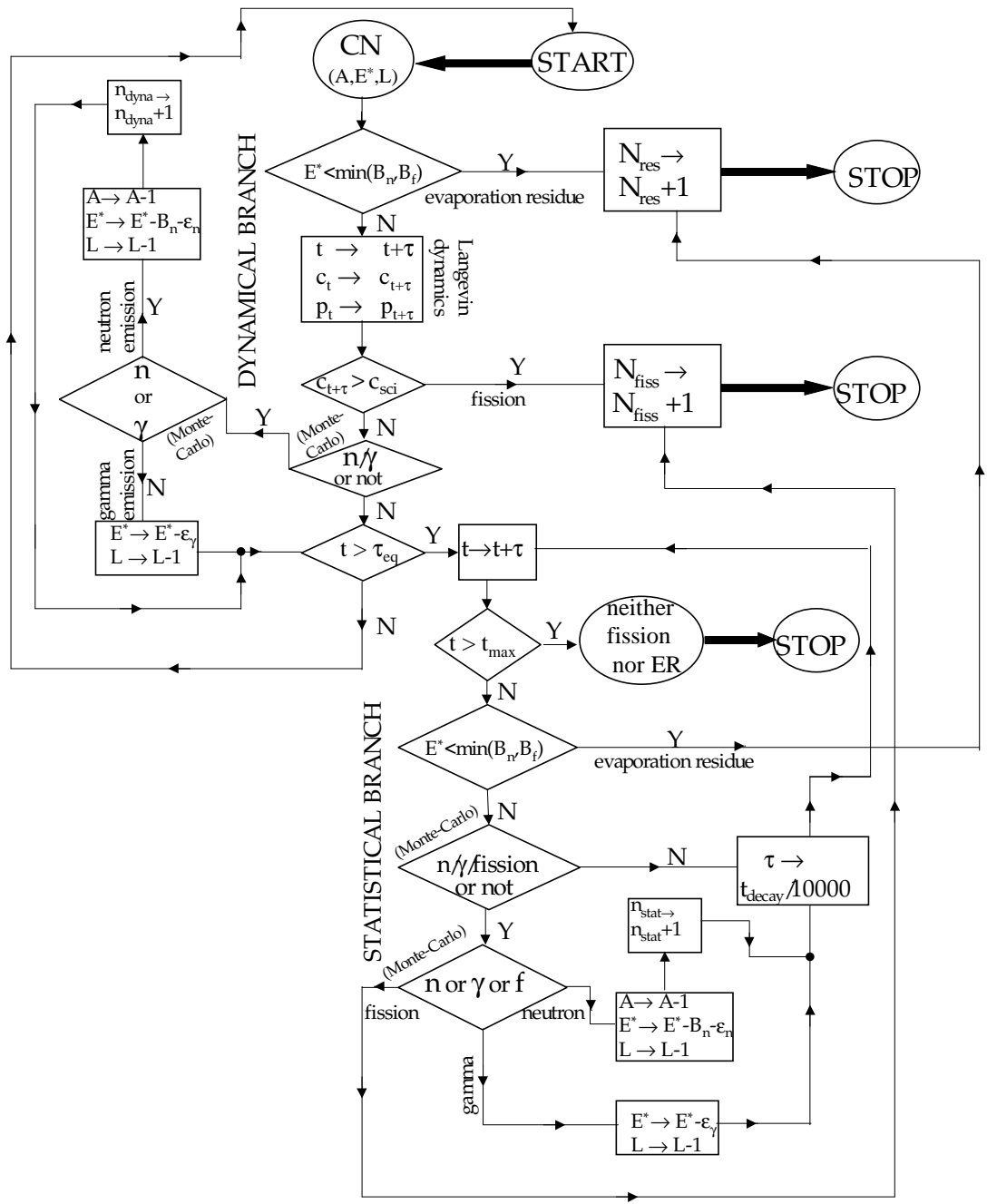,width=16.5cm}}
\caption{Flow chart of the calculational procedure } \label{appg}
\end{figure}

The flow diagram (Fig. \ref{appg}) displays schematically the
logical sequence of the actual calculations described in the
combined dynamical plus statistical model in chapter 4, which
essentially constitutes the code ``DYSTCNF" referred in Appendix F
. The main steps of the algorithm
 used in the calculation are shown in the flow chart,
  others being omitted to make the diagram look simple. The different
symbols and logical steps shown in the figure is being explained
here. The diagram explains the different steps followed by a
trajectory from a compound nucleus till it end up as an
evaporation residue or reaches the fission criteria. The flow
chart starts with a compound nucleus (CN) of mass $A$ with a
particular angular momentum $L$ and excitation energy $E^*$. If
the excitation energy $E^*$ is less than minimum of $B_n$(binding
energy of neutron) and $B_f$(fission barrier), then the compound
nucleus forms an evaporation residue and adds to $N_{res}$ which
gives the total number of evaporation residue formed. The
calculation proceeds if this condition is not satisfied. The
dynamical coordinate and momentum at time $t$ being given by $c_t$
and $p_t$ respectively, evolve in time following Langevin
equations and change over to $c_{t+\tau}$ and $p_{t+\tau}$, $\tau$
being the time step of integration of the Langevin equations.
After each time step, the criteria for scission is checked, i.e,
whether the coordinate $c$ reaches the scission point $c_{sci}$.
The condition being satisfied, it adds to the total number of
fission events which is given by $N_{fiss}$, and the calculation
stops there. If the fission condition is not reached, the
calculation for the trajectory proceeds to check for the emission
of light particles and gamma. The emission of only neutron and
gamma is shown in the flow chart. The Monte Carlo selection
procedure (described in section 4.2.3) is used to decide whether
any emission takes place as well as the type of emission (neutron
or $\gamma$). The energy of the emitted particles is also
calculated by Monte-Carlo procedure (not shown in the flow chart),
using the integrand of the formula for the corresponding decay
width as weight function. The mass number, excitation energy and
angular momentum are adjusted properly after each emission of
neutron or gamma. $\varepsilon$ gives the energy of the emitted
particle. After checking for particle emission, the total elapsed
time $t$ is compared with $\tau_{eq}$. When the total time of
calculation in the dynamical branch exceeds the equilibration time
$\tau_{eq}$ and the fission criteria is not yet reached, one
switches over to the statistical branch as explained in section
4.2.4.  In the statistical branch, the possibility of any
particle/$\gamma$ emission or fission and the kind of emission at
each time step is checked following the same criteria as in the
dynamical branch. The fission width obtained in a separate
calculation is used as input for the statistical branch and
compared  with the particle/$\gamma$ widths to decide the
possibilities. The time step $\tau$ in the statistical branch is
redefined after each step as $\tau=\tau_{decay}/10000$, where
$\tau_{decay}=1/\Gamma_{tot}$, and $\Gamma_{tot}=
\Gamma_n+\Gamma_{\gamma}+\Gamma_f$.
 The final fate of a
trajectory (fission or evaporation residue) may be reached from
either the dynamical part or the statistical one, each adding up
to the respective numbers $N_{fiss}$ or $N_{res}$. The total
number of neutrons emitted in the dynamical branch ($n_{dyna}$) is
added to that emitted from the statistical branch ($n_{stat}$) to
get the total number of neutrons $N_n$. This number is divided by
the total number of fission events $N_{fiss}$ to get the
prescission neutron multiplicity $n_{pre}$. If a trajectory
neither fissions nor reaches the evaporation residue within the
total time $t_{max}$ specified for the combined dynamical plus
statistical calculation, then the calculation is stopped there
with the fate of the trajectory still undecided. However,
$t_{max}$ is taken to be sufficiently long so that the number of
such ``undecided" trajectories is statistically insignificant.


\end{document}